%% file: itc-main.tex
\documentclass[11pt]{article}

\input{preamble}
\usepackage[final]{changes}

\usepackage[strings]{underscore}

\begin{document}

\begin{titlepage}

\title{On the Security of Proofs of Sequential Work in a Post-Quantum World}
\date{\today}

\author{
Jeremiah Blocki\\
Purdue University\\
\texttt{jblocki@purdue.edu}
\and
Seunghoon Lee\\
Purdue University\\
\texttt{lee2856@purdue.edu}
\and
Samson Zhou\\
Carnegie Mellon University\\
\texttt{samsonzhou@gmail.com}
}

\maketitle
\input{abstract}
\setcounter{page}{0}
\end{titlepage}

\input{intro-prelims}
\input{pqrom}
\input{finding-h-seq}
\input{reduction}

\input{conclusion}
\def\shortbib{0}
\bibliographystyle{alpha}
\bibliography{myref}
\FullVersion{
\appendix
\input{seq-bounded}
\input{stddecomp}
\newpage
\input{algorithms}
\newpage
\input{missing-proofs}
\input{usefulbounds}
}{}

\end{document}

%% file: preamble.tex
\usepackage{makeidx}
\usepackage[nointegrals]{wasysym}
\usepackage{amsmath,amssymb}
\usepackage{amsfonts}
\usepackage{dsfont}
\usepackage{latexsym}
\usepackage{subcaption}
\usepackage{graphicx}
\usepackage{fancybox}
\usepackage{hyperref}
\usepackage{thmtools}
\usepackage{color}
\usepackage{tikz}
\usetikzlibrary{shadings}
\usetikzlibrary{arrows,patterns,arrows.meta}
\usepackage{xcolor}
\usepackage[linesnumbered,ruled,vlined,noend]{algorithm2e}

\SetCommentSty{mycommfont}

\usepackage[noend]{algpseudocode}
\usepackage[framemethod=tikz]{mdframed}
\usepackage{xspace}
\usepackage{pgfplots}
\usepackage{framed}
\pgfplotsset{compat=1.5}
\usepackage{standalone}

\usepackage{etoolbox}%
\newcommand{\FullVersion}[2]{%
	\ifstrequal{full}{full}
	{#1}
	{#2}
}

\newif\iflipcs

\iflipcs

\else

\usepackage{fullpage}
\newtheorem{theorem}{Theorem}[section]
\newtheorem{corollary}[theorem]{Corollary}
\newtheorem{lemma}[theorem]{Lemma}

\newtheorem{definition}[theorem]{Definition}

\newtheorem{remark}[theorem]{Remark}
\newtheorem{example}[theorem]{Example}

\newenvironment{proof}{\noindent{\bf Proof : \ }}{\hfill$\Box$\par\medskip}

\fi

\newenvironment{proofof}[1]{\begin{trivlist} \item {\bf Proof
#1:~~}}
  {\qed\end{trivlist}}

\newcommand{\namedref}[2]{\hyperref[#2]{#1~\ref*{#2}}}
\newcommand{\thmlab}[1]{\label{thm:#1}}
\newcommand{\thmref}[1]{\namedref{Theorem}{thm:#1}}
\newcommand{\lemlab}[1]{\label{lem:#1}}
\newcommand{\lemref}[1]{\namedref{Lemma}{lem:#1}}
\newcommand{\claimlab}[1]{\label{claim:#1}}
\newcommand{\claimref}[1]{\namedref{Claim}{claim:#1}}
\newcommand{\corlab}[1]{\label{cor:#1}}
\newcommand{\corref}[1]{\namedref{Corollary}{cor:#1}}
\newcommand{\seclab}[1]{\label{sec:#1}}
\newcommand{\secref}[1]{\namedref{Section}{sec:#1}}
\newcommand{\applab}[1]{\label{app:#1}}
\newcommand{\appref}[1]{\namedref{Appendix}{app:#1}}

\newcommand{\figlab}[1]{\label{fig:#1}}
\newcommand{\figref}[1]{\namedref{Figure}{fig:#1}}
\newcommand{\alglab}[1]{\label{alg:#1}}
\renewcommand{\algref}[1]{\namedref{Algorithm}{alg:#1}}

\newcommand{\deflab}[1]{\label{def:#1}}
\newcommand{\defref}[1]{\namedref{Definition}{def:#1}}

\def \lab    {\mdef{\mathsf{lab}}}

\usetikzlibrary {positioning,chains,fit,shapes,calc,snakes}

\tikzset{
    position/.style args={#1:#2 from #3}{
        at=(#3.#1), anchor=#1+180, shift=(#1:#2)
    }
}

\definecolor{purduedigitalheadlinegold}{HTML}{98700D}
\definecolor{purduecampusgold}{HTML}{C28E0E}
\definecolor{purduecoalgray}{HTML}{4D4038}
\definecolor{purdueevertrueblue}{HTML}{5B6870}
\definecolor{purduemoondustgray}{HTML}{BAA892}
\definecolor{purdueslayterskyblue}{HTML}{6E99B4}
\definecolor{mahogany}{rgb}{0.75, 0.25, 0.0}
\definecolor{darkblue}{rgb}{0.0, 0.0, 0.55}
\definecolor{darkpastelgreen}{rgb}{0.01, 0.75, 0.24}
\definecolor{darkgreen}{rgb}{0.0, 0.2, 0.13}
\definecolor{darkgoldenrod}{rgb}{0.72, 0.53, 0.04}
\definecolor{forestgreen}{rgb}{0.13, 0.55, 0.13}
\definecolor{darkred}{rgb}{0.55, 0.0, 0.0}
\definecolor{tangocolordarkchameleon}{HTML}{4E9A06}

\newcommand\norm[1]{\left\lVert#1\right\rVert}
\newcommand{\COMMENTED}[1]{{}}

\newcommand{\PPr}[1]{\ensuremath{\mathbf{Pr}\left[#1\right]}}

\renewcommand{\o}{\mathbb{O}}

\def \pROM    {\mdef{\mathsf{pROM}}}
\def \indeg    {\mdef{\mathsf{indeg}}}

\def \unique    {\mdef{\mathsf{UNIQUE}}}

\def \bad    {\mdef{\mathsf{BAD}}}
\def \path    {\mdef{\mathsf{PATH}}}
\def \collide    {\mdef{\mathsf{COLLIDE}}}
\def \lucky    {\mdef{\mathsf{LUCKY}}}
\def \pre    {\mdef{\mathsf{PRE}}}
\def \ptr    {\mdef{\mathsf{PTR}}}
\def \gptr    {\mdef{\mathsf{gPTR}}}
\def \spath    {\mdef{\widetilde{\mathsf{PATH}}}}
\def \slucky    {\mdef{\widetilde{\mathsf{LUCKY}}}}
\def \scontain    {\mdef{\widetilde{\mathsf{Contain}}}}
\def \sbad    {\mdef{\widetilde{\mathsf{BAD}}}}
\def \last    {\mdef{\mathsf{LAST}}}

\def \rom    {\mdef{\mathsf{ROM}}}
\def \qrom    {\mdef{\mathsf{qROM}}}
\def \colsubtree    {\mdef{\mathtt{ColorSubTree}}}
\def \coloring    {\mdef{\mathtt{Color}}}
\def \posw    {\mdef{\mathsf{PoSW}}}
\def \colmt    {\mdef{\mathtt{ColoredMT}}}
\def \red    {\mdef{\textcolor{red}{\mathsf{red}}}}
\def \green    {\mdef{\textcolor{forestgreen}{\mathsf{green}}}}

\def \pqrom    {\mdef{\mathsf{pqROM}}}

\def \swap    {\mdef{\mathsf{Swap}}}
\def \contain    {\mdef{\mathsf{Contain}}}
\def \substring    {\mdef{\mathsf{Substring}}}
\def \hseq    {\mdef{\mathsf{HSeq}}}
\def \CPhsO    {\mdef{\mathsf{CPhsO}}}

\def \P    {\mdef{\mathcal{P}}}

\def \niPoSW    {\mdef{\mathsf{niPoSW}}}
\def \Solve    {\mdef{\mathsf{Solve}}}
\def \Verify    {\mdef{\mathsf{Verify}}}
\def \GetChal    {\mdef{\mathsf{GetChallenge}}}
\def \MTReveal    {\mdef{\mathsf{MT.Reveal}}}

\def \negl    {\mdef{\textsf{negl}}}

\newcommand{\A}{\mathcal{A}}

\renewcommand{\H}{\mathcal{H}}

\renewcommand{\O}[1]{\ensuremath{\mathcal{O}\left(#1\right)}}
\newcommand{\qvec}[1]{\ensuremath{|#1\rangle}}

\newcommand{\eps}{\epsilon}
\newcommand{\moment}{L_2}

\def \adv    {\mdef{\mathcal{A}}}

\def \increase    {\mdef{\mathsf{Increase}}}

\def \stddecomp    {\mdef{\mathsf{StdDecomp}}}

\def \scphso    {\mdef{\mathsf{SCPhsO}}}
\def \cphso    {\mdef{\mathsf{CPhsO}}}
\def \CPhsO    {\mdef{\mathsf{CPhsO}}}

\newcommand{\strlen }{\lambda}

\newcommand{\D}{\mathcal{D}}

\newcommand{\mdef}[1]{{\ensuremath{#1}}\xspace}  
\newcommand{\myset}[1]{\mdef{\mathbb{#1}}}       
\newcommand{\myfunc}[1]{\mdef{\mathsf{#1}}}      

\DeclareMathOperator*{\polylog}{polylog}

\newcommand{\superscript}[1]{\ensuremath{^{\mbox{\tiny{\textit{#1}}}}}\xspace}
\def \th {\superscript{th}}     
\def \etal{{\it et~al.}}

\def \N        {\mdef{\mathbb{N}}}                   
\def \I        {\mdef{\myset{I}}}                    

\def \negl     {\mdef{\myfunc{negl}}}                

\newcommand{\ignore}[1]{}

\newif\ifnotes\notestrue
\ifnotes
\newcommand{\samson}[1]{\textcolor{purple}{{\bf (Samson:} {#1}{\bf ) }} \marginpar{\tiny\bf
             \begin{minipage}[t]{0.5in}
               \raggedright SZ:
            \end{minipage}}}            
            \newcommand{\jeremiah}[1]{\textcolor{red}{{\bf (Jeremiah:} {#1}{\bf ) }} \marginpar{\tiny\bf
             \begin{minipage}[t]{0.5in}
               \raggedright JB:
            \end{minipage}}}  	
            \newcommand{\seunghoon}[1]{\textcolor{blue}{{\bf (Seunghoon:} {#1}{\bf ) }} \marginpar{\tiny\bf
             \begin{minipage}[t]{0.5in}
               \raggedright SL:
            \end{minipage}}}						
\else
\newcommand{\samson}[1]{}
\newcommand{\jeremiah}[1]{}
\newcommand{\seunghoon}[1]{}
\fi

\makeatletter
\renewcommand*{\@fnsymbol}[1]{\textcolor{mahogany}{\ensuremath{\ifcase#1\or *\or \dagger\or \ddagger\or
 \mathsection\or \triangledown\or \mathparagraph\or \|\or **\or \dagger\dagger
   \or \ddagger\ddagger \else\@ctrerr\fi}}}
\makeatother

\providecommand{\email}[1]{\href{mailto:#1}{\nolinkurl{#1}\xspace}}

\definecolor{mahogany}{rgb}{0.75, 0.25, 0.0}
\definecolor{darkblue}{rgb}{0.0, 0.0, 0.55}
\definecolor{darkpastelgreen}{rgb}{0.01, 0.75, 0.24}
\definecolor{darkgreen}{rgb}{0.0, 0.2, 0.13}
\definecolor{darkgoldenrod}{rgb}{0.72, 0.53, 0.04}
\definecolor{darkred}{rgb}{0.55, 0.0, 0.0}
\definecolor{tangocolordarkchameleon}{HTML}{4E9A06}
\hypersetup{
     colorlinks   = true,
     citecolor    = mahogany,
		 linkcolor		= olive
}

%% file: abstract.tex
\begin{abstract}
A Proof of Sequential Work (PoSW) allows a prover to convince a resource-bounded verifier that the prover invested a substantial amount of sequential time to perform some underlying computation. PoSWs have many applications including time-stamping, blockchain design, and universally verifiable CPU benchmarks. 
Mahmoody, Moran, and Vadhan (ITCS 2013) gave the first construction of a PoSW in the random oracle model though the construction relied on expensive depth-robust graphs. In a recent breakthrough, Cohen and Pietrzak (EUROCRYPT 2018) gave an efficient PoSW construction that does not require expensive depth-robust graphs. 

\ignore{In each of these constructions, the prover commits to a labeling of a directed acyclic graph $G$ with $N$ nodes and the verifier audits the prover by checking that a small subset of labels are locally consistent, e.g., $\ell_v = \H(v,\ell_{v_1},\ldots,\ell_{v_\delta})$, where $v_1,\ldots,v_\delta$ denote the parents of node $v$. Provided that the graph $G$ satisfies certain structural properties (e.g., depth-robustness) security can be established by arguing that any prover who does not produce a long $\H$-sequence will fail the audit with high probability. An $\H$-sequence $x_0,x_1\ldots x_T$ has the property that $\H(x_i)$ is a substring of $x_{i+1}$ for each $i$, i.e., we can find strings $a_i,b_i$ such that $x_{i+1} = a_i || \H(x_i) || b_i$. \seunghoon{Do we need all the details in the abstract?}
}

In the classical parallel random oracle model, it is straightforward to argue that any successful PoSW attacker must produce a long $\H$-sequence and that any malicious party running in sequential time $T-1$ will fail to produce an $\H$-sequence of length $T$ except with negligible probability. In this paper, we prove that any quantum attacker running in sequential time $T-1$ will fail to produce an $\H$-sequence except with negligible probability -- even if the attacker submits a large batch of quantum queries in each round. The proof is substantially more challenging and highlights the power of Zhandry's recent compressed oracle technique (CRYPTO 2019). We further extend this result to establish post-quantum security of a non-interactive PoSW obtained by applying the Fiat-Shamir transform to Cohen and Pietrzak's efficient construction (EUROCRYPT 2018).

\end{abstract}

%% file: intro-prelims.tex
\section{Introduction}
As we make progress towards the development of quantum computers, it is imperative to understand which cryptographic primitives can be securely and efficiently instantiated in a post-quantum world. 
In this work, we consider the security of proofs of sequential work against quantum adversaries. 

A proof of sequential work (PoSW)~\cite{ITCS:MahMorVad13,EC:CohPie18,AbusalahKKPW19,DottlingLM19} is a protocol for proving that one spent significant sequential computation work to validate some statement $\chi$. 
One motivation for a proof of sequential work is in time-stamping, e.g., if Bob can produce a valid proof $\pi_\chi$ that $N$ sequential steps were spent to validate $\chi$, then Bob can prove that he must have known about $\chi$ at least time $\Omega(N)$ seconds in the past. 
A verifier should be able to validate the proof $\pi_\chi$ quickly, i.e., in time $\polylog(N)$. 

Mahmoody et al.~\cite{ITCS:MahMorVad13} gave the first construction of a proof of sequential work in the random oracle model. 
Their construction was based on labeling a depth-robust graph, i.e., given a random oracle $\H$ and a directed acyclic graph $G = (V=[N],E)$ with $N$ nodes and an initial input $x$, we can compute labels $\ell_1,\ldots,\ell_N$, where the label of the source node is $\ell_1 = \H(\chi, 1, x)$ and an internal node $v$ with parents $v_1,\ldots,v_\delta$ has label $\ell_v = \H(\chi, v, \ell_{v_1},\ldots, \ell_{v_{\delta}})$. 

The prover commits to labels $\ell_1',\ldots, \ell_N'$ (a cheating prover might commit to the wrong labels) and then the verifier selects a random subset $S \subset [N]$ of $|S|=c$ challenge nodes. 
For each challenge node $v \in S$ with parents $v_1,\ldots,v_{\delta}$, the prover reveals $\ell_v'$ along with $\ell_{v_1}',\ldots, \ell_{v_{\delta}}'$ and the verifier checks that $v$ is locally consistent, i.e., $\ell_v' = \H(\chi,v, \ell_{v_1}',\ldots, \ell_{v_{\delta}}')$. 
If we let $R$ denote the subset of locally inconsistent nodes, then the verifier will accept with probability at most $\left(1-|R|/N \right)^c$. 

Mahmoody et al.~\cite{ITCS:MahMorVad13} selected $G$ such that $G$ was $\epsilon$-extremely depth-robust\footnote{A DAG $G$ is said to be \emph{$\epsilon$-extremely depth-robust} if it is $(e,d)$-depth robust for any $e,d>0$ such that $e+d\leq (1-\epsilon)N$ where $N$ is the number of nodes in $G$. Recall that a DAG $G=(V,E)$ is $(e,d)$-depth robust if for any subset $S\subseteq V$ with $|S|\leq e$ there exists a path of length $d$ in $G-S$.}, meaning that for any set $R \subseteq [N]$ of locally inconsistent nodes, there is a directed path of length $T+1=(1-\epsilon)N-R$. 
This path $P = v_0,\ldots, v_T$ corresponds to an $\H$-sequence of length $T$ where an $\H$-sequence is any sequence of strings $x_0,\ldots, x_T$ with the property that $\H(x_{i})$ is a substring of $x_{i+1}$ for each $i < T$. 
Note that the labels $\ell_{v_0}',\ldots, \ell_{v_T}'$ have this property. 
In the classical parallel random oracle model (\pROM), it is relatively straightforward to prove that any algorithm running in $T-1$ rounds and making at most $q$ queries in total fails to produce an $\H$-sequence except with probability $\tilde{\Omega}\left(q^2 2^{-\lambda} \right)$ when $\H:\{0,1\}^{\delta \strlen}\to\{0,1\}^{\strlen}$ outputs binary strings of length $\strlen$~\cite{EC:CohPie18}.

The $\epsilon$-extreme depth-robust graphs used in the construction of Mahmoody et al.~\cite{ITCS:MahMorVad13} were quite expensive, having indegree $\delta = \tilde{\Omega}(\log N)$. 
Alwen et al.~\cite{EC:AlwBloPie18} showed how to construct $\epsilon$-extreme depth-robust graphs with indegree just $\O{\log N}$ though the hidden constants were quite large. 
Cohen and Pietrzak \cite{EC:CohPie18} gave an efficient (practical) construction that avoids depth-robust graphs entirely by cleverly modifying  the Merkle tree structure to obtain a graph $G$ on $N=2^{n+1}-1$ nodes\footnote{The graph $G$ is ``weighted'' depth robust. 
In particular, there is a weighting function $w:V \rightarrow \mathbb{R}_{\geq 0}$ with the property that $\sum_v w(v) \in \O{N \log N}$ and for any subset $S \subseteq V$ with sufficiently small weight $\sum_{v \in S} w(v) \leq cN$ the DAG $G-S$ contains a path of length $\Omega(N)$.}, for any integer $n\geq 1$. 

Both proofs of sequential work can (optionally) be converted into a non-interactive proof by applying the Fiat-Shamir paradigm \added{\cite{C:FiaSha86}}, i.e., given a commitment $c'$ to labels $\ell_1',\ldots, \ell_N'$ we can use public randomness $r= \H(\chi,N+1, c')$ to sample our set of challenge nodes $S$. The non-interactive version could be useful in cases where a prover wants to silently timestamp a statement $\chi$ without even signaling that s/he might have a statement important enough to timestamp, e.g., a researcher who believes they might resolved a famous open problem may wish to timestamp the discovery without signaling the community until s/he carefully double checks the proof.


In all of the above constructions, security relies on the hardness of computing $\H$-sequences of length $T$ in sequential time $T-1$. 
While this can be readily established in the classical parallel random oracle model, proving that this task is in fact hard for a quantum attacker is a much more daunting challenge. 
As Boneh et al.~\cite{AC:BDFLSZ11} pointed out, many of the convenient properties (e.g., extractability, programmability, efficient simulation, rewinding, etc.) that are used in classical random oracle security proofs no longer apply in the quantum random oracle model (\qrom). 
An attacker in the (parallel) quantum random oracle model is able to submit entangled queries, giving the attacker much more power. 
For example, given $y$ a quantum attacker can find a preimage $x'$ such that $\H(x')=y$ with just $\O{2^{ \strlen/2}}$ quantum random oracle queries using Grover's algorithm. 
By contrast, a classical attacker would need at least $\Omega(2^{\strlen})$ queries to a classical random oracle. 
Similarly, a quantum attacker can find hash collisions with at most $\O{2^{\strlen/3}}$ queries, while a classical attacker requires $\Omega(2^{\strlen/2})$ queries. 
In this paper, we explore the post-quantum security of proofs of sequential work in the parallel quantum random oracle model. We aim to answer the following questions: 
\begin{center}
\begin{quote}
\emph{Can a quantum attacker running in $T-1$ sequential rounds produce an $\H$-sequence of length $T$? }
\end{quote}
\begin{quote}
\emph{Can a quantum attacker running in time $T = (1-\alpha)N$ produce a valid non-interactive proof of sequential work with non-negligible probability? }
\end{quote}
\end{center}


\subsection{Our Contributions} 
We answer these questions in the negative, thus confirming the security of proof of sequential work schemes in a post-quantum world. 
We first prove that any quantum attacker making $N-1$ rounds of queries cannot produce an $\H$-sequence of length $N$, except with negligible probability. 

\begin{definition}[$\H$-Sequence]
An \emph{$\H$-sequence} $x_0,x_1,\ldots,x_s\in\{0,1\}^*$ satisfies the property that for each $1\le i\le s$, there exist $a,b\in\{0,1\}^*$ such that $x_i=a|| \H(x_{i-1})|| b$. 
For indexing reasons, we say such an $\H$-sequence has length $s$ (even though there are $s+1$ variables $x_i$). 
\end{definition}

In the classical random oracle model (\rom), it is straightforward to argue that any PoSW prover must find a long $\H$-sequence to pass the audit phase with non-negligible probability. 
Thus, this result already provides compelling evidence that proofs of sequential work are post-quantum secure in the parallel random oracle model. 

Next we consider a non-interactive proof of sequential work  applying the Fiat-Shamir transform to the efficient construction of Cohen and Pietrzak~\cite{EC:CohPie18}, and we prove that this construction is secure in the quantum parallel random oracle model. In particular, we show that any attacker running in sequential time $T=(1-\alpha)N$ will fail to produce a valid proof $\pi_\chi$ for any statement $\chi \in \{0,1\}^{\strlen}$. 

While Cohen and Pietrzak~\cite{EC:CohPie18} proved analogous results in the classical random oracle model, we stress that from a technical standpoint, proving security in the quantum random oracle model is significantly more challenging. In general, there is a clear need to develop new techniques to reason about the security of cryptographic protocols in the quantum random oracle model. 
Most of the techniques that are used in classical random oracle model do not carry over to the (parallel) quantum random oracle model~\cite{AC:BDFLSZ11}. 
For example, if we are simulating a classical attacker, then we can see (extract) all of the random oracle queries that the attacker makes, while we cannot observe a quantum query without measuring it, which would collapse the attacker's quantum state might significantly alter the final output. 

\paragraph*{Warm-Up Problem: Iterative Hashing.} 
As a warm-up, we first prove an easier result in \thmref{thm:seq:main} that an attacker cannot compute $\H^N(x)$ in sequential time less than $N-1$ in the parallel quantum random oracle model, where a similar result was previously proved by Unruh \cite{Unruh15} in the (non-parallel) quantum random oracle model. 
Along the way we highlight some of the key challenges that make it difficult to extend the proof to arbitrary $\H$-sequences. 

\begin{restatable}{theorem}{thmseqmain}
\thmlab{thm:seq:main}
Given a hash function $\H:\{0,1\}^*\to\{0,1\}^{\deleted{4}\strlen}$ and a random input $x$, any quantum attacker that makes up to $q$ queries in each of $N-1$ sequential steps can only compute $\H^N(x)$ with probability at most $\frac{N^2}{2^{\deleted{4}\strlen}}+\frac{1}{2^{\deleted{4}\strlen}-N}+\sqrt{\frac{\replaced{48}{160}\strlen N^4q^2T}{2^{\replaced{\strlen/2}{2\strlen}}}}$ in the quantum parallel random oracle model.
\end{restatable}

The proof of \thmref{thm:seq:main} is straightforward and we defer it to \FullVersion{\appref{app:seq}}{the full version}. Intuitively, iteratively computing $H^N(x)$ induces an $\H$-sequence $x_0,x_1, \ldots, x_n$ with $x_0=x$, $x_N= \H^N(x)$ and $x_{i+1}=\H(x_i)$. One can easily define a sequence of indistinguishable hybrids where in the last hybrid the final output $x_N= \H^N(x) = \H(x_{N-1})$ is information theoretically hidden from the attacker. In general, in hybrid $i$, for each $j \leq i$, the value $x_j=\H^j(x) =\H(x_{j-1})$ remains information theoretically hidden until round $j$. 
In particular, we replace the random oracle $\H$ with a new stateful oracle $\H_i'(\cdot)$ that is almost identical to $\H(\cdot)$, except that for any $j \leq i$ if the query $\H(x_j)$ is submitted to $\H'(\cdot)$ before round $j$ then the response will be a random unrelated $\strlen$-bit string instead of $\H(x_j)$. 

We can argue indistinguishability of hybrids $i$ using a result of \cite{BennettBBV97} because if $j > i$, then $x_j$ is information theoretically hidden up until round $i$ and the total query magnitude of $x_j=\H^j(x)$ during round $i$ is negligible. Here, the total query magnitude of a string $x_j$ during round $i$ is defined as the sum of squared amplitudes on states where the attacker is querying string $x_j$. 
It then follows that except with negligible probability a quantum attacker cannot compute $\H^N(x)$. 
The argument does rely on the assumption that the running time $T$ of the attacker is bounded, e.g., $T \leq 2^{c\strlen}$ for some constant $c>0$. 

\medskip
Our main results are summarized in \thmref{thm:hseq:main} and \thmref{thm:main:POSW}. We show that quantum attackers running in at most $N-1$ sequential steps cannot find an $\H$-sequence of length $N$ with high probability. 
We also show that for any quantum attackers making at most $q$ quantum queries to the random oracle $\H$ over at most $(1-\alpha)N$ rounds will only be able to produce a valid PoSW with negligible probability.

\paragraph*{Technical Challenges: Iterative Hashing vs $\H$-Sequences.}
\ignore{In the appendix we prove an easier result that an attacker cannot compute $\H^N(x)$ in sequential time less than $N-1$ in the parallel quantum random oracle model --- see  \thmref{thm:seq:main} in \appref{app:seq}. Unruh \cite{Unruh15} proved a similar result in the (non-parallel) quantum random oracle model. The proof of  \thmref{thm:seq:main} involves a straightforward hybrid argument and crucially uses the fact that there is only one correct sequence i.e., we can define $H^N(x)=x_{N}$ where $x_0=x$ and $x_{i} = \H(x_{i-1})$ for $i \leq N$.   We can define a series of hybrids where the random oracle $\H$ is replaced with a stateful oracle $H'_i(\cdot)$ which assigns $H'_i(x_k) = 0^\strlen$ in round $j$ for each $j \leq k \leq i$. In the final hybrid the attacker clearly cannot compute $H_{N}(x)=x_N$ as the stateful oracle $H_{N-1}'$ information theoretically hides the final output value $x_N$ through the first $N-1$ rounds. Indistinguishability of hybrids $i$ and $i+1$ follows immediately from  \cite[Theorem 3.3]{BennettBBV97} because for any $j > i$ the total query magnitude on $x_j$ during round $i$ is negligible. }
 
Proving that an attacker cannot find an $\H$-sequence of length $N$ in $N-1$ rounds of parallel queries is significantly more challenging. 
One key difference is that there are exponentially many distinct $\H$-sequences of length $N$ that are consistent with the initial string $x_0$.  
By contrast, when we analyze a hash chain, each value on the chain $\H^j(x_0)$ can be viewed as fixed a priori. 
For $\H$-sequences, it is not clear how one would even define a hybrid where all candidate values of $x_i$ are information-theoretically hidden because these values are not known a priori and there might be exponentially many such candidates.  
In fact, for any $2 \leq i\leq N$ and {\em any} string $y$, it is {\em likely} that there exists an $\H$-sequence $x_0,\ldots,x_N$ such that $y= x_i$. 

Instead, we use a recent idea introduced by \cite{C:Zhandry19} that views the random oracle as a superposition of \emph{databases} rather than queries. 
This view facilitates intuitive simulation of quantum random oracles in a manner similar to classical models, which provides intuitive simulation for queries and circumvents the need to ``record all possible queries'', which would give an exponential number of possible $\H$-sequences in our case. 
We give significantly more intuition in \secref{finding-h-seq}, after formalizing the relevant definitions. 

\begin{restatable}{theorem}{thmhseqmain}
\thmlab{thm:hseq:main}
Let $\H:\{0,1\}^*\to\{0,1\}^{\strlen}$ be a random hash function and let $\delta \geq 1$ be a parameter. 
Let $p$ be the probability that a quantum adversary making at most $q$ queries over $N-1$ rounds outputs $(x_0, y_0),\ldots,(x_{N-1},y_{N-1})$  and $x_N$ s.t. $|x_i|\le\delta\strlen$, $y_i = \H(x_i)$ and $\substring(y_{i-1}, x_i)=1$ for each $i$, i.e., $x_0,\ldots, x_N$ is an $\H$-sequence. 
Then \[ p \leq \frac{64q^3\delta\strlen}{2^\strlen}+\frac{2N}{2^{\strlen}}.\]
\end{restatable}
\noindent Here, $\substring(y_{i-1}, x_i)=1$ means that $y_{i-1}$ is a substring of $x_i$, i.e., there exist $a,b\in\{0,1\}^*$ such that $x_i=a\|y_{i-1}\|b$.

\paragraph*{From $\H$-Sequences to Proof of Sequential Work.}

\thmref{thm:main:POSW} focuses on a non-interactive proof of sequential work obtained by applying the Fiat-Shamir transform to the efficient construction of Cohen and Pietrzak~\cite{EC:CohPie18}. 
This construction is based on a DAG $G$ with $N=2^{n+1}-1$ nodes and maximum indegree $n$. 
Given a random oracle $\H:\{0,1\}^{\strlen (n+2)} \rightarrow \{0,1\}^\strlen$, an honest prover can generate a proof for any statement $\chi \in \{0,1\}^{\strlen}$ in sequential time $\O{N}$. 
We prove that for any constant $\alpha > 0$, an attacker making $q$ queries over $s=N(1-\alpha)$ rounds will fail to produce a valid proof of sequential work for any statement except with negligible probability.
\begin{restatable}{theorem}{thmposwmain}
\thmlab{thm:main:POSW}
Suppose $\A$ makes at most $q$ quantum queries to our random oracle $\H$ over at most $s=N(1-\alpha)$ rounds and let $p$ denote the probability that $\A$ outputs a valid (non-interactive) proof of sequential work. Then \[ p \leq 32q^2 (1-\alpha)^{\lfloor \lambda/n \rfloor } + \frac{2q^3}{2^{\strlen}} +  \frac{64 q^3 (n+2) \strlen}{2^\strlen} +  \frac{2 \lfloor \lambda /n \rfloor (n+2)}{2^{\strlen}}.  \]
\end{restatable}

The main intuition for the proof \thmref{thm:main:POSW} works as follows. Given a quantum database $\D=\{(x_i,y_i):i\geq 1\}$ where $y_i$ encodes the output on input $x_i$ with $\strlen$ bits, we define a set $\lucky_s$ of databases $\D$ based on the graph coloring \FullVersion{(see \defref{def:color})}{(see the full version)}, in which $\D$ does not contain any collision or $\H$-sequence of length $s$, yet still contains a ``lucky'' Merkle tree that has a green path from the challenged node to the root that can be used to extract a proof of sequential work. We show that any attacker making (possibly parallel) $q$ queries can only succeed in measuring a lucky database $\D$ with negligible probability. Finally, we show that any attacker who produces a valid PoSW must measure a database $\D$ that either (1) contains an $\H$-sequence of length $s$, (2) contains a collision, or (3) is a lucky database. Since each of these events has negligible probability, then it follows that with high probability, the attacker cannot produce a valid PoSW.

\subsection{Related Work}
Functions that are inherently sequential to compute are a cryptographic primitive used in many applications, such as proof of sequential work~\cite{ITCS:MahMorVad13}, verifiable delay functions~\cite{BonehBBF18}, and time-lock puzzles~\cite{MahmoodyMV11}. 
The original construction~\cite{ITCS:MahMorVad13} used depth-robust graphs, which have found applications in many areas of cryptography including memory-hard functions (e.g.,~\cite{STOC:AlwSer15,C:AlwBlo16,EC:AlwBloPie17,BlockiZ17,BlockiRZ18,EC:AlwBloPie18,BlockiHKLXZ19}), proofs of replication \cite{EC:Fisch19,CCS:CFMJ19}, and proofs of space~\cite{C:DFKP15,ITCS:Pietrzak19a}. 
Recently, Cohen and Pietrzak~\cite{EC:CohPie18} show that $\H$-sequences are difficult for a classical adversary to compute in the classical parallel random oracle model. 

The Quantum Random Oracle Model (\qrom) was introduced by Boneh \etal~\cite{AC:BDFLSZ11}, who pointed out that for any real world instantiation for the hash function $\H$ (e.g., SHA3), one can build a quantum circuit implementing $\H$. Boneh \etal~\cite{AC:BDFLSZ11} also provided an example of a protocol that is secure in the classical \rom, but not in the \qrom. Quantum attacks and constructions under the quantum random oracle model have been studied in a number of previous settings, such as unclonable public-key quantum money \cite{Aaronson09,AaronsonC13}, quantum Merkle puzzles \cite{BrassardS08,BrassardHKKLS11}, signature schemes \cite{BonehZ13} and construction of random functions \cite{Zhandry12}. 

Security reductions in the classical \rom often exploit properties such as programability and extractability of queries --- properties that are lost in the \qrom. Zhandry introduced compressed oracles \cite{C:Zhandry19} as a way to record quantum queries so that they can be viewed after computation has completed. The new technique has proven to be a useful tool to extend many classical security proofs to the quantum random oracle model, e.g.,~\cite{TCC:BHHHP19,TCC:ChiManSpo19,EC:LiuZha19,EC:AMRS20,hamoudi2020quantum}.   Don et. al. \cite{don2021onlineextractability} recently showed how queries can be extracted on-the-fly in certain settings, e.g., once the algorithm outputs a classical commitment $t$ (e.g., $t=\H(x)$ or $t=\mathsf{Enc}_{pk}(\H(x))$) that is tightly related to the input $x$.

The non-interactive PoSW we consider in this work is obtained by applying the Fiat-Shamir transform to the interactive PoSW construction of Cohen and Pieterzak \cite{EC:CohPie18}. While there is a recent line of work analyzing the security of the Fiat-Shamir transform~\cite{C:FiaSha86} in the quantum random oracle model \cite{EC:KilLyuSch18,C:DFMS19,C:LiuZha19}, applying these results would require us to first establish the security of the interactive PoSW in the (parallel) \qrom. We find it easier to directly show that the non-interactive PoSW construction is secure in the (parallel) \qrom.

There have been a number of work on parallelizing quantum algorithms or considering parallel queries in the quantum random oracle model. 
Zalka \cite{Zalka99} showed that the parallel version of Grover's algorithm is optimal, e.g., in the ideal cipher model, any parallel key-recovery attacker making at most $q=\mathcal{O}(\sqrt{k2^\strlen})$ quantum queries to the ideal cipher must run in sequential time $\Omega(\sqrt{2^\strlen/k})$. 
Grover and Radhakrishnan \cite{Grover2004QuantumSF} generalized Zalka's result in the setting of multiple items to search. 
Jeffery et al. \cite{JefMagWol14} studied the parallel quantum query complexity for the element distinctness and the $k$-sum problem. 
Ambainis et al. \cite{C:AmbHamUnr19} provided an improved one-way to hiding (O2H) theorem in the parallel quantum random oracle model.

In independent work, Chung et al. \cite{ChuFHL20} also studied the problem of finding an $\H$-sequence and non-interactive proofs of sequential work in the parallel quantum random oracle model. 
They gave comparable bounds also using Zhandry's compressed oracle technique \cite{C:Zhandry19}, while leveraging an abstract view of Fourier transforms for arbitrary finite Abelian groups. 
By comparison, our proofs avoid the need for an understanding of abstract algebra, instead using quantum information theory to bound the quantum query complexity through a reduction to classical query complexity.  Thus we believe our techniques to be of independent interest, perhaps appealing to a more general audience while also providing the necessary framework to analyze the security of other classical protocols in a post-quantum world. 

\section{Preliminaries}
Let $\N$ denote the set $\{0,1,\ldots\}$, $[n]$ denote the set $\{1, 2,\ldots,n\}$, and $[a,b]= \{a, a+1, \ldots, b\}$ where $a,b\in \N$ with $a\le b$. 
For a function $f(x)$, we recursively define $f^N(x)=f\circ f^{N-1}(x)$ where $\circ$ is a function/operator composition. 
We say that a non-negative function $\mu(x)$ is \emph{negligible}, if for all polynomials $p(x)$, it holds that $0\le\mu(x)<\frac{1}{p(x)}$ for all sufficiently large $x$. 
\deleted{In this case, we write $\mu(x)=\negl(x)$.} 

Let $\{0,1\}^n$ be the set of all bitstrings of length $n$. Then we define $\{0,1\}^{\leq n}=\cup_{i=0}^n \{0,1\}^i$ to be the set of all bitstrings of length at most $n$ including an empty string $\varepsilon$. We denote $||$ as the concatenation of bitstrings. For a bitstring $x\in\{0,1\}^*$, $x[i]$ denotes its $i$\th bit, and $x[i...j]=x[i]||\ldots||x[j]$.


Given quantum states $|\phi\rangle=\sum\alpha_x|x\rangle$ and $|\psi\rangle=\sum\beta_x|x\rangle$, we define the Euclidean distance between the two states to be the quantity $\sqrt{\sum|\alpha_x-\beta_x|^2}$. 
The \emph{magnitude} of $|x\rangle$ in $|\phi\rangle=\sum\alpha_{x}|x\rangle$ is $\alpha_x$ and the query probability is $|\alpha_x|^2$ -- when we measure the state $|\phi\rangle_x$ we will observe $|x\rangle$ with probability $|\alpha_x|^2$. 

\subsection{Quantum Random Oracle Model}\seclab{compressed}

In the (sequential) quantum random oracle model (\qrom), an adversary is given oracle access to a random hash function $\H:\{0,1\}^m\to\{0,1\}^{\strlen}$. 
The adversary can submit quantum states as queries to the oracle, so that $\H$ takes as input superposition $|\phi_1\rangle,|\phi_2\rangle,\ldots$. 
Each $|\phi_i\rangle$ can be expanded as $|\phi_i\rangle=\sum\alpha_{i,x,y}|x, y\rangle$ so that the output is $\sum\alpha_{i,x,y}|x,y\oplus \H(x)\rangle$. 
Note that when the initial state is of the form $|\phi\rangle=\sum\alpha_x|x, 0^w\rangle$, then the output state will be of the form $\sum\alpha_x|x,\H(x)\rangle$.

\paragraph*{Compressed Oracle Technique in the Sequential \qrom.}
Here we introduce the compressed oracle representation introduced by  Zhandry \cite{C:Zhandry19}, which is equivalent to the standard oracle in function. 
However, the difference between the compressed oracle and the regular oracle is in the encodings of the oracle and query registers as queries are made to the oracles. 
We will extend the ideas of this technique to the parallel \qrom later on.

First, we formally define a database $\D$. 
A database $\D$ is defined by $\D=\{(x_i,y_i):i\geq 1\}$ where we write $\D(x_i)=y_i$ to denote that $y_i$ encodes the output on input $x_i$ with $\strlen$ bits. 
When $\D=\{\}$ is empty, it is equivalent to viewing the random oracle as being in superposition of all possible random oracles. 
After $q$ queries, the state can be viewed as $\sum_{x,y,z,\D}\alpha_{x,y,z}|x,y,z\rangle\otimes|\D\rangle$, where $\D$ is a compressed dataset of at most $q$ input/output pairs, $x,y$ are the query registers, and $z$ is the adversary's private storage. 

Formally, the compressed oracle technique for the sequential \qrom works as follows. Let $\H:\{0,1\}^m\rightarrow\{0,1\}^\strlen$ be a random hash function and suppose an adversary is given an oracle access to $\H$. Then we have the following observations:
\begin{itemize}
\item It is equivalent to view the usual random oracle mapping $|x,y\rangle\mapsto|x,y\oplus \H(x)\rangle$ (denote as $\mathsf{StO}$) as the \emph{phase} oracle $\mathsf{PhsO}$ that maps $|x,y\rangle$ to $(-1)^{y\cdot \H(x)}|x,y\rangle$ by applying Hadamard transforms before and after the oracle query.\footnote{Notice that both $\mathsf{StO}$ and $\mathsf{PhsO}$ are unitary matrices and $\mathsf{StO}=(I^m\otimes \H^{\otimes \strlen})\mathsf{PhsO}(I^m\otimes \H^{\otimes\strlen})$ where $I^m$ is the identity matrix on the first $m$ qubits and $\H^{\otimes \strlen}$ is the Hadamard transform on the $\strlen$ output qubits.}
\item It is also equivalent to view the oracle $\H$ as being in (initially uniform) superposition $\sum_\H|\H\rangle$ where we can encode $\H$ as a binary vector of length $2^m\times \strlen$ encoding the $\strlen$-bit output for each $m$-bit input string. Under this view the oracle maps the state $\qvec{\phi}=\sum_{x,y}\alpha_{x,y}|x,y\rangle\otimes\sum_\H|\H\rangle$ to $\sum_{x,y}\alpha_{x,y}|x,y\rangle\otimes\sum_\H|\H\rangle(-1)^{y\cdot \H(x)}$.
\end{itemize}

If the attacker makes at most $q$ queries, then we can compress the oracle $\H$ and write $\qvec{\phi}=\sum_{x,y}\alpha_{x,y}|x,y\rangle\otimes\sum_{\D} |\D\rangle$, where each dataset $\D \in \{0,1\}^{\strlen \times 2^m}$ is sparse, i.e., $\D(x) \neq \bot$ for at most $q$ entries. 
Intuitively, when $\D(x) = \bot$, we view the random oracle as being in a uniform superposition over potential outputs. 
Moreover, we can think of the basis state $|\D\rangle$ as corresponding to the superposition $\sum_{\H \in \H_\D} |\H\rangle$ where $\mathcal{H}_\D \subseteq \{0,1\}^{2^m \lambda}$ denote the set of all random oracles that are consistent with $\D$, i.e., if $\H \in\mathcal{H}_\D$ then for all inputs $x$ we either have $\D(x) = \bot$ or $\D(x)=\H(x)$. 
When viewed in this way, the basis state $|\D\rangle$ encodes prior queries to the random oracle along with the corresponding responses. 
We can use a compressed phase oracle \cphso (described below) to model a phase oracle.

\paragraph*{Compressed Phase Oracle.}
To properly define a compressed phase oracle \cphso in the sequential \qrom, a unitary local decompression procedure $\stddecomp_x$ that acts on databases was first defined in~\cite{C:Zhandry19}. 
Intuitively, $\stddecomp_x$ decompresses the value of the database at position $x$ when the database $\D$ is not specified on $x$ and there is a room to expand $\D$, and $\stddecomp_x$ does nothing when there is no room for decompression. 
If $\D$ is already specified on $x$, then we have two cases: if the corresponding $y$ registers are in a state orthogonal to a uniform superposition, then $\stddecomp_x$ is the identity (no need to decompress). 
If the $y$ registers are in the state of a uniform superposition, then $\stddecomp_x$ removes $x$ from $\D$. 
We refer to \FullVersion{\appref{sec:stddecomp}}{the full version} for a full description of $\stddecomp_x$. 
Now we define $\stddecomp,\increase,\cphso'$ on the computational basis states as
\begin{align*}
\stddecomp\left(|x,y\rangle\otimes|\D\rangle\right) &= |x,y\rangle\otimes\stddecomp_x|\D\rangle,\\
\increase\left(|x,y\rangle\otimes|\D\rangle\right) &= |x,y\rangle\otimes|\D\rangle|(\bot,0^{\strlen})\rangle,\text{ and}\\
\cphso'\left(|x,y\rangle\otimes|\D\rangle\right) &= (-1)^{y\cdot\D(x)}|x,y\rangle\otimes|\D\rangle,
\end{align*}
where the procedure \increase appends a new register $|(\bot,0^\strlen)\rangle$ at the end of the database. 
Note that $|\D\rangle|(\bot,0^\strlen)\rangle$ is a database that computes the same partial function as $\D$, but the upper bound on the number of points is increased by 1. 
Here, we remark that we define $\bot\cdot y=0$ when defining $\cphso'$, which implies that $\cphso'$ does nothing if $(x,y)$ has not yet been added to the database $\D$. 
Finally, the compressed phase oracle \cphso can be defined as follows:
\[ \cphso = \stddecomp\circ\cphso'\circ\stddecomp\circ\increase, \]
which means that when we receive a query, we first make enough space by increasing the bound and then decompress at $x$, apply the query, and then re-compress the database. 
We remark that $\cphso$ successfully keeps track of positions that are orthogonal to the uniform superposition only because if $(x,y)$ was already specified in $\D$ and the $y$ registers are in the state of a uniform superposition, then $\stddecomp$ removes $x$ from $\D$ so that $\cphso'$ does nothing as explained before and the second $\stddecomp$ in $\cphso$ will revert $(x,y)$ back to the database.

\subsection{Useful Lemmas for Compressed Oracles} \seclab{app:usefullemmas}
Next we introduce some useful lemmas given by Zhandry \cite{C:Zhandry19} that are helpful for proving our main result. 
We first introduce the following variant of Lemma 5 from \cite{C:Zhandry19}, which is still true for $\cphso$ because $\mathsf{StO}$ and $\mathsf{PhsO}$ are perfectly indistinguishable by applying a Hadamard transform before and after each query.
\begin{lemma}[\cite{C:Zhandry19}]\lemlab{lem:zhandryFive}
Consider a quantum algorithm $\adv$ making queries to a random oracle $H$ and outputting tuples $(x_1,\ldots,x_k,y_1,\ldots,y_k,z)$. 
Let $R$ be a collection of such tuples. 
Suppose with probability $p$, $\adv$ outputs a tuple such that (1) the tuple is in $R$, and (2) $\H(x_i)=y_i$ for all $i$. Now consider running $\adv$ with the oracle $\cphso$, and suppose the database $\D$ is measured after $\adv$ produces its output. Let $p'$ be the probability that (1) the tuple is in $R$, and (2) $\D(x_i)=y_i$ for all $i$ (and in particular $\D(x_i)\ne\bot$). Then $\sqrt{p}\leq \sqrt{p'} + \sqrt{k/2^n}$.
\end{lemma}
We say that a database $\D$ contains a collision if we have $(x,y),(x',y)\in\D$ for $x\ne x'$. We use the notation $\collide$ to denote the set of all databases that contain a collision. Zhandry upper bounded the probability of finding collisions in a database after making queries to a compressed oracle in the following lemma.
\begin{lemma}[\cite{C:Zhandry19}]\lemlab{lem:zhandryCollision}
For any adversary making $q$ queries to $\cphso$ and an arbitrary number of database read queries, if the database $\D$ is measured after the $q$ queries, the resulting database will contain a collision with probability at most $q^3/2^\strlen$.
\end{lemma}

%% file: pqrom.tex
\section{Parallel Quantum Random Oracle Model}
\seclab{sec:pqrom}
\ignore{Cohen and Pietrzak~\cite{EC:CohPie18} provide a construction for proofs of sequential work and showed that any attacker must produce a long $\H$-sequence in order to successfully fool a verification algorithm by claiming a false proof of sequential work with non-negligible probability under this construction. 

For example, consider the example of using the following labeling rule to obtain labels for nodes of a directed acyclic graph. 
\begin{definition}[Labeling]
\deflab{def:lab}
Let $\Sigma = \{0,1\}^w$.  
Given a directed acyclic graph $G=(V,E)$, we define the labeling of a node $v$ with input $x$ by 
\[\lab_v(x)=
\begin{cases}
\H(1, x),&\indeg(v)=0 \\
\H(v, \lab_{v_1}(x)\circ\cdots\circ\lab_{v_d}(x)),& 0< \indeg(v)=d,
\end{cases}\]
where $v_1,\ldots,v_d$ are the parents of vertex $v$ in $G$, according to some predetermined topological order. 
We omit the dependency on $x$ if the context is clear. 
\end{definition}
Cohen and Pietrzak~\cite{EC:CohPie18} provide a construction for proofs of sequential work by fixing labels $\ell_1,\ell_2,\ldots,\ell_v$, i.e. via Merkle tree, and then checking the labels for local consistency. 
\begin{definition}[Green/red node]
For a fixed labeling $\ell_1,\ell_2,\ldots,\ell_v$, a node $v\in V$ with parents $v_1,\ldots, v_d$ is {\em green} if $\ell_{v}=\H(v,\ell_{v_1}\circ \ell_{v_2}\circ \cdots \circ \ell_{v_d})$. 
A node that is not green is a {\em red} node.
\end{definition}
Similar techniques for using green/red nodes to check for local consistency were also used in~\cite{BlockiGGZ19}. 
Note that correctness of the fixed labeling $\ell_i$ for an input $x$ is not required, i.e. we do not require $\ell_i=\lab_i(x)$. 

The local testing procedure of Cohen and Pietrzak~\cite{EC:CohPie18} rejects with high probability if there is no long path of $\Omega(n)$ green nodes, but accepts if all nodes are green. 
\begin{lemma}[\cite{EC:CohPie18}]
For a fixed labeling $\ell_1,\ell_2,\ldots,\ell_v$, any path of green nodes corresponds to an $\H$-sequence.
\end{lemma}
Hence, a quantum adversary that can construct a fictitious proof of sequential work could also construct an $\H$-sequence.

We now define a sequential trace for an adversary working in the \pqrom, even though the adversary must make batches of queries. 
\begin{definition}[Sequential trace for \pqrom]\deflab{trace}
Let $q_1,\ldots,q_{N-1}$ be a sequence of batches of queries, so that $q_i$ is a set of superpositions that represents the batch of queries made in round $i$. 
Let $r_i=|q_i|$ denote the number of queries made in round $i$. 
Then we define a \emph{sequential trace} for the first round of queries to be the sequence $\psi_0,\psi_1,\ldots,\psi_{r_1}$, where $\psi_0=|(x_1,y_1),\ldots,(x_{r_1},y_{r_1})\rangle$, $\psi_1=|(x_1,y_1\oplus \H(x_1)),\ldots,(x_{r_1},y_{r_1})\rangle$ and so forth, so that $\psi_{r_1}=|(x_1,y_1\oplus \H(x_1)),\ldots,(x_{r_1},y_{r_1}\oplus \H(x_{r_1}))\rangle$. 
In each subsequent round $i$, the sequential trace contains the states $\psi_{z+1},\ldots,\psi_{z+s_i}$, where $z=\sum_{j=1}^{i-1} s_j$. 
Then we have 
\[\psi_{z+1}=|(x_1,y_1\oplus \H(x_1)),\ldots,(x_z,y_z\oplus \H(x_z)),(x_{z+1},y_{z+1}),\ldots,(x_{z+s_i},y_{z+s_i})\rangle,\]
and so forth.
\end{definition}}

\paragraph*{Quantum Query Bounds with Compressed Dataset.}
As a warm-up, we review how Zhandry \cite{C:Zhandry19} used his compressed oracle technique to provide a greatly simplified proof that Grover's algorithm is asymptotically optimal. 
\thmref{Zha19:Grover} proves that the final measured database $\D$ will not contain a pre-image of $0^{\strlen}$ except with probability $\O{q^2/2^{\strlen}}$. 
We sketch some of the key ideas below as a warm-up and to highlight some of the additional challenges faced in our setting.

\begin{theorem}[\cite{C:Zhandry19}]\thmlab{Zha19:Grover}
For any adversary making $q$ queries to \cphso and an arbitrary number of database read queries, if the database $\D$ is measured after the $q$ queries, the probability it contains a pair of the form $(x,0^{\strlen})$ is at most $\O{q^2/2^{\strlen}}$.
\end{theorem}

We can view \thmref{Zha19:Grover} as providing a bound for amplitudes of basis states with a database $\D$ in a set $\bad$ that is defined as 
\[\bad = \{\D:\D\text{ contains a pair of the form }(x,0^{\strlen})\}. \]
Given a basis state $|x,y,z\rangle\otimes|\D\rangle$ with $\D\not\in\bad$ and $x \not \in \D$, then the random oracle \CPhsO maps this basis state to
\[\qvec{\psi}=|x,y,z\rangle\otimes\frac{1}{2^{\strlen/2}}\sum_w (-1)^{y\cdot w}|\D\cup(x,w)\rangle,\]
where the amplitude on states with the corresponding database $\D$ in $\bad$ is just $2^{-\strlen/2}$. 
We use the following notation to generalize this approach for other purposes:
\begin{definition}
For a collection of basis states $\widetilde{\mathcal{S}}$ and $\qvec{\psi}=\sum_X \alpha_X|X\rangle$, we define
\[\moment(\qvec{\psi},\widetilde{\mathcal{S}})=\sqrt{\sum_{X\in\widetilde{\mathcal{S}}}|\alpha_X|^2}\] 
to denote \emph{the root of the sum of the squared magnitudes of the projection of $\psi$ onto the set of basis states \replaced{$\widetilde{\mathcal{S}}$}{$\mathcal{S}$}}. If $\mathcal{S}$ is a set of databases and $\qvec{\psi}=\sum_{x,y,z,\D}\alpha_{x,y,z,\D}|x,y,z\rangle\otimes|\D\rangle$ is a state we define $\moment(\qvec{\psi},\mathcal{S}) =\sqrt{ \sum_{x,y,z} \sum_{\D\in\mathcal{S}} \left|\alpha_{x,y,z,\D}  \right|^2}$.
\end{definition}
An equivalent way to define $\moment(\qvec{\psi},\widetilde{\mathcal{S}})$ is $\moment(\qvec{\psi},\widetilde{\mathcal{S}}) = \norm{P_{\widetilde{\mathcal{S}}} ( \qvec{\psi})}_2$ where $P_{\widetilde{\mathcal{S}}}$ projects $\qvec{\psi}$ onto the space spanned by $\widetilde{\mathcal{S}}$ e.g., if $\qvec{\psi}=\sum_{x,y,z,\D}\alpha_{x,y,z,\D}|x,y,z\rangle\otimes|\D\rangle$ then \[P_{\widetilde{\mathcal{S}}} ( \qvec{\psi}) = \sum_{\replaced{\qvec{x,y,z}\otimes\qvec{\D}}{|x,y,z,\D \rangle} \in \widetilde{\mathcal{S}}}  \alpha_{x,y,z,\D}|x,y,z\rangle\otimes|\D\rangle.\]  

Using this notation, we can view the proof of \thmref{Zha19:Grover} as bounding $\moment(\cphso\qvec{\psi}, \bad) \allowbreak- \moment(\qvec{\psi},\bad)$, i.e., the increase in root of squared amplitudes on bad states after each random oracle query. 

There are a number of challenges to overcome when directly applying this idea  to analyze $\H$-sequences. 
First, we note that \thmref{Zha19:Grover} works in the sequential \qrom, which means that the attacker can make only one query in each round. 
In our setting, the quantum attacker is allowed to make more than $T$ queries provided that the queries are submitted in parallel batches over $T-1$ rounds. 
Without the latter restriction, an attacker that makes $T$ total queries will trivially be able to find an $\H$-sequence, even if the attacker is not quantum, by computing $\H^T(x)$ over $T$ rounds. 
We formalize the \emph{parallel quantum random oracle model} \pqrom in \secref{pqrom} to model an attacker who submits batches $(x_1,y_1),\ldots$ of random oracle queries in each round.

The second primary challenge is that we can not find a static (a priori fixed) set $\bad$. 
A na\"ive approach would fix $\bad$ to be the set of databases that contain an $\H$-sequence of length $T$, but this would not allow us to bound $\moment(\cphso \qvec{\psi}, \bad) - \moment(\qvec{\psi},\bad)$. 
In particular, if our state $\qvec{\psi}$ after round $r$ has non-negligible squared amplitudes on datasets $\D$ that contain an $\H$-sequence of length $r+1$, then it is likely that the attacker will be able to produce an $\H$-sequence of length $T$ after round $T-1$ --- in this sense a bad event has already occurred. 
In our setting, the bad sets must be defined carefully in a round-dependent fashion $r$. 
Intuitively, we want to show that $\moment(\cphso^k \qvec{\psi}, \bad_{r+1}) - \moment(\qvec{\psi},\bad_r)$ is small, where $\bad_{r}$ contains datasets $\D$ that contain an $\H$-sequence of length $r+1$ and $\cphso^k$ denotes a parallel phase oracle that processes $k \geq 1$ queries in each round. 
However, reasoning about the behavior of  $\cphso^k$ introduces its own set of challenges when $k > 1$. 
We address these challenges by carefully defining sets $\bad_{r,j}$, i.e., the bad set of states after the first $j$ (out of $k$) queries in round $j$ have been processed. 
See \secref{finding-h-seq} for more details.

\subsection{Parallel Quantum Random Oracle Model \pqrom}\seclab{pqrom}

Recall that in the sequential quantum random oracle model (\qrom), an adversary can submit quantum states as queries to the oracle, so that a random hash function $H$ takes as input superposition $\qvec{\phi_1},\qvec{\phi_2},\ldots,$ and so on. 
Each $\qvec{\phi_i}$ can be expanded as $\qvec{\phi_i}=\sum\alpha_{i,x,y}|x, y\rangle$ so that the output is $\sum\alpha_{i,x,y}|x,y\oplus \H(x)\rangle$. 
Note that when the initial state is of the form $\qvec{\phi}=\sum\alpha_x|x, 0^w\rangle$, then the output state will be of the form $\sum\alpha_x|x,\H(x)\rangle$. 
In the \emph{parallel} quantum random oracle model (\pqrom), the adversary can make a batch of queries $q_1,q_2,\ldots$ each round and receives the corresponding output for each of the queries. 
More precisely, if $r_i$ is the number of queries made in round $i$, then the input takes the form $|(x_1,y_1),\ldots,(x_{r_i},y_{r_i})\rangle$ and the corresponding output is $|(x_1,y_1\oplus \H(x_1)),\ldots,(x_{r_i},y_{r_i}\oplus \H(x_{r_i}))\rangle$.

We also remark that the equivalence of the standard and phase oracles that we discussed in \secref{compressed} remains true with parallelism (by applying Hadamard transforms before and after the query); therefore, we will only consider extending the compressed oracles only on the $\cphso$ by convenience.

\paragraph*{Extending Compressed Oracle Technique to \pqrom.}
As mentioned before, we need to extend $\cphso$ to be able to handle parallel queries. 
To make the analysis simpler, we have an approach that essentially sequentially simulates a batch of parallel queries, which as a result, is equivalent to process parallel queries at once. 
Consider the following example; given a state $|B_i\rangle=|(x_1,y_1),\ldots,(x_k,y_k),z\rangle\otimes|\D\rangle$, let $\qvec{\psi_1}$ be the state after processing the first query $(x_1,y_1)$, and let $\qvec{\psi_2}$ be the state after processing the second query $(x_2,y_2)$ so that $\qvec{\psi_2}=\cphso\qvec{\psi_1}$. 
However, recall that $\cphso$ only acts on the first coordinate. 
Thus to handle the parallel query sequentially, we would need to switch the order of the coordinates to process the second query properly.
Hence, we make a slight modification and redefine $\qvec{\psi_2}=\swap_{1,2}\circ\cphso\circ\swap_{1,2}\qvec{\psi_1}$, where
\[\swap_{1,2} |(x_1,y_1),(x_2,y_2),\ldots\rangle = |(x_2,y_2),(x_1,y_1),\ldots\rangle,\]
and similarly $\swap_{i,j}(\cdot)=\swap_{j,i}(\cdot)$ swaps the positions of queries $x_i$ and $x_j$. 
Thus, we now define our \emph{parallel} version of a $\cphso$ oracle, called as $\cphso^i$, which handle $i$ parallel queries, recursively as
\[ \cphso^i = \swap_{1,i}\circ\cphso\circ\swap_{1,i}\circ\cphso^{i-1}, \]
where $\cphso^1=\cphso$. 
For a compact notation, we define \[\scphso_i:=\swap_{1,i}\circ\cphso\circ\swap_{1,i}. \]
That is, we can interpret $\cphso^i$ as applying $\scphso_j$ (essentially) sequentially for $j=1,\ldots,i$, i.e., $\cphso^i = \prod_{j=1}^i \scphso^j = \scphso^i\circ\cdots\circ\scphso^1$.

We remark that there are other possible approaches in extending $\cphso$ to the \pqrom. 
For example, see \FullVersion{\appref{pqrom}}{the full version} for further details regarding an additional approach to extending $\cphso$ to the \pqrom. 

%% file: finding-h-seq.tex

\section{Finding $\H$-Sequences in the \pqrom}\seclab{finding-h-seq}
In this section, we show that quantum adversaries cannot find $\H$-sequences of length $N$ using fewer than $N-1$ steps, thereby showing the security of a construction for proof of sequential work in the \emph{parallel} quantum random oracle model in the next section. 
Recall that an $\H$-sequence $x_0,x_1,\ldots,x_s\in\{0,1\}^*$ satisfies the property that for each $0\le i\le s-1$, there exist $a,b\in\{0,1\}^*$ such that $x_{i+1}=a|| \H(x_i)|| b$. 
Note that the sequence $\H(x),\H^2(x),\ldots,\H^N(x)$ is an $\H$-sequence, so in fact, this proves that quantum adversaries are even more limited than being unable to compute $\H^N(x)$ for a given input $x$ in fewer than $N-1$ steps.
In our analysis, we require that $\H$ outputs a $\strlen$-bit string but permit each term $x_i$ in the $\H$-sequence to have length $\delta\strlen$, for some variable parameter $\delta\ge1$.

We begin by introducing some helpful notation. Given a database $\D=\{(x_1,y_1),\ldots,\allowbreak(x_q,y_q)\}$ in the compressed standard oracle view, we can define a directed graph $G_{\D}$ on $q$ nodes ($v_{x_1},\ldots,v_{x_q}$) so that there is an edge from node $v_{x_i}$ to node $v_{x_j}$ if and only if there exist strings $a,b$ such that $x_j = a || y_i || b$. 
Thus, the graph $G_{\D}$ essentially encodes possible $\H$-sequences by forming edges between nodes $v_{x_i}$ and $v_{x_j}$ if and only if $y_i$ is a substring of $x_j$. 
More precisely, given a path $P=(v_{x_{i_0}},v_{x_{i_1}},\ldots,v_{x_{i_k}})$ in $G_{\D}$, we define
\begin{itemize}
\item $\hseq(P):=(x_{i_0},x_{i_1},\ldots,x_{i_k})$ denotes a corresponding $\H$-sequence of length $k$, and
\item $\last(P):=v_{x_{i_k}}$ denotes the endpoint of the path $P$ in $G_{\D}$ that corresponds to the output of the last label in the corresponding $\H$-sequence.
\end{itemize}
We also define a predicate $\substring(x,y)$ where $\substring(x,y)=1$ if and only if $x$ is a substring of $y$, i.e., there exist $a,b\in\{0,1\}^*$ such that $y=a||x|| b$, and $\substring(x,y)=0$ otherwise. 

\begin{example}
Suppose that $\D=\{(x_1,y_1),\ldots,(x_8,y_8)\}$ where $(x_1,y_1)=(00000,000)$, $(x_2,y_2)=(00010,001)$, $(x_3,y_3)=(00101,010)$, $(x_4,y_4)=(00110,011)$, $(x_5,y_5)=(01001,100)$, $(x_6,y_6)=(01100,101)$, $(x_7,y_7)=(10010,110)$, and $(x_8,y_8)=(11001,111)$. 
We observe that the graph $G_\D$ induced from the database $\D$ should include the edge $(v_1,v_2)$ since $x_2 = 00010 = y_1|| 10$, and so forth. 
Then we have the following graph $G_{\D}$ \FullVersion{in \figref{fig:example-h}}{(see the full version)}, which includes an $\H$-sequence of length $s=5$. 
In this example, we can say that for a path $P=(v_1,v_2,v_3,v_5,v_7,v_8)$ of length $5$, we have a corresponding $\H$-sequence $\hseq(P)=(x_1,x_2,x_3,x_5,x_7,x_8)$ of length $5$ since we have $x_2=y_1||10$, $x_3=y_2||01$, and so on. 
Note that in this case we have $\last(P)=v_8$.
\end{example}

%

\begin{definition}\deflab{def:path}
We define $\path_s$ to be the set of the databases (compressed random oracles) $\D$ such that $G_{\D}$ contains a path of length $s$.
\[ \path_s := \{\D: G_{\D}\text{ contains a path of length }s\}. \]
Note that $\path_s$ intuitively corresponds to an $\H$-sequence of length $s$. We also define $\spath_s$ to be the set of basis states with $\D$ in $\path_s$ as follows:
\[ \spath_s := \{|(x_1,y_1),\ldots,(x_k,y_k),z\rangle\otimes|\D\rangle:\D\in\path_s\}. \]
\end{definition}

\paragraph*{Challenges of Quantum Query Bounds on Finding an $\H$-Sequence.}
To bound the probability that a single round of queries finds an $\H$-sequence of length $s+1$ conditioned on the previous queries not finding an $\H$-sequence of length $s$, we consider the set of basis states $\{|B_i\rangle\}_i$ of the form $|(x_1,y_1),\ldots,(x_k,y_k),z\rangle\otimes|\D\rangle$, where $\D$ contains at most $q-k$ entries and $\D\notin\path_s$. 
Let $\qvec{\psi}=\sum\alpha_i|B_i\rangle$ be an arbitrary state that is a linear combination of $\{|B_i\rangle\}_i$ and let $\qvec{\psi'}=\cphso^k\qvec{\psi}$. 
Then we would like to bound $\moment(\qvec{\psi'},\path_{s+1})$, but there are substantial challenges in computing $\moment(\qvec{\psi'},\path_{s+1})$ directly. 

For example, given a decomposition of $\qvec{\psi} = \sum_{B} \alpha_B |B\rangle$ into basis states, we might like to compute $\eta_B =\cphso^k(|B\rangle)$ for each basis state $\qvec{B}$ and then decompose $\qvec{\psi'} = \sum_B \alpha_B \eta_B$. However, each term $\eta_B$ is no longer a basis state making it difficult to describe the state $\qvec{\psi'}$ in a helpful way so that we can bound $\moment(\qvec{\psi'},\spath_{s+1})$.
The challenges are amplified as $\qvec{\psi'}$ is the result of $k$ parallel queries.

\paragraph*{Overcoming the Challenges.}
Our approach is to consider an intermediate sequence of states $\qvec{\psi_0}=\qvec{\psi},\ldots,\qvec{\psi_k} = \qvec{\psi'}$, where $\qvec{\psi_i}$ intuitively encodes the state after the $i$\th query (in the block of parallel queries) is processed. 
Then from the definition of $\cphso^i$, we have $\qvec{\psi_{i+1}}=\swap_{1,i+1}\circ\cphso\circ\swap_{1,i+1}\qvec{\psi_i}=\scphso_{i+1}\qvec{\psi_i}$ for all $i\in[k]$. 
This approach presents a new subtle challenge. 
Consider a basis state $\qvec{B}=|(x_1,y_1),\ldots,(x_k,y_k)\rangle \otimes |\D \rangle$, where the longest path in $G_\D$ (the $\H$-sequence) has length $s-1$. 
We can easily argue that $\moment(\scphso_1 \qvec{B}, \spath_{s+1})$ is negligible since the initial basis state $\qvec{B} \not \in \spath_{s}$. 
Now we would like to argue that $\moment(\scphso_2\circ \scphso_1 \qvec{B}, \spath_{s+1})$ is negligibly small, but it is unclear how to prove this since we might have $\moment(\scphso_1 \qvec{B}, \spath_{s}) = 1$, e.g., all of the datasets $\D$ found in the support of $\scphso_1 \qvec{B}$ have paths of length $s$.

Overcoming this barrier requires a much more careful definition of our ``bad'' states. We introduce some new notions to make the explanations clearer. Suppose that a database $\D\notin\path_s$. If $\D$ has no $\H$-sequence of length $s$, it may not be the case that $\qvec{\psi_i}$ has no $\H$-sequence of length $s$. 
However, the intuition is that since the queries $x_1,\ldots,x_k$ are made in the same round, then it is acceptable to have an $\H$-sequence of length $s$, provided that the last entry in the $\H$-sequence involves some $(x_i,y_i)$. 
Thus we define $\path_{s,i}(x_1,\ldots,x_k)$ to be a set of the databases with the induced graph $G_{\D}$ having a path of length $s$ that does not end in a term that contains $\H(x_i)$:
\begin{align*}
\path_{s,i}(x_1,\ldots,x_k):= \{\D:~&G_{\D}\text{ contains a path }P\text{ of length }s\text{ and }\\
&\last(P)\notin\{v_{x_1},\ldots,v_{x_i}\}\},
\end{align*}
where we recall that $\last(P)$ denotes the endpoint of the path $P$ in $G_{\D}$, which corresponds to the output of the last label in the corresponding $\H$-sequence as defined before. 
We then define 
\[ \spath_{s,i} := \{|(x_1,y_1),\ldots,(x_k,y_k),z\rangle\otimes|\D\rangle : \D\in\path_{s,i}(x_1,\ldots,x_k)\},\]
which denotes the set of the states where the corresponding database $\D$ is in the set $\path_{s,i}(x_1,\ldots,x_k)$. 

Now we define a set $\contain_{s,i}$, which intuitively represents the set of databases so that the queries correspond to the guesses for preimages:
\begin{align*}
\contain_{s,i}(x_1,\ldots,x_k):= \{\D~:~&\exists j\le k \text{ s.t. }\substring(\D(x_i),x_j)=1\text{ and }\\
&G_{\D}\text{ contains a path of length }s\text{ ending at }x_i\}.
\end{align*}
We then define 
\[\scontain_{s,i}:= \{|(x_1,y_1),\ldots,(x_k,y_k),z\rangle\otimes|\D\rangle : \D\in\contain_{s,i}(x_1,\ldots,x_k)\},\]
which denotes the set of the states where the corresponding database $\D$ is in the set $\contain_{s,i}(x_1,\ldots,x_k)$. 
Therefore, we define $\bad_{s,i}(x_1,\ldots,x_k)$ to be the set of databases that are not in $\path_s$ but upon the insertion of $(x_1,y_1),\ldots,(x_i,y_i)$, is a member of $\path_{s+1}$:
\[\bad_{s,i}(x_1,\ldots,x_k):=\path_{s,i}(x_1,\ldots,x_k)\cup\bigcup_{j=1}^i\contain_{s,j}(x_1,\ldots,x_k)\]
Finally, we define 
\[{\sbad_{s,i}}:= \{|(x_1,y_1),\ldots,(x_k,y_k),z\rangle\otimes|\D\rangle : \D\in\bad_{s,i}(x_1,\ldots,x_k)\}\]
to represent the set of the states where the corresponding database $\D$ is in the set\\$\bad_{s,i}(x_1,\ldots,x_k)$.

We now process each query $(x_1,y_1),\ldots,(x_k,y_k)$ sequentially and argue that the mass projected onto $\path_{s+1}$ by each step $\cphso^i$ is negligible. 
To prove this, we argue that $\moment(\qvec{\psi_i}, \sbad_{s,i})$ is negligible for all $i \leq k$. 
We use the convention that $\qvec{\psi_0}$ is the initial state and $\qvec{\psi_i}=\scphso_i\qvec{\psi_{i-1}}$ for all $i\in[k]$ so that $\qvec{\psi_k}$ is the last state, after all the queries have been processed. 
Similarly, we use the convention that $\sbad_{s,0}=\spath_s$. 

We first give the following lemma. 
\begin{lemma}
\lemlab{lem:bad}
Suppose that $\D'$ is a database such that $\D'(x_{i+1})=\bot$. If $\D=\D'\cup(x_{i+1},w)\notin\bad_{s,i}(x_1,\ldots,x_k)$, then $\D\notin\bad_{s,i+1}(x_1,\ldots,x_k)$.
\end{lemma}
\begin{proof}
Since $\D\notin\path_{s,i}(x_1,\ldots,x_k)$, any path of length $s$ must end at one of the vertices $v_{x_1},\ldots,v_{x_i}$ in the graph $G_{\D}$. 
Hence, no path of length $s$ ends at $v_{x_{i+1}}$ unless we have a duplicate query $x_j=x_{i+1}$ for some $j<i+1$. 
Now we have two cases:
\begin{enumerate}
\item 
If $x_{i+1}$ is distinct from $x_j$ for all $j<i+1$, then by the previous observation we immediately have that $\D\notin\path_{s,i+1}(x_1,\ldots,x_k)$. 
Furthermore, $G_{\D}$ contains no path of length $s$ ending at node $v_{x_{i+1}}$ since any path of length $s$ must end at one of $v_{x_1},\ldots,v_{x_i}$. 
Hence, $\D\notin\contain_{s,i+1}(x_1,\ldots,x_k)$. 
Taken together, we have that $\D\notin\bad_{s,i+1}(x_1,\ldots,x_k)$.
\item 
If $x_{i+1}=x_j\,(j<i+1)$ is a duplicate query, then there might be a path of length $s$ ending at $v_{x_{i+1}}=v_{x_j}$ in $\D$. 
However, due to the duplicate we have $\{v_{x_1},\ldots,v_{x_i}\}=\{v_{x_1},\ldots,v_{x_{i+1}}\}$. 
Therefore, $\D\notin\path_{s,i+1}(x_1,\ldots,x_k)$. 
Now we want to argue that $\D\notin\contain_{s,i+1}(x_1,\ldots,x_k)$. Note that $\D(x_{i+1})=\D(x_j)$ for some $j<i+1$, which implies that 
$\substring(\D(x_j),x_l)=1 \Leftrightarrow \substring(\D(x_{i+1}),x_l)=1$ 
for all $l\in[k]$. 
If $\D\in\contain_{s,i+1}(x_1,\ldots,x_k)$, then there exists a path of length $s$ ending at $x_j$ and $\substring(\D(x_j),x_l)=1$ for some $l\leq k$. 
This implies that $\D\in\contain_{s,j}(x_1,\ldots,x_k)$ and therefore $\D\in\bad_{s,i}(x_1,\ldots,x_k)$, which is a contradiction. 
Hence, we have that $\D\notin\contain_{s,i+1}(x_1,\ldots,x_k)$ and therefore $\D\notin\bad_{s,i+1}(x_1,\ldots,x_k)$ in this case as well.
\end{enumerate}
Taken together, we can conclude that if $\D\notin\bad_{s,i}(x_1,\ldots,x_k)$, then it is also the case that $\D\notin\bad_{s,i+1}(x_1,\ldots,x_k)$.
\end{proof}

\begin{lemma}
\lemlab{lem:step:negl}
For each $i\in\{0,1,\ldots,k-1\}$, 
\[\moment(\qvec{\psi_{i+1}},\sbad_{s,i+1})-\moment(\qvec{\psi_i},\sbad_{s,i})\le\frac{4\sqrt{q\delta\strlen+k\delta\strlen}}{2^{\strlen/2}}.\] 
\end{lemma}
\begin{proof}
To argue that the projection onto $\sbad_{s,i}$ increases by a negligible amount for each query, we use a similar argument to \cite{C:Zhandry19}. 
Recall that $\scphso_{i+1}=\swap_{1,i+1}\circ\CPhsO\circ\swap_{1,i+1}$. 
Namely, we consider the projection of $\qvec{\psi_{i+1}}=\scphso_{i+1}\qvec{\psi_i}$ onto orthogonal spaces as follows: 
\begin{itemize}
\item We first define $P_i$ (resp. $P$) to be the projection onto the span of basis states $|(x_1, y_1) ,\ldots, \allowbreak(x_k, y_k),z\rangle\otimes|\D\rangle\in\sbad_{s,i}$ (resp. basis states in $\sbad_{s,i+1}$). 
\item Next we define $Q_i$ to be the projection onto states $|(x_1, y_1) ,\ldots, (x_k, y_k),z\rangle \otimes|\D\rangle\not\in \sbad_{s,i}$ such that $y_{i+1}\neq 0$ and $\D(x_{i+1})=\bot$. Intuitively, $Q_i$ represents the projection onto states that are not bad where $\scphso_{i+1} |(x_1, y_1) ,\ldots, (x_k, y_k),z\rangle \otimes|\D\rangle = |(x_1, y_1) ,\ldots, (x_k, y_k),z\rangle \sum_{w}  \otimes|\D \cup (x_{i+1},w)\rangle $ will add a new tuple $(x_{i+1},w)$ to the dataset.  
\item We then define $R_i$ to be the projection onto states $|(x_1, y_1) ,\ldots, (x_k, y_k),z\rangle \otimes|\D\rangle$ such that $|(x_1, y_1) ,\allowbreak\ldots, (x_k, y_k),z\rangle \otimes|\D\rangle\notin\sbad_{s,i}$, $y_{i+1}\neq 0$ and $\D(x_{i+1})\neq\bot$, so that the value of $x_{i+1}$ has been specified in databases corresponding to these states. 
\item Finally, we define $S_i$ to be the projection onto states $|(x_1, y_1) ,\ldots, (x_k, y_k),z\rangle\otimes|\D\rangle$ such that $|(x_1, y_1) ,\ldots, (x_k, y_k),z\rangle \otimes|\D\rangle  \notin \sbad_{s,i}$ and $y_{i+1}=0$. 
\end{itemize}
Since $P_i,Q_i,R_i,S_i$ project onto disjoint states that span the entirety of $\qvec{\psi_{i+1}}$ then we have $P_i+Q_i+R_i+S_i=\I$, where $\I$ denotes the identity operator. 
We analyze how $P$ acts on these components separately. For the component $P_i\qvec{\psi_i}$, it is easy to verify that $\norm{P\circ\scphso_{i+1}(P_i\qvec{\psi_i})}_2 \le \norm{\scphso_{i+1}(P_i\qvec{\psi_i})}_2  \le\norm{P_i\qvec{\psi_i}}_2$. See \FullVersion{\appref{app:missing}}{the full version} for the formal proof of \lemref{lem:PPi}.
\begin{restatable}{lemma}{lemPPi}
\lemlab{lem:PPi} $\norm{P\circ\scphso_{i+1}(P_i\qvec{\psi_i})}_2\le\norm{P_i\qvec{\psi_i}}_2$.
\end{restatable}
\noindent
Next, to analyze how the projection $P$ acts on $\scphso_{i+1}(Q_i\qvec{\psi_i})$, we note that $\scphso_{i+1}( \allowbreak|(x_1, y_1) ,\ldots, \allowbreak(x_k, y_k),z\rangle \otimes|\D\rangle) = |(x_1, y_1) ,\ldots, (x_k, y_k),z\rangle  \otimes \sum_{w} 2^{-\lambda/2}(-1)^{w \cdot y_{i+1}}|\D \cup (x_{i+1},w)\rangle $ for any basis state in the support of $Q_i\qvec{\psi_i}$. 
We then use a classical counting argument to upper bound the number of strings $w$ such that $|(x_1, y_1) ,\ldots, (x_k, y_k),z\rangle  \otimes \sum_{w} |\D \cup (x_{i+1},w)\rangle \in \sbad_{s,i+1}$ by decomposing the databases in $\bad_{s,i+1}(x_1,\ldots,x_k)$ into the databases in $\path_{s,i+1}(x_1,\ldots,x_k)$ and the databases in $\bigcup_{j=1}^{i+1}\contain_{s,j}(x_1,\ldots,x_k)$. Intuitively, since $\D \not \in \bad_{s,i}(x_1,\ldots,x_k)$ the only way to have $\D \cup (x_{i+1},w) \in \path_{s,i+1}(x_1,\ldots,x_k)$ is if $w$ is a substring of some $x_j$ with $j\leq k$ or $w$ is the substring of some other input $x$ in the database $\D$. 
We bound the number of databases in $\path_{s,i+1}(x_1,\ldots,x_k)$ by noting that \emph{any} string $x\in \{0,1\}^{\delta \strlen}$ contains {\em at most} $\delta\strlen$ unique contiguous substrings of length $\strlen$, so there are at most $\delta\strlen$ values of $w$ such that $\substring(w,x)=1$. 
Since $|\D\cup(x_{i+1},w)\rangle$ contains at most $q$ entries, then by a union bound, there are at most $q\delta\strlen$ strings $w$ such that exists $x \in\{0,1\}^*$ such that $\substring(w,x)=1$ and $\D(x) \neq \bot$ or $x=x_{i+1}$. Thus, $\left| \left\{ w: D \cup (x_{i+1},w) \in \bad_{s,i+1}(x_1,\ldots,x_k) \right\} \right| \leq q \delta \strlen$.   
We similarly bound the number of databases in $\bigcup_{j=1}^{i+1}\contain_{s,j}(x_1,\ldots,x_k)$ by noting that if $\D \not \in \bad_{s,i}(x_1,\ldots,x_k)$ then the only way for $\D \cup (x_{i+1}, w)$ to be in  $\bigcup_{j=1}^{i+1}\contain_{s,j}(x_1,\ldots,x_k)$ is if for some $j \leq k$ the string $w$ is a substring of $x_j$ i.e., $\substring(w,x_j)=1$. A similar argument shows that the number of strings $w$ such that $\D \cup (x_{i+1}, w) \in \bigcup_{j=1}^{i+1}\contain_{s,j}(x_1,\ldots,x_k)$ is at most $k \delta \lambda$. 

We thus show the following \FullVersion{in \appref{app:missing}}{(see the full version for the full proof)}:
\begin{restatable}{lemma}{lemPQi}
\lemlab{lem:PQi}
$\norm{P\circ\scphso_{i+1}(Q_i\qvec{\psi_i})}^2_2\le\frac{q\delta\strlen+k\delta\strlen}{2^{\strlen}}.$
\end{restatable}

We next consider how $P$ acts upon the basis states of $\scphso_{i+1}( R_i\qvec{\psi_i})$. 
We first relate this quantity to $\scphso_{i+1}(|x,y,z\rangle\otimes|\D'\cup(x_{i+1},w)\rangle)$, where $\D'$ is the database $\D$ with $x_{i+1}$ removed. 
Since $\D=|\D' \cup (x_{i+1},w)\rangle \not \in \bad_{s,i}(x_1,\ldots,x_k)$, then we again use a similar classical counting argument to upper bound the number of strings $w'$ such that $\D'\cup(x_{i+1},w')\in\bad_{s,i+1}(x_1,\ldots,x_k)$. \lemref{lem:PRi} then follows from algebraic manipulation similar to \cite{C:Zhandry19}. See \FullVersion{\appref{app:missing}}{the full version} for the formal proof.
\begin{restatable}{lemma}{lemPRi}
\lemlab{lem:PRi}
$\norm{P\circ\scphso_{i+1}(R_i\qvec{\psi_i})}_2^2\le\frac{9(q\delta\strlen+k\delta\strlen)}{2^{\strlen}}.$
\end{restatable}

\noindent
Finally, we bound the projection of $P$ onto the states of $\scphso_{i+1}(S_i\qvec{\psi_i})$:

\begin{restatable}{lemma}{lemPSi}
\lemlab{lem:PSi}
$\norm{P \circ \scphso_{i+1} (S_i \qvec{\psi_i}) }_2 = 0.$
\end{restatable}
\begin{proof}
We observe that $P\circ\scphso_{i+1}(S_i\qvec{\psi_i})=0$, since for any basis state $|(x_1,y_1),\ldots, \allowbreak(x_k,y_k),z \rangle \otimes |\D\rangle$ state in the support of $S_i\qvec{\psi_i}$, we have \[ \scphso_{i+1} |(x_1,y_1),\ldots, (x_k,y_k),z \rangle \otimes |\D\rangle = |(x_1,y_1),\ldots, (x_k,y_k),z \rangle \otimes |\D\rangle, \] i.e., $\scphso_{i+1} (S_i \qvec{\psi_i}) =  S_i \qvec{\psi_i}$. 
We also note that since $\D \not \in \bad_{s,i}(x_1,\ldots,x_k)$ and $x_{i+1}$ is not being inserted into the dataset that $\D \not \in \bad_{s,i+1}(x_1,\ldots,x_k)$. 
Hence, we have that $\norm{P \circ \scphso_{i+1} (S_i \qvec{\psi_i}) }_2 = 0$. 
\end{proof}

\noindent Thus from \lemref{lem:PPi}, \lemref{lem:PQi}, \lemref{lem:PRi}, \lemref{lem:PSi}, and by triangle inequality, we have 
\begin{align*}
\norm{P\circ\scphso_{i+1}\qvec{\psi_i}}_2&\le\norm{P\circ\scphso_{i+1}(P_i\qvec{\psi_i})}_2+\norm{P\circ\scphso_{i+1}(Q_i\qvec{\psi_i})}_2\\
&+\norm{P\circ\scphso_{i+1}(R_i\qvec{\psi_i})}_2+\norm{P\circ\scphso_{i+1}(S_i\qvec{\psi_i})}_2\\
&\le\norm{P_i\qvec{\psi_i}}_2+\frac{3\sqrt{q\delta\strlen+k\delta\strlen}}{2^{\strlen/2}}+\frac{\sqrt{q\delta\strlen+k\delta\strlen}}{2^{\strlen/2}}\\
&\le\norm{P_i\qvec{\psi_i}}_2+\frac{4\sqrt{q\delta\strlen+k\delta\strlen}}{2^{\strlen/2}}.
\end{align*}
Since we have $\norm{P\circ\scphso_i\qvec{\psi_i}}_2=\moment(\qvec{\psi_{i+1}},\sbad_{s,i+1})$ and $\norm{P_i\qvec{\psi_i}}_2=\moment(\qvec{\psi_i},\sbad_{s,i})$, we can conclude that $\moment(\qvec{\psi_{i+1}},\sbad_{s,i+1})-\moment(\qvec{\psi_i},\sbad_{s,i})\le\frac{4\sqrt{q\delta\strlen+k\delta\strlen}}{2^{\strlen/2}}$.
\end{proof}

We now bound the probability that a single round of queries finds an $\H$-sequence of length $s+1$ conditioned on the previous queries not finding an $\H$-sequence of length $s$. 

\begin{lemma}
\lemlab{lem:each:step}
Let $\qvec{\psi}$ be an initial state and let $\qvec{\psi'} = \cphso^k\qvec{\psi}$. Then we have $\moment(\qvec{\psi'},\spath_{s+1})- \moment(\qvec{\psi},\spath_{s}) \le\frac{4k\sqrt{q\delta\strlen+k\delta\strlen}}{2^{\strlen/2}}$.
\end{lemma}
\begin{proof}
We consider the sequence of states $\qvec{\psi}=\qvec{\psi_0},\ldots,\qvec{\psi_k}=\qvec{\psi'}$ with  $\qvec{\psi_i} = \cphso^i\qvec{\psi}$.
By \claimref{pathsubsetbad} it suffices to bound $\moment(\qvec{\psi_k},\sbad_{s,k})$ as $\moment(\qvec{\psi_k},\spath_{s+1}) \leq \moment(\qvec{\psi_k},\sbad_{s,k})$. See \FullVersion{\appref{app:missing}}{the full version} for the proof of \claimref{pathsubsetbad}. 

\begin{restatable}{claim}{claimpathbad}
\claimlab{pathsubsetbad}
$\spath_{s+1} \subseteq \sbad_{s,k}$.
\end{restatable}

Recall that we use the convention $\sbad_{s,0}=\spath_s$ and $\qvec{\psi_0}$ is the initial state so that by \lemref{lem:step:negl}, 
\begin{align*}
\moment(\qvec{\psi_k},\sbad_{s,k})&= \moment(\qvec{\psi_0},\sbad_{s,0})+\sum_{i=0}^{k-1}\big[\moment(\qvec{\psi_{i+1}},\sbad_{s,i+1})-\moment(\qvec{\psi_i},\sbad_{s,i})\big]\\
&\le  \moment(\qvec{\psi_0},\sbad_{s,0})+ \sum_{i=0}^{k-1}\frac{4\sqrt{q\delta\strlen+k\delta\strlen}}{2^{\strlen/2}}\\
&= \moment(\qvec{\psi_0},\spath_s)+\frac{4k\sqrt{q\delta\strlen+k\delta\strlen}}{2^{\strlen/2}}.
\end{align*}
Hence, $\moment(\qvec{\psi_k},\spath_{s+1}) \le  \moment(\qvec{\psi_k},\sbad_{s,k})\le  \moment(\qvec{\psi_0},\spath_s)+\frac{4k\sqrt{q\delta\strlen+k\delta\strlen}}{2^{\strlen/2}}$ which implies that $\moment(\qvec{\psi'},\spath_{s+1})- \moment(\qvec{\psi},\spath_{s}) \le\frac{4k\sqrt{q\delta\strlen+k\delta\strlen}}{2^{\strlen/2}}$.
\end{proof}

We now show that a quantum adversary that makes up to $q$ rounds over $N-1$ rounds can only find an $\H$-sequence of length $N$ with negligible probability. 
\begin{lemma}
\lemlab{lem:oracle:reveal}
Suppose that in each round $i\in[N-1]$, the adversary $\A$ makes a query to the parallel oracle $\cphso^{k_i}$ and that the total number of queries is bounded by $q$, i.e., $\sum_{i=1}^{N-1} k_i \leq q$. 
Then $\A$ measures a database in $\path_N$ with probability at most $\frac{32q^3\delta\strlen}{2^\strlen}$. 
\end{lemma}
\begin{proof}
Let $\qvec{\psi_0}$ be the initial state and let $U_r$ represent a unitary transform applied by $\A$ in between batches of queries to the quantum oracle. Then we define $\qvec{\psi_r}= U_r \circ \cphso^{k_r}\qvec{\psi_{r-1}}$ for each round $r \in[N-1]$. 
Thus, the attacker $\A$ yields a sequence of states $\qvec{\psi_0},\ldots, \qvec{\psi_{N-1}}$.  
We remark that $U_r$ may only operate on the state $|x,y,z\rangle$ and cannot impact the compressed oracle $\D$, e.g., $U_r \left( |x',y',z'\rangle \otimes | \D \rangle \right) = \sum_{x,y,z} \alpha_{x,y,z} |x,y,z\rangle \otimes |\D\rangle$. 
Thus, 
\[\moment(U_r \circ \cphso^{k_r}\qvec{\psi_{r-1}}, \spath_{r+1}) = \moment(\cphso^{k_r}\qvec{\psi_{r-1}}, \spath_{r+1}),\] 
so we can effectively ignore the intermediate unitary transform $U_r$ in our analysis below. Now we can apply the previous lemma to conclude that 
  \[ \moment(\qvec{\psi_{i}}, \spath_{i+1}) \leq  \frac{4k_i\sqrt{2q\delta\strlen}}{2^{\strlen/2}} +  \moment(\qvec{\psi_{i-1}}, \spath_{i})  \ . \] 
By the triangle inequality we have  \[ \moment(\qvec{\psi_{N-1}}, \spath_{N}) \leq \sum_{i=0}^{N-1} \frac{4k_i\sqrt{2q\delta\strlen}}{2^{\strlen/2}} \leq \frac{\sqrt{32q^3\delta\strlen}}{2^{\strlen/2}}  \ . \]
Hence, \deleted{we can conclude that }$\A$ measures a database in $\path_N$ with probability at most $\left(\frac{\sqrt{32q^3\delta\strlen}}{2^{\strlen/2}}\right)^2=\frac{32q^3\delta\strlen}{2^\strlen}$.
\end{proof}

Thus we have shown that a quantum adversary that makes $N-1$ rounds of parallel queries should generally not find an $\H$-sequence of length $N$ within their queries. 
Then we bound the probability that the quantum adversary outputs an $\H$-sequence of length $N$ by a standard approach, e.g.~\cite{EC:CohPie18,C:Zhandry19} of additionally the probability that the quantum adversary makes a ``lucky guess''. 

\thmhseqmain*
\begin{proof}
By \lemref{lem:zhandryFive} the probability $p$ is upper bounded by $2p' + 2N2^{-\strlen}$ where $p'$ denotes the probability that an attacker measures $\D \in \path_N$. By \lemref{lem:oracle:reveal} we have $p' \leq \frac{32q^3\delta\strlen}{2^\strlen}$ and the result follows immediately. 
\end{proof}

%% file: reduction.tex

\section{Security of PoSW in the \pqrom}\seclab{reduction}
In this section, we show the security of a construction for proofs of sequential work in the \emph{parallel} quantum random oracle model (\pqrom). 
Here, we focus on the non-interactive version of PoSW, which can be obtained by applying a Fiat-Shamir transform to the PoSW defined in \cite{EC:CohPie18}. 
Note that the PoSW defined in both \cite{ITCS:MahMorVad13} and \cite{EC:CohPie18} are interactive, in which a statement $\chi$ is randomly sampled from the verifier and the prover constructs a proof based on the input statement $\chi$.

\subsection{The Definition of Non-Interactive PoSW}
We first formally define the non-interactive PoSW in the random oracle model.
\begin{definition}[Non-Interactive PoSW]\deflab{posw}
A \emph{Non-Interactive Proof of Sequential Work} (\niPoSW) consists of polynomial-time oracle algorithms $\Pi_\niPoSW = \left( \Solve, \Verify\right)$ that use public parameters, as defined below.
\begin{itemize}
\item \textbf{Public Parameters.} Security parameter $\strlen\in\mathbb{N}$ and a random oracle $\H:\{0,1\}^*\rightarrow \{0,1\}^{\strlen}$.
\item \textbf{Solve.} Given a time parameter $T\in\mathbb{N}$, the statement $\chi$,  \P computes a solution $\pi\gets\Solve^{\H(\cdot)}(1^\lambda, T,\chi)$. The final proof is $(\chi,\pi)$.
\item \textbf{Verify.} \P can verify that the proof is genuine by running $\{0,1\}\gets\allowbreak\Verify^{\H(\cdot)}(1^\lambda, T,\chi,\pi)$. 
\end{itemize}
\end{definition}
We require the following properties:
\begin{enumerate}
\item \textbf{Correctness.} For any $\chi\in\{0,1\}^\strlen$, $\lambda, T\in\mathbb{N}$ we have
\[\Verify^{\H(\cdot)}(1^\lambda, T,\chi,\Solve^{\H(\cdot)}(1^\lambda,T,\chi))=1.\]
That is, an honest prover should always produce a valid proof with probability $1$, regardless of the choice of the statement $\chi\in\{0,1\}^*$,  running in time parameter $T$ and security parameter $\lambda$.
\item \textbf{Efficiency.} \Solve should run in time $T\cdot\mathsf{poly}(\strlen)$ and \Verify should run in time $\mathsf{poly}(\lambda, \log T)$. Similarly, the solution $\pi$ must have size $\mathsf{poly}(\lambda, \log T)$.
\item \textbf{Security.} We say that a construction $\Pi_{\niPoSW}$ is \emph{$\left(T(\cdot),q(\cdot),\epsilon(\cdot)\right)$-secure (resp. $\Pi_{\niPoSW}$ is $\left(T(\cdot),q(\cdot),\epsilon(\cdot)\right)$-quantum secure)} if any algorithm $\adv$ running in sequential time at most $T=T(\lambda)$ in the \pROM (resp. \pqrom) and making at most $q=q(\lambda)$ total queries to the random oracle should fail to produce a valid proof for any statement $\chi\in\{0,1\}^\strlen$ (selected by the adversary) except with a negligible probability $\eps(\lambda)$, i.e., if $(\pi', \chi) \gets\adv^{\H(\cdot)}(1^\lambda,T)$ denotes the solution generated from $\adv$, then
\[ \Pr_{\H(\cdot)}\left[ \Verify^{\H(\cdot)}(1^\lambda, T,\chi,\pi')=1 \right]\leq\eps(\lambda), \]
where the probability is taken over the randomness of the random oracle $\H$.
\end{enumerate}

\ignore{
\begin{remark}
Our definition does not consider precomputations, however, since a statement $\chi$ is sample randomly, we can easily generalize it to the setting where the attacker performs precomputation. In particular, this can be done by using Lemma 6 in \cite{Unruh15} where the Lemma shows that it is indistinguishable whether to have additional (arbitrary number of) offline random oracle queries or not. \seunghoon{added remark from reviewer C's comment}
\end{remark}
}
\ignore{
We will build a DAG-based construction of the non-interactive PoSW later in the section. Hence, in the DAG-based setting, the randomized algorithm $\GetChal$ will be defined accordingly as follows. Given a DAG $G=(V,E)$ on $N$ nodes, we define $\GetChal_G(k;R)$ to be a randomized algorithm that takes as input the parameter $k$ and a uniformly random string $R$ as a coin and outputs a subset $C\subseteq V$ of $k$ nodes in $G$. Now we can define the $(T,k,\eps)$-security of the pair $(G,\GetChal_G)$ which guarantees the security requirement in the definition of the non-interactive PoSW.

\begin{definition}\deflab{securitychal}
Given a DAG $G=(V,E)$ on $N$ nodes, the pair $(G,\GetChal_G)$ is said to be \emph{$(T,k,\eps)$-secure} if for all subgraphs $G'$ which do not contain a directed path of length $T$ we have $\Pr[\GetChal_G(k;R)\subseteq V(G')]\leq \eps$.
\end{definition}
Intuitively, 
}

\subsection{The Underlying DAG $G_n^{\mathsf{PoSW}}$ (\cite{EC:CohPie18})}\seclab{sec:dag}
We will use the same DAG in our construction of the non-interactive PoSW as the DAG defined in \cite{EC:CohPie18}. Here, we briefly recall their construction of the DAG $G_n^{\mathsf{PoSW}}$ \FullVersion{as shown in \figref{fig:cp18}}{(the figure is given in the full version)}. For $n\in\mathbb{N}$, let $N=2^{n+1}-1$ and first construct a complete binary tree $B_n=(V,E')$ of depth $n$, where $|V|=N$ and all the directed edges go from the leaves towards the root. We identify the $N$ nodes $V=\{0,1\}^{\leq n}$ with the binary strings of length at most $n$ except for the root, and we let the root be the empty string $\varepsilon$. Below the root we add directed edges from nodes $0$ and $1$ to node $\varepsilon$. Similarly, for each node $v$ we add directed edges from nodes $(v \| 0)$ and $(v \| 1)$ to $v$. Here, $\|$ denotes the concatenation of the strings. Now we define the DAG $G_n^{\mathsf{PoSW}}=(V,E)$ by starting with $B_n=(V,E')$ and appending some edges as follows:
\begin{itemize}
\item For all leaf nodes $u\in\{0,1\}^n$, add an edge $(v,u)$ for any $v$ that is a left sibling of a node on the path from $u$ to the root $\varepsilon$.
\end{itemize}
For example, for $u=0110$, the path from $u$ to the root is $0110\rightarrow 011\rightarrow 01\rightarrow 0 \rightarrow \varepsilon$ and the left siblings of the nodes on this path are $010$ and $00$. Hence, we add the edges $(010,0110)$ and $(00,0110)$ to $E'$. We refer to \cite{EC:CohPie18} for the full description of $G_n^{\mathsf{PoSW}}$.


\subsection{The Non-Interactive Version of \cite{EC:CohPie18} Construction}
Applying the Fiat-Shamir transform to the interactive PoSW construction \cite{EC:CohPie18}, we have the following algorithms \Solve and \Verify in the non-interactive PoSW construction. For the notational simplicity, let $G_n=G_n^{\mathsf{PoSW}}$ be the underlying DAG from \cite{EC:CohPie18}.

\vspace{0.5em}
\noindent$\Solve^{\H(\cdot)}(1^\lambda, T,\chi,C)$:
\begin{itemize}
\item Compute the labels of the graph $G_n$ with $n =1+\lceil \log T \rceil$, i.e., compute $\ell_i = \H_\chi(i,\ell_{p_1^i},\ldots,\allowbreak \ell_{p_{d_i}^i}),1\leq i\leq N$, where $p_1^i,\ldots,p_{d_i}^i$ denotes the parents\footnote{Given a DAG $G$ and a directed edge $(u,v)$ we say that node $u$ is a parent of node $v$. While this is the standard notion of parent in a DAG it can be counter-intuitive since the ``tree" edges in our DAG are directed towards the root i.e., nodes $011$ and $010$ are both parents of node $01$.} of node $i$ and $\H_\chi(\cdot):=\H(\chi,\cdot)$.
\item  Compute $R=\H_\chi(N+1,\ell_{\varepsilon})$ and parse $R$ to get $k=\lfloor \lambda/n \rfloor$ strings $c_1,\ldots,c_k \in \{0,1\}^n$. Compute the challenges $C=\{c_1,\ldots,c_k\}$ where each $n$-bit string $c_i$ corresponds to a leaf node in $G_n$.
\item Prove that everything on the path from node $c_i$ to the root is locally consistent. This can be done by a Merkle tree reveal $\MTReveal$, which reveals the labels of all the siblings on path from node $c_i$ to the root. More precisely, for a node $y\in\{0,1\}^{\leq n}$, we define \[\MTReveal(y)=\{\ell_{y[0\ldots j-1]\|(y[j]\oplus 1)}\}_{j=1}^{k},\]
where we recall that $y[0\ldots j]$ denotes the substring of $y$ to the $j$\th bit and $y[0\ldots 0]$ denotes the empty string. 
In this way, we can reveal the labels of all the siblings on path from $x$ to the root $\varepsilon$. 
In particular, a solution $\pi$ consists of the following:
\[ \pi=\{\ell_{\varepsilon},c_i, \MTReveal(c_i) \text{ for }1\leq i\leq k\}. \]
\item Output $(\chi,\pi)$.
\end{itemize}

\noindent$\Verify^{\H(\cdot)}(1^\lambda, T,\chi,\pi)$:
\begin{itemize}
\item Parse $\pi$ to extract $\ell_{\varepsilon}'$ and $c_1',\ldots,c_k'$. Set $R' =\H_\chi(N+1, \ell_{\varepsilon}')$ and split $R'$ to obtain $k=\lfloor \lambda / n \rfloor$ challenges $c_1'',\ldots,c_k''$ each of length $n$. Output $0$ if $c_i'' \neq c_i'$ for any $i\leq k$. Otherwise, let $({p_1^i}',\ldots,{p_{d_i}^i}')=\mathsf{parents}(c_i')$ for each $i$. 
\item Extract $\ell_{c_i'}'$ and $\ell_{{p_j^i}'}'$ from $\pi$ for each $i \leq k$ and $j \leq d_i'$.  
\item Check that each leaf node $c_i'$ is locally consistent. That is, one checks that $\ell_{c_i'}'$ is correctly computed from its parent labels:
\[ \ell_{c_i'}'\stackrel{?}{=}\H_\chi(c_i',\ell_{{p_1^i}'}',\ldots,\ell_{{p_{d_i}^i}'})\text{ where }({p_1^i}',\ldots,{p_{d_i}^i}')=\mathsf{parents}(c_i').\]
(Note that in $G_n$ all of $c_i'$'s parents are siblings of nodes on the path from $c_i'$ to the root $\epsilon$).
\item Check the Merkle tree openings for consistency. That is, for $i=n-1,n-2,\ldots,0$, check that
\[ \ell_{c_i'[0...i]} := \H_\chi(c_i'[0...i],\ell_{c_i'[0...i]\|0},\ell_{c_i'[0...i]\|1}),\]
and verify that the computed root $\ell_{c_i'[0...0]}$ is equal to $\ell_{\varepsilon}\in\pi$ that we received in the proof.
\end{itemize}

\subsection{Security}
We argue that for any fixed constant $\alpha > 0$ the non-interactive proof of sequential work outline above is $(T=(1-\alpha)N,q,\eps)$-secure for $\epsilon(\lambda) = 32q^2 (1-\alpha)^{\lfloor \lambda/n \rfloor } + 2q^32^{-\strlen} +  64 q^3 (n+2) \strlen 2^{-\strlen} +  2 \lfloor \lambda /n \rfloor (n+2) 2^{-\strlen}$. We first define a set $\lucky_s$ of databases $\D$ in which $\D$ contains no collision or $\H$-sequence of length $s$, yet the dataset $\D$ contains a lucky Merkle tree that can be used to extract a proof of sequential work for some statement $\chi$. We then argue that {\em any} attacker making $q$ queries fails to measure such a lucky dataset $\D$ except with negligible probability. This is true even if the attacker is not restricted to run in sequential time $s$. Finally, to complete the argument we argue that any cheating attacker who produces a valid proof of sequential work can be converted into an algorithm that measures a dataset $\D$ that either (1) contains an $\H$-sequence of length $s$, (2) contains a collision or (3) is in $\lucky_s$. Assuming that our attacker is sequentially bounded and makes at most $q$ queries, the probability of any of these three outcomes must be negligible. Thus, the PoSW construction must be secure against any sequentially bounded attacker.


\paragraph*{Coloring the Graph $G^n_{\posw}$.}
Given a database $\D$, a statement $\chi$, and a candidate PoSW solution $y=\ell_{\epsilon}$ we define an algorithm $\colmt_\D(\chi,y)$ \FullVersion{(\algref{alg:colmt})}{} which returns a copy of the DAG $G_n^{\posw}$ in which each node is colored $\red$ or $\green$\FullVersion{ based on \defref{def:color}}{}. Intuitively, green nodes indicate that the corresponding labels are locally consistent with the database $\D$ while red nodes are locally inconsistent. If the PoSW solution $\ell_{\epsilon}$ is entirely consistent with $\D$ then every node in $G^n_\posw$ will be colored green. On the other hand, if there is no entry of the form $(x,y) \in \D$ then the root node in $G^n_{\posw}$ would be colored red along with every other node below it. To define $\colmt_\D(\chi,y)$ we use a recursive helper function $\colsubtree_\D$\FullVersion{ (\algref{alg:colsubtree})}{} which outputs a colored subgraph rooted at an intermediate node $v$. We briefly introduce how these algorithms work below and refer to \FullVersion{\appref{sec:algorithms}}{the full version} for the complete descriptions.


\begin{figure}[ht]
\centering
\includestandalone[width=0.8\textwidth]{fig-colsubtree}
\caption{A succinct illustration of $\colsubtree_\D(\chi,\varepsilon,x_\varepsilon,y_\varepsilon)$ where $\varepsilon$ is an empty string and the database $\D$ is defined as above. Note that since $n=3$, the nodes on the lowest layer are leaf nodes and the dashed edges are shown to help understand how the coloring works (we do not actually draw the edges in the algorithm). As described above, we have the following cases to color the node to red: (1) in node $1$, there exists $x_1$ such that $(x_1,y_1)\in\D$, however, $x_1$ cannot be parsed properly, i.e., $x_1\ne \chi \| \textcolor{red}{1} \| \ell_{10} \|\ell_{11}$ (parsing fail\protect\footnotemark), (2) in node $00$, there is no $x_{00}$ such that $(x_{00},y_{00})\in\D$ (undefined entry in $\D$), and (3) in node $010$ -- which is a leaf node -- we have an entry $(x_{010},y_{010})\in\D$, but when parsing $x_{010}$ into $\chi \| 010 \| \ell_{00}'$, we observe that the predefined label $\ell_{00}$ of node $00$ and the value $\ell_{00}'$ does not match (local inconsistency).}
\figlab{fig:colsubtree}
\end{figure}
\footnotetext{Note that if we have another entry $(x_1',y_1)\in\D$ satisfying $x_1' = \chi \| 1 \| \ell_{10} \|\ell_{11}$ then we have a collision in the database and we do not have a parsing fail here. Similar argument holds for another parsing fail case in (3) as well. We will deal with the case that the database has collisions separately and assume that we do not have any collisions so that a unique Merkle subtree is generated as output.}

\begin{enumerate}
\item The algorithm $\colsubtree_\D(\chi,v,x_v,y_v)$ generates a subset of nodes that consists of a Merkle subtree along with the coloring of each node in the set.
\begin{itemize}
\item The algorithm takes as input $(\chi,v,x_v,y_v)$ where $\chi\in\{0,1\}^*$ is a statement, $v\in\{0,1\}^{\leq n}$ denotes a node in $G_\D$\footnote{We remark that $v$ corresponds to the identification of a node that is a binary string of length at most $n$, different from the label of the node $\ell_v$.}, and $x_v\in\{0,1\}^*, y_v\in\{0,1\}^\strlen$ are the bitstrings. Here, $y_v$ is a potential candidate to be the label of node $v$. It outputs a subset of nodes $V'\subseteq V(G_\D)$, which consists of a Merkle subtree with root node $v$ and the corresponding coloring set $\coloring(V'):=\{\coloring(v):v\in V'\}$.
\item Recall that a node $v$ is \emph{green} if it is locally consistent; for example, let $(x,y)\in\D$ and for node $v$ with label $\ell_v=y$, if $v_0, v_1$ with labels $y_0, y_1$ are the parents of $v$ then $v$ is locally consistent if and only if it satisfies $\H_\chi(v,y_0,y_1)=y$. Since we satisfies $\H(x)=y$ as $(x,y)\in\D$, one would need to satisfy $x=\chi \|v\|y_0\|y_1$ for $v$ to be locally consistent. 
\item Hence, we start parsing $x_v$ into $\chi \|v'\|y_{v\|0}\|y_{v\|1}$ and see if it succeeds. That is, check if $v'=v$. If it fails, then we say it as ``parsing fail'', which is illustrated in \figref{fig:colsubtree} (node $1$). In this case, we color the node red and stop generating the subtree.
\item If we succeed to parse $x_v$, then we color $v$ to green and can proceed to its parents and see if there is an entry in the database $\D$ with having its $y$-coordinate as the label of its parent node. If it fails, then it becomes our second fail and we color the node red and stop generating the subtree. It is illustrated in \figref{fig:colsubtree} (node $00$).
\item When we color the leaf nodes, we follow the same procedure except that we could have more than two parents based on the edge structure in \cite{EC:CohPie18}, and we have another possibility of ``parsing fail'' because the labels of its parents should be predefined by construction. If this kind of parsing fail occurs (node $010$ in \figref{fig:colsubtree}), then we color the node to red.
\end{itemize}
\item $\colmt_\D(\chi, y)$ generates a complete Merkle tree rooted at a node $\varepsilon$ with label $\ell_\varepsilon=y$ and appends the edges as shown in \cite{EC:CohPie18}. The algorithm works simple; find an entry $(x,y)\in\D$ and call $\colsubtree_\D(\chi,\varepsilon,x,y)$. Fill the missing nodes with label $\bot$ and color them all red. Then we add the edges as described in \secref{sec:dag}. If there is no such $(x,y)$ in $\D$ then we abort the entire algorithm. We refer to \figref{fig:colmt} for an example of running the algorithm.
\end{enumerate}

\begin{figure}[ht]
\centering
\includestandalone[width=0.7\textwidth]{fig-coloredmt}
\caption{One example of $\colmt_\D(\chi, y)$ where $n=3$ and $(x,y)\in\D$. On the left side is the output of $\colsubtree_\D(\chi,\varepsilon,x,y)$ (different from \figref{fig:colsubtree}) and we fill the undefined nodes colored red and add the edges on the right side. Note that newly added nodes and edges are shown in blue.}
\figlab{fig:colmt}
\end{figure}

\paragraph*{Notations.}
Recall that in \defref{def:path}, we defined $\path_s$ to be the set of the databases (compressed random oracles) $\D$ such that $G_{\D}$ contains a path of length $s$, which corresponds to the $\H$-sequence of length $s$. Now we define the set $\collide$ to be the set of the databases that contains a collision:
\[\collide:=\{\D:\D\text{ contains pairs }(x,y),(x',y)\text{ such that }x\ne x'\}.\] 
Given a node (string) $v = (v_1\|\ldots\| v_n) \in \{0,1\}^n$, we use $v_{\leq i} \in \{0,1\}^i$ to denote the substring $v_1\|\ldots\| v_i$, $v_{\leq 0}:=\varepsilon$, and we use $\ptr(v,\chi) = \{ v_{\leq i}~: 0\leq i \leq n\}$ to denote the set of all nodes on the direct path from $v$ to the root of a Merkle tree constructed from $\colmt_\D(\chi,\cdot)$. 
Given a coloring of the Merkle tree, we define the predicate $\gptr(v,\chi)$, which verifies that every node on the path from $v$ to the root is green.\footnote{Here, $\ptr$ stands for ``\textsf{P}ath \textsf{T}o the \textsf{R}oot" and $\gptr$ stands for ``\textsf{g}reen \textsf{P}ath \textsf{T}o the \textsf{R}oot".}
That is, $\gptr(v,\chi)=1$ if and only if $\coloring(v')=\green$ for all $v'\in\ptr(v,\chi)$ and $0$ otherwise. For example, in \figref{fig:colmt}, we have $\gptr(011,\chi)=1$ because the color of nodes in $\ptr(011,\chi)=\{011,01,0,\varepsilon\}$ are all green. On the other hand, we observe that $\gptr(000,\chi)=0$ despite the node $000$ is green as we have an intermediate red node $00$ in $\ptr(000,\chi)$.

Now we define $\lucky(\D,\chi,y)$ to be the set of $\strlen$-bit strings that produce lucky challenges for the Merkle tree rooted at $y$:
\begin{align*}
\lucky(\D, \chi, y):= \{ w\in&\{0,1\}^\strlen : \,w=w_1 \| \ldots \| w_k \| z\text{ where }k=\lfloor \strlen/n\rfloor,\\ 
&|w_i|=n \text{, and }\gptr(w_i,\chi)=1\,\,\forall 0\leq i\leq k  \}.
\end{align*}
Then the set $\lucky_s$ is defined to be the set of databases that contains lucky challenges not in $\collide$ and $\path_s$, i.e., 
\begin{align*}
\lucky_s := &\{ \D:\exists (x,y)\in\D\text{ s.t. } x=\chi\| N+1\| \ell_\eps ~\wedge~  y\in\lucky(\D,\chi, \ell_{\eps}  ) \}\\
&\qquad\setminus \left(\collide \cup \path_s \right).
\end{align*}
We also define $\slucky_s$ to be the set of basis states with $\D$ in $\lucky_s$ as follows:
\[ \slucky_s := \{ \qvec{x,y,z}\otimes\qvec{\D}:\D\in\lucky_s\}. \]
We remark that when defining the set $\lucky_s$, we need to assume that $\D\not\in\collide$ to ensure that we get a unique Merkle tree rooted at $\ell_\varepsilon$ and we additionally need to assume that $\D\not\in\path_s$ otherwise the set $\{v\in\{0,1\}^n:\gptr(v,\chi)=0\}$ may not be large, i.e., if all nodes are green.

We also define $\pre(\D)$ to be the set of $\strlen$-bit strings $w$ that become a preimage of a hash value. It happens when $w$ is a substring of $x$ where $(x,y)\in\D$.
\[ \pre(\D) := \{ w\in\{0,1\}^\strlen : \exists(x,y)\in\D\text{ s.t. }\substring(w,x)=1\}. \]

\paragraph*{Security of PoSW against Quantum Attackers.}
Let $G_n=(V,E),\coloring(V)\gets\colmt_\D(\chi,\cdot)$. 
We first introduce a helper lemma that is a classical argument (not quantum) and comes immediately from rephrasing the intermediate claim in the proof of \cite[Theorem 1]{EC:CohPie18}. 
\begin{lemma}[\cite{EC:CohPie18}]\lemlab{lem:rednodes}
If $\D\not\in\collide$ and $\D\not\in\path_T$ with $T=(1-\alpha)N$ for some constant $0<\alpha<1$, then \[ \left| \{ v\in\{0,1\}^n : \gptr(v,\chi)=0 \} \right| \geq \alpha 2^n \ , \] i.e., at least $\alpha 2^n$ out of $2^n$ challenges (leaf nodes) in $G_n$ must fail to respond correctly.
\end{lemma}
We immediately have the following corollary which bounds the number of tuples $(v_1,\ldots, v_k, y)$ such that all challenges $v_i$ are lucky i.e., $ \gptr(v_i,\chi)=1 ~ \forall i\leq k$. Here, $y \in \{0,1\}^{k'}$ is an auxiliary string. 

\begin{corollary}\corlab{cor:greenchal}
If $v_1,\ldots,v_k$ are the leaf nodes in $G_n$ (i.e., $v_i\in V,|v_i|=n\,\,\forall i\leq k$), then we have that 
\[ \big| \{ (v_1,\ldots,v_k,y) : \gptr(v_i,\chi)=1 ~ \forall i\leq k, \,y\in\{0,1\}^{k'} \} \big| \leq 2^{nk+k'} (1-\alpha)^k. \]
\end{corollary}

\begin{restatable}{lemma}{lemlucky}
\lemlab{lem:lucky}
Let $\alpha$ be any constant satisfying $q \leq \frac{2^\lambda (1-\alpha)^{\lfloor \lambda/n \rfloor}}{(n+1)\strlen}$ then for any state \[ \qvec{\phi} = \sum_{x,y,z, \D : |\D| \leq q} \alpha_{x,y,z,\D} |x,y,z \rangle \otimes \qvec{\D} \] whose database register is a superposition of databases with at most $q$ entries, we have
\[ \moment(\cphso\qvec{\phi},\slucky_s) - \moment(\qvec{\phi},\slucky_s)\leq 4(1-\alpha)^{\frac{\lfloor \lambda/n \rfloor}{2} }. \]
\end{restatable}
\begin{proof}
\added{(Sketch) The proof of }\lemref{lem:lucky}\added{ is similar to }\lemref{lem:step:negl}\added{. We provide a complete proof in }\FullVersion{\appref{app:missing}}{the full version} and sketch the high level details here. 
We consider the projection of $\qvec{\phi'}=\cphso\qvec{\phi}$ onto orthogonal spaces $P,Q,R,S$ where (1) $P$ projects onto the span of basis states $\qvec{x,y,z}\otimes\qvec{\D}\in\slucky_s$, (2) $Q$ projects onto states $\qvec{x,y,z}\otimes\qvec{\D}$ such that $\qvec{x,y,z}\otimes\qvec{\D}\not\in\slucky_s$, $y\ne 0$, and $\D(x)=\bot$, (3) $R$ projects onto states $\qvec{x,y,z}\otimes\qvec{\D}$ such that $\qvec{x,y,z}\otimes\qvec{\D}\not\in\slucky_s$, $y\ne 0$, and $\D(x)\ne\bot$, and (4) $S$ projects onto states $\qvec{x,y,z}\otimes\qvec{\D}$ such that $\qvec{x,y,z}\otimes\qvec{\D}\not\in\slucky_s$ and $y=0$. Then since $P,Q,R,S$ project onto disjoint states that span the entirety of $\qvec{\phi'}$ then we have $P+Q+R+S=\mathbb{I}$, where $\mathbb{I}$ denotes the identity operator. We analyze how $P$ acts on these components separately and have that $\norm{P\circ\cphso(P\qvec{\phi})}_2\leq \norm{P\qvec{\phi}}_2$\FullVersion{ (see \lemref{lem:PP})}{}, $\norm{P \circ\cphso (Q \qvec{\phi})}_2^2  \leq (1-\alpha)^{\lfloor \lambda/n \rfloor }$\FullVersion{ (see \lemref{lem:PQ})}{}, $\norm{P\circ\cphso(R\qvec{\phi})}_2^2\leq 9(1-\alpha)^{\lfloor \strlen/n\rfloor}.$\FullVersion{ (see \lemref{lem:PR})}{}, and $\norm{P\circ\cphso(S\qvec{\phi})}_2=0$\FullVersion{ (see \lemref{lem:PS})}{}. Then by triangle inequality, we have that 
\begin{align*}
\norm{P\circ\cphso\qvec{\phi}}_2 &\leq \norm{P\circ\cphso(P\qvec{\phi})}_2 + \norm{P\circ\cphso(Q\qvec{\phi})}_2\\
&\qquad+ \norm{P\circ\cphso(R\qvec{\phi})}_2 + \norm{P\circ\cphso(S\qvec{\phi})}_2 \\
&\leq \norm{P\qvec{\phi}}_2 + (1-\alpha)^{\frac{\lfloor \lambda/n \rfloor}{2} } + 3(1-\alpha)^{\frac{\lfloor \strlen/n\rfloor}{2}} \\
&\leq \moment(\qvec{\phi},\slucky_s) +  4(1-\alpha)^{\frac{\lfloor \lambda/n \rfloor}{2} }.
\end{align*}
Since we have that $\norm{P\circ\cphso\qvec{\phi}}_2 = \moment(\cphso\qvec{\phi},\slucky_s)$, we complete the proof.
\end{proof}

\added{From \lemref{lem:lucky}, we have the following Lemma. The proof of \lemref{lem:lucky:reveal} can be found in \FullVersion{\appref{app:missing}}{the full version}.}

\begin{restatable}{lemma}{lemreveal}
\lemlab{lem:lucky:reveal}
Suppose that our quantum attacker $\A$ makes at most $q$  queries to $\cphso$ then the probability $p'$ of measuring a database $\D \in \lucky_s$ for $s=N(1-\alpha)$ is at most $16q^2(1-\alpha)^{\lfloor \lambda/n \rfloor}$. 
\end{restatable}

\thmposwmain*
\begin{proof}
Suppose that $\A$ make queries to a random oracle $\H$ and outputs tuples $((x_1,y_1),\ldots,\allowbreak(x_k,y_k),z)$ and let $R$ be a collection of such tuples that contain a valid PoSW for some statement $\chi \in \{0,1\}^\strlen$. Recall that with probability $p$, the algorithm $\A$ outputs a tuple such that (1) the tuple is in $R$ (contains a valid PoSW), and (2) $\H(x_i)=y_i$ for all $i$.  Now consider running $\A$ with the oracle $\cphso$ (applying the Hadamard Transform before/after each query) and measuring the database $\D$ after $\A$ outputs. We first observe that if the final tuple is in $R$ and $\D(x_i)=y_i$ for all $i$, then we must have $\D \in \lucky_{s+1} \cup \path_{s+1} \cup \collide$. In particular, if $\D$ does not contain an $\H$-sequence of length $s+1$ or a collision, then we must have $\D \in \lucky_{s+1}$ since the proof of sequential work is valid. 
 
However, the probability of measuring a dataset $\D \in \lucky_{s+1} \cup \path_{s+1} \cup \collide$ can be upper bounded by $16q^2(1-\alpha)^{\lfloor \lambda/n \rfloor }  + 32 q^3 (n+2) \strlen 2^{-\strlen} + q^32^{-\strlen}$ by applying \lemref{lem:lucky:reveal} (Lucky Merkle Tree), \lemref{lem:oracle:reveal} (Long $\H$-sequence), and \lemref{lem:zhandryCollision} to upper bound the probability that $\D \in \lucky_{s+1}$, $\D \in \path_{s+1}$ and $\D \in \collide$, respectively.

We also observe that the number of input/output pairs in our PoSW is $k=\lfloor \strlen/n \rfloor(n+2)$ since we have $\lfloor \strlen/n \rfloor$ challenges where each challenge consists of a statement $\chi$, a node itself, and at most $n$ parents. Hence, by applying \lemref{lem:zhandryFive}, we have that
\[ \sqrt{p} \leq \sqrt{16q^2(1-\alpha)^{\lfloor \lambda/n \rfloor }  + \frac{32 q^3 (n+2) \strlen}{2^\strlen} + \frac{q^3}{2^{\strlen}}} + \sqrt{\frac{\lfloor \strlen/n \rfloor(n+2) }{2^{\strlen}}}, \]
which implies that
\[  p \leq 32q^2 (1-\alpha)^{\lfloor \lambda/n \rfloor } + \frac{2q^3}{2^{\strlen}} +  \frac{64 q^3 (n+2) \strlen}{2^\strlen} +  \frac{2 \lfloor \lambda /n \rfloor (n+2)}{2^{\strlen}},\]
since $\sqrt{a}\leq \sqrt{b}+\sqrt{c}$ implies $a\leq b+c+2\sqrt{bc}\leq 2(b+c)$ for any $a,b,c>0$.
\end{proof}


%% file: fig-colsubtree.tex
\begin{tikzpicture}
[node distance=1cm,auto,font=\small,
    every node/.style={node distance=2cm},
    greennode/.style ={ circle, draw, pattern=north east lines, pattern color=tangocolordarkchameleon!40,inner sep=-0.5pt, minimum width =0.8 cm},
    rednode/.style ={ circle, draw, pattern=north east lines, pattern color=red!40,inner sep=-0.5pt, minimum width =0.8 cm},
    fail/.style={rectangle,draw,red,inner sep=1pt,minimum width=0.8cm},
    rec/.style={rectangle,draw,inner sep=1pt,minimum width=0.8cm}
]

\node[rec, align=center] (D) at (0,1.5) {$\begin{aligned}
    \D = \{ &(x_\varepsilon = \chi \| \ell_0 \| \ell_1, y_\varepsilon), (x_0 = \chi \| 0 \| \ell_{00} \|\ell_{01}, y_0), \\
         &(x_1 = \chi \| 0 \| \ell_{10} \| \ell_{11}, y_1), (x_{01} = \chi \| 01 \| \ell_{010} \| \ell_{011}, y_{01}),\\
         &(x_{010} = \chi \| 010 \| \ell_{00}' , y_{010}), (x_{011} = \chi \| 011 \| \ell_{00} \| \ell_{010}, y_{011})\}
  \end{aligned}$};
\node[greennode] (v) at (0,0) {$\varepsilon$};
\node[greennode] (v0) at (-1.5,-1.5) {$0$};
\node[rednode] (v1) at (1.5,-1.5) {$1$};
\node[rednode] (v00) at (-2.8,-3) {$00$};
\node[greennode] (v01) at (-0.2,-3) {$01$};
\node[rednode] (v010) at (-1.3,-4.5) {$010$};
\node[greennode] (v011) at (0.9,-4.5) {$011$};

\path[draw,-{Latex[length=2mm]},dashed] (v0) to (v);
\path[draw,-{Latex[length=2mm]},dashed] (v1) to (v);
\path[draw,-{Latex[length=2mm]},dashed] (v00) to (v0);
\path[draw,-{Latex[length=2mm]},dashed] (v01) to (v0);
\path[draw,-{Latex[length=2mm]},dashed] (v010) to (v01);
\path[draw,-{Latex[length=2mm]},dashed] (v011) to (v01);
\path[draw,-{Latex[length=2mm]},dashed] (v00) to (v011);
\path[draw,-{Latex[length=2mm]},dashed] (v00) to (v010);
\path[draw,-{Latex[length=2mm]},dashed] (v010) to (v011);

\node[anchor=west] (lv) at ($(v) + (0.5,0)$) {$\ell_\varepsilon=y_\varepsilon$};

\node[anchor=west] (lv1) at ($(v1) + (0.5,0)$) {$\ell_{1}=y_{1}$};
\node[fail,anchor=west] (fail1) at ($(v1) + (0.6,-0.6)$) {\textcolor{red}{$x_1 \ne \chi \| 1 \| \ell_{10} \| \ell_{11}$, parsing fail!}};
\path[-{Latex[length=1mm]},color=red] (fail1.west) edge[bend left] node{} (v1.south);

\node[anchor=east] (lv0) at ($(v0) + (-0.5,0)$) {$\ell_{0}=y_{0}$};

\node[anchor=east] (lv00) at ($(v00) + (-0.5,0)$) {$\ell_{00}=y_{00}$};
\node[fail,anchor=east] (fail2) at ($(v00) + (-0.5,-0.6)$) {$^{\not\exists} x_{00}\text{ s.t. }(x_{00},y_{00})\in\D$};
\path[-{Latex[length=1mm]},color=red] (fail2.east) edge[bend right] node{} (v00.south west);

\node[anchor=west] (lv01) at ($(v01) + (0.5,0)$) {$\ell_{01}=y_{01}$};

\node[anchor=east] (lv010) at ($(v010) + (-0.5,0)$) {$\ell_{010}=y_{010}$};
\node[fail,anchor=east] (fail3) at ($(v010) + (-0.5,-0.9)$) {$\begin{aligned}&(x_{010},y_{010})\in\D\text{, but}\\&\ell_{00}' \ne\ell_{00},\text{ local inconsistency!}\end{aligned}$};
\path[-{Latex[length=1mm]},color=red] (fail3.east) edge[bend right] node{} (v010.south);

\node[anchor=west] (lv011) at ($(v011) + (0.5,0)$) {$\ell_{011}=y_{011}$};

\end{tikzpicture}

%% file: fig-coloredmt.tex
\begin{tikzpicture}
[node distance=1cm,auto,font=\small,
    every node/.style={node distance=2cm},
    greennode/.style ={ circle, draw, pattern=north east lines, pattern color=tangocolordarkchameleon!40,inner sep=-0.5pt, minimum width =0.6 cm},
    rednode/.style ={ circle, draw, pattern=north east lines, pattern color=red!40,inner sep=-0.5pt, minimum width =0.6 cm},
    fail/.style={rectangle,draw,red,inner sep=0.5pt,minimum width=0.6cm},
    newred/.style ={ circle, draw,blue, pattern=north east lines, pattern color=red!40,inner sep=-0.5pt, minimum width =0.6 cm}
]

\node[greennode] (v) at (0,0) {$\varepsilon$};
\node[greennode] (v0) at (-1,-1) {$0$};
\node[rednode] (v1) at (1,-1) {$1$};
\node[rednode] (v00) at (-1.8,-2) {$00$};
\node[greennode] (v01) at (-0.2,-2) {$01$};
\node[greennode] (v000) at (-2.25,-3) {$000$};
\node[rednode] (v001) at (-1.35,-3) {$001$};
\node[rednode] (v010) at (-0.65,-3) {$010$};
\node[greennode] (v011) at (0.25,-3) {$011$};

\node (rightarr) at (2.5,-1.8) {$\Rightarrow$};

\node[greennode] (nv) at (7,0) {$\varepsilon$};
\node[greennode] (nv0) at (5,-1) {$0$};
\node[rednode] (nv1) at (9,-1) {$1$};
\node[rednode] (nv00) at (4,-2) {$00$};
\node[greennode] (nv01) at (6,-2) {$01$};
\node[newred] (nv10) at (8,-2) {$10$};
\node[newred] (nv11) at (10,-2) {$11$};
\node[greennode] (nv000) at (3.5,-3) {$000$};
\node[rednode] (nv001) at (4.5,-3) {$001$};
\node[rednode] (nv010) at (5.5,-3) {$010$};
\node[greennode] (nv011) at (6.5,-3) {$011$};
\node[newred] (nv100) at (7.5,-3) {$100$};
\node[newred] (nv101) at (8.5,-3) {$101$};
\node[newred] (nv110) at (9.5,-3) {$110$};
\node[newred] (nv111) at (10.5,-3) {$111$};

\path[draw,blue,-{Latex[length=1.5mm]}] (nv0) to (nv);
\path[draw,blue,-{Latex[length=1.5mm]}] (nv1) to (nv);
\path[draw,blue,-{Latex[length=1.5mm]}] (nv00) to (nv0);
\path[draw,blue,-{Latex[length=1.5mm]}] (nv01) to (nv0);
\path[draw,blue,-{Latex[length=1.5mm]}] (nv10) to (nv1);
\path[draw,blue,-{Latex[length=1.5mm]}] (nv11) to (nv1);
\path[draw,blue,-{Latex[length=1.5mm]}] (nv000) to (nv00);
\path[draw,blue,-{Latex[length=1.5mm]}] (nv001) to (nv00);
\path[draw,blue,-{Latex[length=1.5mm]}] (nv010) to (nv01);
\path[draw,blue,-{Latex[length=1.5mm]}] (nv011) to (nv01);
\path[draw,blue,-{Latex[length=1.5mm]}] (nv100) to (nv10);
\path[draw,blue,-{Latex[length=1.5mm]}] (nv101) to (nv10);
\path[draw,blue,-{Latex[length=1.5mm]}] (nv110) to (nv11);
\path[draw,blue,-{Latex[length=1.5mm]}] (nv111) to (nv11);
\path[draw,blue,-{Latex[length=1.5mm]}] (nv000) to (nv001);
\path[draw,blue,-{Latex[length=1.5mm]}] (nv010) to (nv011);
\path[draw,blue,-{Latex[length=1.5mm]}] (nv100) to (nv101);
\path[draw,blue,-{Latex[length=1.5mm]}] (nv110) to (nv111);
\path[draw,blue,-{Latex[length=1.5mm]}] (nv00) to (nv010);
\path[draw,blue,-{Latex[length=1.5mm]}] (nv00) to (nv011);
\path[draw,blue,-{Latex[length=1.5mm]}] (nv10) to (nv110);
\path[draw,blue,-{Latex[length=1.5mm]}] (nv10) to (nv111);
\path[draw,blue,-{Latex[length=1.5mm]}] (nv0) to (nv100);
\path[draw,blue,-{Latex[length=1.5mm]}] (nv0) to (nv101);
\path[draw,blue,-{Latex[length=1.5mm]}] (nv0) to (nv110);
\path[draw,blue,-{Latex[length=1.5mm]}] (nv0) to (nv111);

\end{tikzpicture}

%% file: conclusion.tex

\section{Conclusion} 
We have shown that any attacker in the parallel quantum random oracle model making  $q \ll 2^{\strlen/3}$ total queries cannot find an $\H$-sequence of length $N$ in $N-1$ sequential rounds except with negligible probability. Using this result as a building block, we then prove that the non-interactive proof of sequential work of Cohen and Pietrzak~\cite{EC:CohPie18} is secure against any attacker making $q \ll 2^{\strlen / n}$ queries and running in sequential time $T=(1-\alpha)N$. We leave it as an open question whether or not the $\strlen/n$ term from this lower bound is inherent or whether the construction could be tweaked to establish security when $q \ll 2^{\strlen / n}$. The main technical hurdle is extracting more than $\strlen/n$ challenges from a single random oracle output or adapting the proof to handle a modified construction where we extract challenges from multiple random oracle outputs. An alternative approach would be to introduce a second random oracle with longer outputs, which could be used to extract $\Omega(\lambda)$ challenges.
 
 Our results also highlight the power of the recent compressed random oracle technique of Zhandry \cite{C:Zhandry19} and raises a natural question about whether or not these techniques could be extended to establish the security of important cryptographic primitives such as memory-hard functions or proofs of space in the quantum random oracle model. Alwen and Serbinenko~\cite{STOC:AlwSer15} previously gave a classical pebbling reduction in the classical parallel random oracle model showing that the cumulative memory complexity of a data-independent memory hard function is tightly characterized by the pebbling complexity of the underlying graph. Would it be possible to establish the post-quantum security of memory hard functions through a similar reduction in the parallel quantum random oracle model?
 

%% file: seq-bounded.tex
\section{Warm-up: Iterative Hashing}
\applab{app:seq}
In this section, we show that a quantum attacker with a specific upper bound on running time cannot compute the output of a sequential function in significantly fewer steps, with high probability. 

\ignore{
\begin{restatable}{theorem}{thmseqmain}
\thmlab{thm:seq:main}
Given a hash function $\H:\{0,1\}^*\to\{0,1\}^{\strlen}$ and a random input $x$, any quantum attacker that makes up to $q$ queries in each of $N-1$ sequential steps can only compute $\H^N(x)$ with probability at most $\frac{N^2}{2^{\strlen}}+\frac{1}{2^{\strlen}-N}+\sqrt{\frac{48\strlen N^4q^2T}{2^{\strlen/2}}}$ in the parallel quantum random oracle model.
\end{restatable}}

Namely, we show that for a hash function $\H$ and a difficulty parameter $N$, a quantum adversary that can make $q$ quantum queries to the random oracle cannot compute $\H^N(x)$ in fewer than $N-1$ steps, with high probability. 
To make this argument, we construct a sequence of hybrids, so that in each hybrid only differs on a small subset of inputs from the previous hybrid, so that the Euclidean distance between the final states of a sequence of operations of the hybrids must be small. 
It then follows that with high probability, the first hybrid (the real world sampler), with high probability, cannot be distinguished from the final hybrid, in which the adversary information theoretically cannot find $\H^N(x)$.

We first describe the real-world sampler.
\begin{mdframed}
\begin{center}
\textbf{Real-World Sampler}
\end{center}
\vskip 0.5em
\noindent Let $\strlen>0$ and $\H:\{0,1\}^*\to\{0,1\}^{\strlen}$ be a uniform random hash function. 

\noindent\textbf{Input}: $|x,y\rangle$, where $x$ and $y$ have the same length and are up to $4\strlen$ qubits.
\begin{enumerate}
\item
Output $|x,y\oplus \H(x)\rangle$.
\end{enumerate}
\end{mdframed}
Given an input $x$, let $\ell_i(x)=\H^i(x)$ for $1\le i\le N$. 
Define a series of hybrids as follows.
We first replace the real world sampler with a hybrid that does not return any collisions among $\ell_1,\ldots,\ell_N$.
\begin{mdframed}
\begin{center}
\textbf{Hybrid 0}
\end{center}
\vskip 0.5em
\noindent Fix a distinguished $x'$ and let $\H$ be a hash function such that $\{\ell_1,\ell_2,\ldots,\ell_N\}$ are distinct.

\noindent\textbf{Input}: $|x,y\rangle$, where $x$ and $y$ have the same length and are up to $\strlen$ qubits.
\begin{enumerate}
\item
Output $|x,y\oplus \H(x)\rangle$.
\end{enumerate}
\end{mdframed}
In the first hybrid, we replace $\H(x)$ in the first round, but otherwise use $\H(x)$ in the remaining rounds.
\begin{mdframed}
\begin{center}
\textbf{Hybrid 1}
\end{center}
\vskip 0.5em
\noindent Fix a distinguished $x'$ and let $\H$ be a hash function such that $\{\ell_1,\ell_2,\ldots,\ell_N\}$ are distinct. Let $r_1,r_2,\ldots,r_N$ be a set of distinct random strings uniformly drawn from $\{0,1\}^{\strlen}$. Let $G_1(x)$ be defined as follows:
\[
G_1(x)=
\begin{cases}
\H(x),\qquad &x\notin\{\ell_1,\ell_2,\ldots,\ell_N\}\\
r_i,\qquad &x=\ell_i, 1\le i\le N.
\end{cases}
\]
\noindent\textbf{Input}: $|x,y\rangle$, where $x$ and $y$ have the same length and are up to $\strlen$ qubits.
\begin{enumerate}
\item
If query is made in Round 1, output $|x,y\oplus G_1(x)\rangle$.
\item
If query is made in Round $i$, where $i>1$, output $|x,y\oplus \H(x)\rangle$.
\end{enumerate}
\end{mdframed}
We then similarly define a sequence of hybrids in the following manner:
\begin{mdframed}
\begin{center}
\textbf{Hybrid $i$}
\end{center}
\vskip 0.5em
\noindent Fix a distinguished $x'$ and let $\H$ be a hash function such that $\{\ell_1,\ell_2,\ldots,\ell_N\}$ are distinct. Let $r_1,r_2,\ldots,r_N$ be a set of distinct random strings uniformly drawn from $\{0,1\}^{\strlen}$. For each $1\le j\le i$, define function $G_j(x)$ by:
\[
G_j(x)=
\begin{cases}
\H(x),\qquad &x\notin\{\ell_i,\ell_{i+1},\ldots,\ell_N\}\\
r_j,\qquad &x=\ell_j, i\le j\le N.
\end{cases}
\]
\noindent\textbf{Input}: $|x,y\rangle$, where $x$ and $y$ have the same length and are up to $\strlen$ qubits.
\begin{enumerate}
\item
If query is made in Round $j$, where $j\le i$, output $|x,y\oplus G_j(x)\rangle$.
\item
Otherwise, output $|x,y\oplus \H(x)\rangle$.
\end{enumerate}
\end{mdframed}
\subsection{Indistinguishability of Hybrids}
In this section, we bound the Euclidean distance between the final states of the hybrids. 
The crucial observation is that by design, the hybrids only differ on a small subset of inputs, so then the Euclidean distance between the final states of a sequence of operations of the hybrids must be small. 
Thus with high probability, the real world sampler cannot be distinguished from the hybrids.
\begin{lemma}
\lemlab{lem:unique}
For each $i\in[N]$, define $\unique_i$ as the event that labels $\ell_1,\ldots,\ell_i$ are all distinct and $\unique$ as the event $\unique_N$. 
Then $\PPr{\unique}\ge 1-\frac{N^2}{2^{\strlen}}$.
\end{lemma}
\begin{proof}
Consider the conditional probability $\PPr{\unique_{i+1} ~|\unique_i} = 1-\frac{i}{2^{\strlen}}$ for any fixed $i$. 
Thus,
\begin{align*}
\PPr{\unique_{N}} &= \prod_{i=1}^{N-1} \PPr{\unique_{i+1} ~|\unique_i} = \prod_{i=1}^{N-1} \left( 1-\frac{i}{2^{\strlen}}\right)\\
&\geq 1- \sum_{i=1}^{N-1} \frac{i}{2^{\strlen}}\ge 1-\frac{N^2}{2^{\strlen}}.
\end{align*}
\end{proof}
First, note that the real world sampler is indistinguishable from hybrid $0$ conditioned on the event $\unique$. 

\begin{lemma}
\lemlab{lem:single:hybrid}
Fix $1\le j\le N-1$. 
Let $\mathcal{A}$ be any attacker in the parallel quantum random oracle model. 
Then conditioned on the event $\unique$, the probability that in hybrid $j$, $\mathcal{A}$ returns $\ell_N$ in round $j$ is at most $\frac{1}{2^{\strlen}-N}$.
\end{lemma}
\begin{proof}
Note that in hybrid $j$, the label $\ell_{N-1}$ remains hidden to $\mathcal{A}$ in an information theoretic sense, e.g., even if the attacker has access to the full truth table of $G_j$. 
Conditioned on the event that $\ell_i\neq\ell_j$ for any $i\neq j$, the attacker $\mathcal{A}$ only knows that $\ell_{N-1}$ is not one of the labels $\ell_1,\ldots,\ell_{N-2}$ and thus only knows that hybrid $j$ is one of $2^{\strlen}-N+2$ strings, so the probability that a measurement of a single quantum query of hybrid $j$ matches $\ell_N$ is less than $\frac{1}{2^{\strlen}-N}$.
\end{proof}

\begin{lemma}
\lemlab{lem:magnitude:hybrids}
Fix $1\le j\le n-3$. 
Let $\qvec{\phi}=\sum\alpha_x|x\rangle$ be the superposition of a quantum query that an adversary $\mathcal{A}$ chooses. 
Let $\xi_x$ be the squared magnitude of a state $x$ in which hybrid $j$ answers differently than hybrid $j+1$. 
Then over the choices of all hash functions $\H$, 
\[\PPr{\sum \xi_x>\frac{1}{2^{\strlen/2}}}<\frac{3\strlen}{2^{\strlen/2}}.\]
\end{lemma}
\begin{proof}
The proof follows similarly to the proof of \lemref{lem:single:hybrid}. 
Observe that hybrid $j$ and hybrid $j+1$ only answers differently on input $\ell_{j+1}$. 
Since $\ell_{j+1}$ is a string in $\{0,1\}^{\strlen}$, the probability that $G_j(x)=\ell_{j+1}$ for a specific (classical) input $x$ is $\frac{1}{2^{\strlen}}$. 
Given a $\strlen$ length qubit $\qvec{\psi}$, let $S$ be the set of states with probability at least $\frac{1}{2^{\strlen/2}}$. 
Observe that if $G_j(x)=\ell_{j+1}$ for any $x\in S$, then the magnitude of $x$ is greater than $\frac{1}{2^{\strlen/2}}$ by definition of $S$. 
But since $|S|<2^{\strlen/2}$, the probability that there exists some $x\in S$ such that $G_j(x)=\ell_{j+1}$ is at most $\frac{2^{\strlen/2}}{2^{\strlen}}=\frac{1}{2^{\strlen/2}}$ by a simple union bound.

On the other hand, Chernoff bounds imply that there are greater than $9\strlen\added{/4}$ values $y\in\{0,1\}^{\strlen}$ such that $G_j(y)=\ell_{j+1}$ only with probability less than $\frac{1}{2^{\strlen/2}}$. 
By definition, any state not in $S$ has probability at most $\frac{1}{2^{\strlen/2}}$, and so the sum of the magnitudes of any $y$ with $G_j(y)=\ell_{j+1}$ but $y\notin S$ is at most $\frac{9\strlen/4}{2^{\strlen/2}}$.

Thus,
\[\PPr{\sum \xi_x>\frac{1}{2^{\strlen/2}}}<\frac{9\strlen/4+1}{2^{\strlen/2}}\le\frac{3\strlen}{2^{\strlen/2}}.\]
\end{proof}

From applying \lemref{lem:dist} and \lemref{lem:magnitude:hybrids}, we can obtain an upper bound on the Euclidean distance between the final states of hybrid $i$ and hybrid $i+1$. 
\begin{lemma}
\lemlab{lem:adjacent:hybrids}
Conditioned on the event $\unique$, the total variation distance between the final states of hybrid $i$ and hybrid $i+1$ is at most $\sqrt{\frac{48\strlen T}{2^{\strlen/2}}}$.
\end{lemma}
\begin{proof}
Let $r$ be a state in which hybrid $j$ answers differently than hybrid $j+1$. 
By \lemref{lem:magnitude:hybrids}, $\sum_{(t,\cdot)\in S} q_S(|\phi_t\rangle)\le\frac{3\strlen}{2^{\strlen/2}}$ for a fixed $t$. 
Summing over all $j$ timesteps, it follows that $\sum_{(t,\cdot)\in S} q_S(|\phi_t\rangle)\le\frac{3\strlen N}{2^{\strlen/2}}$.
Then by applying \lemref{lem:dist}, the Euclidean distance between the final states of hybrid $i$ and hybrid $i+1$ is at most $\sqrt{\frac{3\strlen T}{2^{\strlen/2}}}$. 
Thus by \lemref{lem:tvd}, the total variation distance is at most $\sqrt{\frac{48\strlen T}{2^{\strlen/2}}}$.
\end{proof}

\begin{lemma}
\lemlab{lem:final:distance}
Conditioned on $\unique$, the total variation distance between the final states of the Real World Sampler and hybrid $i$ is at most $\sqrt{\frac{48\strlen N^2T}{2^{\strlen/2}}}$ for any $1\le i\le N-1$.
\end{lemma}
\begin{proof}
Conditioned on $\unique$, the total variation distance between the final states of hybrid $i$ and hybrid $i+1$ is at most $\sqrt{\frac{48\strlen T}{2^{\strlen/2}}}$ for all $0\le i\le N-3$ by \lemref{lem:adjacent:hybrids}. 
By repeatedly invoking the triangle inequality, the total variation distance between the final states of the real world sampler and hybrid $i$ is at most $\sqrt{\frac{48\strlen N^2T}{2^{\strlen/2}}}$ for any $1\le i\le N-1$, conditioned on $\unique$.
\end{proof}

\thmseqmain*
\begin{proof}
First, observe that $\unique$ occurs with probability at least $1-\frac{N^2}{2^{\strlen}}$ by \lemref{lem:unique}. 
Conditioned on the event $\unique$, any history revealing attacker in hybrid $N-1$ outputs $\ell_N$ with probability at most $\frac{1}{2^{\strlen}-N}$ by \lemref{lem:single:hybrid}. 
Moreover, we claim that conditioned on $\unique$ occurring, the real world sampler is indistinguishable from any hybrid. 

Specifically, by \lemref{lem:final:distance}, the real world sampler and hybrid $N$ have total variation distance at most $\sqrt{\frac{48\strlen N^2T}{2^{\strlen/2}}}$ and thus only differ on $\sqrt{\frac{48\strlen N^2T}{2^{\strlen/2}}}$ fraction of inputs. 
Since hybrid $N-1$ makes at most $qN$ queries in total, the real world sampler differs in at most $\sqrt{\frac{48\strlen N^4q^2T}{2^{\strlen/2}}}$ outputs. 

Therefore, with probability at least $1-\frac{N^2}{2^{\strlen}}-\frac{1}{2^{\strlen}-N}-\sqrt{\frac{48\strlen N^4q^2T}{2^{\strlen/2}}}$, in the $\qrom$, any quantum attacker making $q$ queries per step (modeled by the real world sampler) requires at least $N-1$ steps.
\end{proof}

%% file: stddecomp.tex
\section{Full Description of $\stddecomp_x$}
\applab{sec:stddecomp}
Here we give a formal definition of the local decompression procedure $\stddecomp_x$ for a given $x$~\cite{C:Zhandry19}. Let $\D$ be a database with a collection of $(x,y)$ pairs and $|\tau\rangle$ be a uniform superposition. If $(x,y)\in\D$, we write $\D(x)=y$ and if no such pair is in $\D$ for an input $x$, then we write $\D(x)=\bot$. Given an upper bound $t$ on the number of set points, $\stddecomp_x$ is defined based on the following cases:
\begin{enumerate}
\item If $\D(x)=\bot$ and $|\D|<t$, then
\[ \stddecomp_x|\D\rangle = \frac{1}{\sqrt{2^{\strlen}}}\sum_y|\D\cup(x,y)\rangle, \]
inserting the pair $(x,|\tau\rangle)$ into $\D$ which means that $\stddecomp_x$ decompresses the value of $\D$ at $x$.
\item If $\D(x)=\bot$ and $|\D|=t$, then
\[ \stddecomp_x|\D\rangle = |\D\rangle, \]
which means that $\stddecomp_x$ does nothing when there is no room for decompression.
\item For a $\D'$ such that $\D'(x)=\bot$ and $|\D'|<t$,
\[ \stddecomp_x\left( \frac{1}{\sqrt{2^n}}\sum_y (-1)^{z\cdot y}\qvec{\D'\cup(x,y)} \right) = \left\{ \begin{array}{ll} \frac{1}{\sqrt{2^n}}\sum_y (-1)^{z\cdot y}\qvec{\D'\cup(x,y)} & \text{if }z\ne 0,\\ \qvec{\D'} & \text{if }z=0. \end{array} \right. \]
That means that if $\D$ is already specified on $x$ and the corresponding $y$ registers are in a state orthogonal to a uniform superposition ($z\ne 0$) then $\stddecomp_x$ does nothing since there is no need to decompress. If $y$ registers are in the state of a uniform superposition ($z=0$) then $\stddecomp_x$ removes $(x,y)$ from $\D$. 
\end{enumerate}

\section{Alternative Compressed Oracle Technique to the \pqrom}
\applab{pqrom}
In this section, we introduce an alternative approach to extend $\cphso$ to the \pqrom. Recall that to extend the compressed oracle technique \cite{C:Zhandry19} to the \pqrom, we need to redefine \stddecomp and \cphso that can handle multiple queries as input. For the input that takes $k$ queries in a round, i.e., the input that takes the form $|(x_1,y_1),\ldots,(x_k,y_k)\rangle$, the \emph{parallel} local decompression procedure $\stddecomp^k$ can be defined as
{\small\[ \stddecomp^{k}|(x_1,y_1),\ldots,(x_k,y_k)\rangle\otimes|\D\rangle=|(x_1,y_1),\ldots,(x_k,y_k)\rangle\otimes\left(\prod_{j=1}^{k}\stddecomp_{x_j}\right)|\D\rangle,\]}
such that we apply $\stddecomp$ for each $x_i$'s and make a product on it. 
Similarly, the \emph{parallel} compressed phase oracle $\cphso^k$ is defined over the computational basis states as
\[ \cphso'^k|(x_1,y_1),\ldots,(x_k,y_k)\rangle\otimes|\D\rangle = (-1)^{\sum_{j=1}^{k}y_j\cdot\D(x_j)} |(x_1,y_1),\ldots,(x_k,y_k)\rangle\otimes|\D\rangle \]
and
\[\cphso^k=\stddecomp^k\circ\cphso'^k\circ\stddecomp^{k}\circ\mathsf{Increase}^{k},\]
where $\mathsf{Increase}^k$ is the procedure which initializes $k$ new registers $|(\bot,0^n),\ldots,(\bot,0^n)\rangle$ and appends it to the end of each database in a parallel query round.

%% file: algorithms.tex
\newpage
\section{Algorithms to Generate $G_n^{\posw}$ with Colorings}
\applab{sec:algorithms}
\begin{algorithm}[H]
\DontPrintSemicolon
\SetAlgoLined
\KwIn{$(\chi,v,x_v,y_v)$ where $\chi\in\{0,1\}^*$ is a statement, $v\in\{0,1\}^{\leq n}$ is a node in $G_\D$ and $x_v\in\{0,1\}^*,y_v\in\{0,1\}^\strlen$ are the bitstrings}
\KwOut{A subset of nodes $V'\subseteq V(G_\D)$ that consists the Merkle subtree with root node $v$ and the coloring set $\coloring(V'):=\{\coloring(v):v\in V'\}$}
\vspace{0.3em}
Initialize $V'=\varnothing$ and $\coloring(V')=\varnothing$\;
Add $v$ to $V'$ with label $\ell_v=y_v$\;
 \If(\tcp*[f]{$v$ is not a leaf node}){$|v|<n$}{
 	Parse $x_v=\chi \|v'\|y_v^0\|y_v^1$ where $y_v^0,y_v^1\in\{0,1\}^\strlen$\;
	\If(\tcp*[f]{parse fail}){$v'\ne v$}{
		$\coloring(v)=\red$ and add $\coloring(v)$ to $\coloring(V')$
	}
	\Else{
		Add nodes $(v\|0)$ and $(v\|1)$ to $V'$ with labels $\ell_{v\|0}=y_v^0$ and $\ell_{v\|1}=y_v^1$\;
		$\coloring(v)=\green$ and add $\coloring(v)$ to $\coloring(V')$\;
		\If{$\exists\, x_v^{0,1},\ldots,x_v^{0,m}$ s.t. $(x_v^{0,1},y_v^0),\ldots,(x_v^{0,m},y_v^0)\in\D$}{
			Pick the smallest $x_v^{0,i}\in\{x_v^{0,1},\ldots,x_v^{0,m}\}$ in a lexicographical order\footnotemark\;
			$(V_0,\coloring(V_0))\gets\colsubtree_\D(\chi,v\|0,x_v^{0,i},y_v^0)$\tcp*[f]{recursion on node $v\|0$}\;
			$V'=V'\cup V_0$\;
			$\coloring(V')=\coloring(V')\cup\coloring(V_0)$
		}
		\Else{
			$\coloring(v\|0)=\red$\;
			Add $\coloring(v\|0)$ to $\coloring(V')$\; 
		}
		\If{$\exists\, x_v^{1,1},\ldots,x_v^{1,m'}$ s.t. $(x_v^{1,1},y_v^1),\ldots,(x_v^{1,m'},y_v^1)\in\D$}{
			Pick the smallest $x_v^{1,j}\in\{x_v^{1,1},\ldots,x_v^{1,m'}\}$ in a lexicographical order\;
			$(V_1,\coloring(V_1))\gets\colsubtree_\D(\chi,v\|1,x_v^{1,j},y_v^1)$\tcp*[f]{recursion on node $v\|1$}\;
			$V'=V'\cup V_1$\;
			$\coloring(V')=\coloring(V')\cup\coloring(V_1)$
		}
		\Else{
			$\coloring(v\|1)=\red$\;
			Add $\coloring(v\|1)$ to $\coloring(V')$\;
		}
	}
 }
 \ElseIf(\tcp*[f]{$v$ is a leaf node}){$|v|=n$}{
 	Parse $x_v=\chi \|v'\|y_{p_1}\|\cdots\|y_{p_{d_v}}$ where $\{p_1,\ldots,p_{d_v}\}=\mathsf{parents}(v)$\;
	\If(\tcp*[f]{parse fail}){$v'\ne v$}{
		$\coloring(v)=\red$ and add $\coloring(v)$ to $\coloring(V')$
	}
	\ElseIf(\tcp*[f]{another parse fail}){$\exists \,i$ s.t. $y_{p_i}\neq\ell_{p_i}$}{
		$\coloring(v)=\red$ and add $\coloring(v)$ to $\coloring(V')$
	}
	\Else{
		$\coloring(v)=\green$ and add $\coloring(v)$ to $\coloring(V')$
	}
 }
 \Return{$V',\coloring(V')$}
 \caption{Merkle tree generation and coloring $\colsubtree_\D$}\alglab{alg:colsubtree}
\end{algorithm}
\footnotetext{Note that if $\D$ has no collision, then there exists a \emph{unique} $x_v^0$ such that $(x_v^0,y_v^0)\in\D$. The same observation holds for line 18 of \algref{alg:colsubtree} and line 2 of \algref{alg:colmt}.}
\newpage
\begin{algorithm}[H]
\DontPrintSemicolon
\SetAlgoLined
\KwIn{$\chi\in\{0,1\}^*, y\in\{0,1\}^\strlen$ where $\chi$ is a statement}
\KwOut{A graph $G_n^{\posw}=(V,E)$ with the coloring set $\coloring(V)$}
\vspace{0.3em}
Initialize a graph $G=(V,E)$ with $V=\varnothing, E=\varnothing$, and $\coloring(V)=\varnothing$\;
\If{$\exists x_1,\ldots,x_m\text{ s.t. }(x_1,y),\ldots,(x_m,y)\in\D$}{
	Pick the smallest $x_j\in\{x_1,\ldots,x_m\}$ in a lexicographical order\;
	$(V_0,\coloring(V_0))\gets\colsubtree_\D(\chi, \varepsilon,x_j,y)$\;
	$V=V\cup V_0$\;
	$\coloring(V)=\coloring(V)\cup\coloring(V_0)$\;
	\ForEach{$v\in\{0,1\}^{\leq n}$}{
		\If(\tcp*[f]{fill up the undefined nodes}){$v\not\in V$}{
		Add a node $v$ to $V$ with label $\ell_{v}=\bot$\;
		$\coloring(v)=\red$ and add $\coloring(v)$ to $\coloring(V)$\;
		}
	}
	\For(\tcp*[f]{add edges to the graph as in \cite{EC:CohPie18}}){$0\leq i<n$}{
		Add an edge $(x_j\|b,x_j)$ to $E$ for each $b\in\{0,1\}$ and $x_j\in\{0,1\}^i$
	}
	\ForEach(\tcp*[f]{add additional edges to the leaf nodes}){$v\in\{0,1\}^n$}{
		\ForEach{$a\text{ s.t. }v=a\|1\|a'$}{
			Add an edge $(a\|0,v)$ to $E$
		}
	}
}
\Else{
\Return{$\bot$}
}
\Return{$G=(V,E), \coloring(V)$}
 \caption{Generating $G_n^{\posw}$ with coloring $\colmt_\D$}\alglab{alg:colmt}
\end{algorithm}

\section{Missing Figures}
\applab{app:fig}
\vspace{-1.5em}
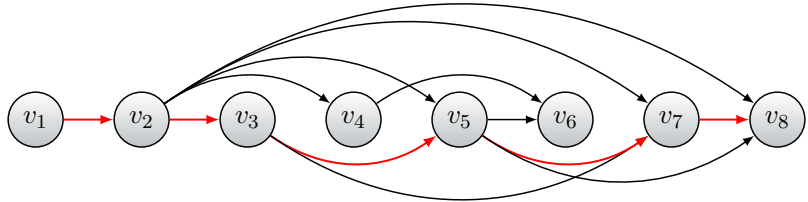
\begin{figure}[!htb]
\centering
\resizebox{0.65\textwidth}{!}{
\begin{tikzpicture}[scale=0.88,-latex, auto ,node distance =3 cm and 2cm,on grid,semithick ,
live/.style ={ circle ,top color =white , bottom color = purdueevertrueblue!40 ,
draw,black, text=black , minimum width =0.7 cm},
deleted/.style ={ circle ,top color = purduemoondustgray!20 , bottom color = purduemoondustgray!50,
draw,purduemoondustgray!50, text=white , minimum width =0.7 cm}]
\node[live] (1) {$v_1$};
\node[live] (2) [right=1.5cm of 1] {$v_2$};
\node[live] (3) [right=1.5cm of 2] {$v_3$};
\node[live] (4) [right=1.5cm of 3] {$v_4$};
\node[live] (5) [right=1.5cm of 4] {$v_5$};
\node[live] (6) [right=1.5cm of 5] {$v_6$};
\node[live] (7) [right=1.5cm of 6] {$v_7$};
\node[live] (8) [right=1.5cm of 7] {$v_8$};
\path[color=red,thick] (1) edge node {} (2);

\path[color=red,thick] (2) edge node {} (3);
\path (2) edge[bend left=35] node {} (4);
\path (2) edge[bend left=35] node {} (5);
\path (2) edge[bend left=35] node {} (7);
\path (2) edge[bend left=35] node {} (8);

\path[color=red,thick] (3) edge[bend right=35] node {} (5);
\path (3) edge[bend right=35] node {} (7);

\path (4) edge[bend left=35] node {} (6);

\path (5) edge node {} (6);
\path[color=red,thick] (5) edge[bend right=35] node {} (7);
\path (5) edge[bend right=35] node {} (8);

\path[color=red,thick] (7) edge node {} (8);
\end{tikzpicture}}
\caption{A directed acyclic graph $G_\D$ induced from the database $\D$. We remark that an $\H$-sequence of length $5$ is $v_1\rightarrow v_2\rightarrow v_3\rightarrow v_5\rightarrow v_7\rightarrow v_8$, which is the longest path in the graph. We also remark that in this example, $q=8=2^{\strlen}$ implies $\strlen=3$ and $\delta\strlen = 5$ where $\delta=5/3>1$.}
\figlab{fig:example-h}
\end{figure}

\begin{figure}[ht]
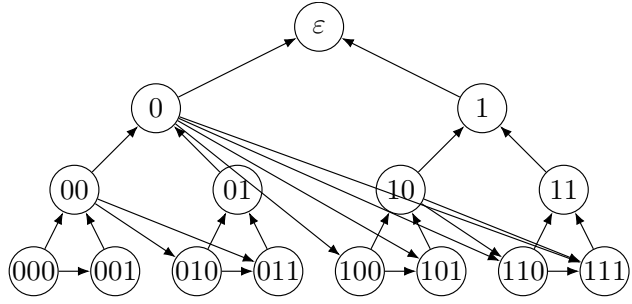

\centering
\includestandalone[width=0.5\textwidth]{fig-cp18graph}
\caption{Illustration of $G_3^{\posw}$. Note that on the complete Merkle tree, we add the edges from any node that is a left sibling of a node on the path from a leaf node to the root, i.e., for node $111$, we add edges $(110,111), (10,111),(0,111)$ since the path from node $111$ to the root is $111\rightarrow 11 \rightarrow 1\rightarrow \varepsilon$ and the corresponding left siblings are $110, 10,$ and $0$.}
\figlab{fig:cp18}
\end{figure}

%% file: fig-cp18graph.tex
\begin{tikzpicture}
[node distance=1cm,auto,font=\small,
    every node/.style={node distance=2cm},
    newnode/.style ={ circle, draw, inner sep=-0.5pt, minimum width =0.6 cm},
    greennode/.style ={ circle, draw, pattern=north east lines, pattern color=tangocolordarkchameleon!40,inner sep=-0.5pt, minimum width =0.6 cm},
    rednode/.style ={ circle, draw, pattern=north east lines, pattern color=red!40,inner sep=-0.5pt, minimum width =0.6 cm},
    fail/.style={rectangle,draw,red,inner sep=0.5pt,minimum width=0.6cm},
    newred/.style ={ circle, draw,blue, pattern=north east lines, pattern color=red!40,inner sep=-0.5pt, minimum width =0.6 cm}
]

\node[newnode] (nv) at (7,0) {$\varepsilon$};
\node[newnode] (nv0) at (5,-1) {$0$};
\node[newnode] (nv1) at (9,-1) {$1$};
\node[newnode] (nv00) at (4,-2) {$00$};
\node[newnode] (nv01) at (6,-2) {$01$};
\node[newnode] (nv10) at (8,-2) {$10$};
\node[newnode] (nv11) at (10,-2) {$11$};
\node[newnode] (nv000) at (3.5,-3) {$000$};
\node[newnode] (nv001) at (4.5,-3) {$001$};
\node[newnode] (nv010) at (5.5,-3) {$010$};
\node[newnode] (nv011) at (6.5,-3) {$011$};
\node[newnode] (nv100) at (7.5,-3) {$100$};
\node[newnode] (nv101) at (8.5,-3) {$101$};
\node[newnode] (nv110) at (9.5,-3) {$110$};
\node[newnode] (nv111) at (10.5,-3) {$111$};

\path[draw,-{Latex[length=1.5mm]}] (nv0) to (nv);
\path[draw,-{Latex[length=1.5mm]}] (nv1) to (nv);
\path[draw,-{Latex[length=1.5mm]}] (nv00) to (nv0);
\path[draw,-{Latex[length=1.5mm]}] (nv01) to (nv0);
\path[draw,-{Latex[length=1.5mm]}] (nv10) to (nv1);
\path[draw,-{Latex[length=1.5mm]}] (nv11) to (nv1);
\path[draw,-{Latex[length=1.5mm]}] (nv000) to (nv00);
\path[draw,-{Latex[length=1.5mm]}] (nv001) to (nv00);
\path[draw,-{Latex[length=1.5mm]}] (nv010) to (nv01);
\path[draw,-{Latex[length=1.5mm]}] (nv011) to (nv01);
\path[draw,-{Latex[length=1.5mm]}] (nv100) to (nv10);
\path[draw,-{Latex[length=1.5mm]}] (nv101) to (nv10);
\path[draw,-{Latex[length=1.5mm]}] (nv110) to (nv11);
\path[draw,-{Latex[length=1.5mm]}] (nv111) to (nv11);
\path[draw,-{Latex[length=1.5mm]}] (nv000) to (nv001);
\path[draw,-{Latex[length=1.5mm]}] (nv010) to (nv011);
\path[draw,-{Latex[length=1.5mm]}] (nv100) to (nv101);
\path[draw,-{Latex[length=1.5mm]}] (nv110) to (nv111);
\path[draw,-{Latex[length=1.5mm]}] (nv00) to (nv010);
\path[draw,-{Latex[length=1.5mm]}] (nv00) to (nv011);
\path[draw,-{Latex[length=1.5mm]}] (nv10) to (nv110);
\path[draw,-{Latex[length=1.5mm]}] (nv10) to (nv111);
\path[draw,-{Latex[length=1.5mm]}] (nv0) to (nv100);
\path[draw,-{Latex[length=1.5mm]}] (nv0) to (nv101);
\path[draw,-{Latex[length=1.5mm]}] (nv0) to (nv110);
\path[draw,-{Latex[length=1.5mm]}] (nv0) to (nv111);

\end{tikzpicture}

%% file: missing-proofs.tex
\section{Missing Proofs}
\applab{app:missing}

\lemPPi*
\begin{proof}
We first note that 
\[\scphso_{i+1}=\swap_{1,i+1}\circ\stddecomp\circ\cphso'\circ\stddecomp\circ\increase\circ\swap_{1,i+1},\]
where every operation is unitary with the exception of the operator $\increase$. Thus,
\[\norm{\scphso_{i+1}(P_i\qvec{\psi_i})}_2=\norm{\swap_{1,i+1}\circ\stddecomp\circ\cphso'\circ\stddecomp(\qvec{Y_i})}_2,\]
where $\qvec{Y_i}:=\increase(\qvec{X_i})$ and $\qvec{X_i}:=\swap_{1,i+1}(P_i\qvec{\psi_i})$. Since $\swap_{1,i+1}\circ\stddecomp\circ\cphso'\circ\stddecomp$ is unitary, we have 
\[\norm{\swap_{1,i+1}\circ\stddecomp\circ\cphso'\circ\stddecomp(\qvec{Y_i})}_2\leq\norm{\qvec{Y_i}}_2,\] 
and since $\swap_{1,i+1}$ is unitary, we have $\norm{\qvec{X_i}}_2=\norm{\swap_{1,i+1}(P_i\qvec{\psi_i})}_2\leq\norm{P_i\qvec{\psi_i}}_2$. Finally, we can observe that the operator $\increase$ (though not unitary) preserves norms, i.e., for any $\qvec{Z}=\sum_{\vec{x},\vec{y},z,\D}\alpha_{\vec{x},\vec{y},z,\D}\qvec{\vec{x},\vec{y},z}\otimes\qvec{\D}$, we have
\[ \increase(\qvec{Z}) = \sum_{\vec{x},\vec{y},z,\D}\alpha_{\vec{x},\vec{y},z,\D}\qvec{\vec{x},\vec{y},z}\otimes\qvec{\D}\qvec{\bot,0^\strlen}\text{, and}\]  
\[\norm{\increase(\qvec{Z})}_2^2 = \sum_{\vec{x},\vec{y},z,\D}\alpha_{\vec{x},\vec{y},z,\D}^2 = \norm{\qvec{Z}}_2^2 \ .   \]
Plugging in $\qvec{Z}=\qvec{X_i}$ we have $\norm{\qvec{Y_i}}_2=\norm{\increase(\qvec{X_i})}_2 = \norm{\qvec{X_i}}_2$. Taken together, we have \[ \norm{P\circ\scphso_{i+1}(P_i\qvec{\psi_i})}_2 \le \norm{\scphso_{i+1}(P_i\qvec{\psi_i})}_2  \le \norm{\qvec{Y_i}}_2=\norm{\qvec{X_i}}_2 \le \norm{P_i\qvec{\psi_i}}_2 \ . \]
\end{proof}

\lemPQi*
\begin{proof}
Recall that $P$ is the projection onto the span of basis states in $\sbad_{s,i+1}$. 
Note that $\scphso_{i+1}$ maps any basis states in the support of $Q_i\qvec{\psi_i}$ to \[|(x_1,y_1),\ldots,(x_k,y_k),z\rangle\otimes\sum_w 2^{-\lambda/2}(-1)^{w\cdot y_{i+1}}|\D \cup (x_{i+1},w)\rangle.\] 
We thus consider a classical counting argument to analyze the number of strings $w$ such that the states above are in $\sbad_{s,i+1}$. 
Recall that
\[\bad_{s,i+1}(x_1,\ldots,x_k):=\path_{s,i+1}(x_1,\ldots,x_k)\cup\bigcup_{j=1}^{i+1}\contain_{s,j}(x_1,\ldots,x_k).\]
Thus, we decompose the databases in $\bad_{s,i+1}(x_1,\ldots,x_k)$ into the databases in $\path_{s,i+1}(x_1,\allowbreak\ldots,x_k)$ and the databases in $\bigcup_{j=1}^{i+1}\contain_{s,j}(x_1,\ldots,x_k)$ and count the number of such strings $w$ separately as follows.
\begin{itemize}
\item
We first consider the databases that have a path of length $s$ that includes but does not end at $v_{x_{i+1}}$, i.e., $\D\cup(x_{i+1},w)\in\path_{s,i+1}(x_1,\ldots,x_k)$. Intuitively, since $\D\not\in\bad_{s,i}(x_1,\ldots,x_k)$ (due to the definition of $Q_i$), the only way to have $\D\cup(x_{i+1},w)\in\path_{s,i+1}$ is if $w$ is a substring of $x_{i+1}$ or $w$ is a substring of some other input $x$ in the database.
For a particular string $x\in \{0,1\}^{\delta \strlen}$ (e.g., any random oracle input s.t. $\D(x) \neq \bot$) , $x$ contains {\em at most} $\delta\strlen$ unique contiguous substrings of length $\strlen$, so there are at most $\delta\strlen$ values of $w$ such that $\substring(w,x)=1$.  
Since $|\D\cup(x_{i+1},w)\rangle$ consists of databases with at most $q$ entries, then by a union bound, there are at most $q\delta\strlen$ such $w$ in total. 
\item
We now bound the number of databases in $\bigcup_{j=1}^{i+1}\contain_{s,j}(x_1,\ldots,x_k)$ by noting that if $\D \not \in \bad_{s,i}(x_1,\ldots,x_k)$ then the only way for $\D \cup (x_{i+1}, w)$ to be in  $\bigcup_{j=1}^{i+1}\contain_{s,j}(x_1,\ldots\allowbreak,x_k)$ is if for some $j \leq k$ the string $w$ is a substring of $x_j$ i.e., $\substring(w,x_j)=1$. 
Recall that the queries $x_1,\ldots,x_{i+1}$ are parallel, so that we can consider $x_1,\ldots,x_{i+1}$ independently, which would not be the case if the queries were sequential.   
For a fixed $j$, there are at most $\delta\strlen$ unique substrings of length $\strlen$ in $x_j$. 
Thus taking a union bound over all indices $j\in[i+1]$, there are at most $k\delta\strlen$ values of $w$ such that $\substring(w,x_j)=1$.  
\end{itemize}
\vskip 0.5em
Hence, we have \[\norm{P\circ\scphso_{i+1}(Q_i\qvec{\psi_i})}^2_2\le\frac{q\delta\strlen+k\delta\strlen}{2^{\strlen}}.\]
\renewcommand{\Box}{}
\end{proof}

\lemPRi*
\begin{proof}
From algebraic manipulation similar to \cite{C:Zhandry19}, we have that if $\D'$ is the database $\D$ with $x_i$ removed, then $\scphso_{i+1}(|x,y,z\rangle\otimes|\D'\cup(x_{i+1},w)\rangle)$ can be written as 
\begin{align*}
|(x_1,y_1),\ldots,&(x_{k},y_k),z\rangle\otimes\Big((-1)^{y_{i+1}\cdot w}\left(|\D' \cup (x_{i+1},w)\rangle + \frac{1}{2^{\strlen/2}}|\D'\rangle\right) \\
&+ \frac{1}{2^\strlen}\sum_{w'}(1-(-1)^{y_{i+1}\cdot w}-(-1)^{y_{i+1}\cdot w'})|\D'\cup(x_{i+1},w')\rangle\Big).
\end{align*}
Observe that since $\D=|\D' \cup (x_{i+1},w)\rangle \not \in \bad_{s,i}(x_1,\ldots,x_k)$ then we have $\D \not \in \bad_{s,i+1}(x_1,\allowbreak\ldots,x_k)$ by \lemref{lem:bad}. 
Similarly, it follows that $\D' \not \in \bad_{s,i+1}(x_1,\ldots,x_k)$. 
Thus, we can simplify our above equation after applying the projection $P$:

\begin{align*}
P &\circ
\scphso_{i+1} (|x,y,z\rangle\otimes|\D'\cup(x_{i+1},w)\rangle) =  
|(x_1,y_1),\ldots,(x_{k},y_k),z\rangle\otimes \\& \Big( \frac{1}{2^\strlen}\sum_{\substack{w':\D'\cup(x_{i+1},w')\rangle \in \\ \bad_{s,i+1}(x_1,\ldots,x_k)}}(1-(-1)^{y_{i+1}\cdot w}-(-1)^{y_{i+1}\cdot w'})|\D'\cup(x_{i+1},w')\rangle\Big) \ .
\end{align*}

\noindent It is helpful to write \[R_i \qvec{\psi_i} = \sum_{\vec{x},\vec{y},z, \D' ,w} \alpha_{\vec{x},\vec{y},z, \D' ,w} |(x_1,y_1),\ldots, (x_k,y_k),z\rangle  \otimes |\D' \cup (x_{i+1},w) \rangle \] where we use $\vec{x} = (x_1,...,x_k)$ and $\vec{y} = (y_1,\ldots, y_k)$ to simplify notation in the subscripts. 
Thus we have that $\norm{P\circ\scphso_{i+1}(R_i\qvec{\psi_i})}_2^2$ can be upper bounded by:
\begin{align*}
\frac{1}{4^\strlen}\sum_{\vec{x},\vec{y},z,\D'} \sum_{\substack{w': \D' \cup (x_{i+1},w')\in\\ \bad_{s,i+1}(x_1,\ldots,x_k)}}\norm{\sum_w\alpha_{\vec{x},\vec{y},z,\D',w} (1-(-1)^{y_{i+1}\cdot w}-(-1)^{y_{i+1}\cdot w'})}_2^2,
\end{align*}
since changing either $\vec{x},\vec{y},z,\D'$ or $w'$ results in a distinct basis state. 
Now we once again use a classical counting argument to upper bound the number of strings $w'$ such that $\D'\cup(x_{i+1},w')\in\bad_{s,i+1}(x_1,\ldots,x_k)$. Similar to the argument from \lemref{lem:PQi}, we decompose the databases in $\bad_{s,i+1}(x_1,\ldots,x_k)$ into the databases in $\path_{s,i+1}(x_1,\ldots,x_k)$ and in $\bigcup_{j=1}^{i+1}\contain_{s,j}(x_1,\ldots,x_k)$ and count the number of such strings $w'$ separately as follows.

\begin{itemize}
\item We first bound the databases $\D'$ that have a path of length $s$ that does not end at $w'$, i.e., $\D'\cup(x_{i+1},w')\in\path_{s,i+1}(x_1,\ldots,x_k)$. Intuitively, since $\D'\not\in\bad_{s,i}(x_1,\ldots,x_k)$ (due to the definition of $R_i$), the only way to have $\D'\cup(x_{i+1},w')\in\path_{s,i+1}$ is if $w'$ is a substring of $x_{i+1}$ or $w'$ is a substring of some other input $x$ in the database. 
Any fixed $x\in\D'$ contains at most $\delta\strlen$ unique contiguous substrings of length $\strlen$, so there are at most $\delta\strlen$ values of $w'$ such that $\substring(w',x)=1$. 
Taking a union bound over at most $q$ possible values of $x\in\D'$, there are at most $q\delta\strlen$ such $w'$ in total. 
\item We next bound the databases $\D'$ in $\bigcup_{j=1}^{i+1}\contain_{s,j}(x_1,\ldots,x_k)$ by noting that if we have that $\D' \not \in \bad_{s,i}(x_1,\ldots,x_k)$, then the only way for $\D' \cup (x_{i+1}, w')$ to be in $\bigcup_{j=1}^{i+1}\contain_{s,j}(x_1,\ldots,x_k)$ is if for some $j \leq k$ the string $w'$ is a substring of $x_j$ i.e., $\substring(w',x_j)=1$. 
For a fixed $j$, there are at most $\delta\strlen$ unique contiguous substrings of length $\strlen$ in $x_j$. 
Taking a union bound over at most $i\le k$ indices $j$, there are at most $k\delta\strlen$ values $w'$ such that $\substring(w',x_j)=1$. 
\end{itemize}
\vskip 0.5em
Hence, by Cauchy-Schwarz inequality on the sum of $w$, we have
\begin{align*}
\norm{P\circ\scphso_{i+1}(R_i\qvec{\psi_i})}_2^2&\le\frac{1}{4^\strlen}\sum_{x,y,z,\D'}(q\delta\strlen+k\delta\strlen)\norm{\sum_w\alpha_{x,y,z,\D',w}\cdot 3}_2^2\\
&\le\frac{(q\delta\strlen+k\delta\strlen)}{4^\strlen}\sum_{x,y,z,\D',w}2^\strlen\norm{\alpha_{x,y,z,\D',w}\cdot 3}_2^2\\
&= \frac{9(q\delta\strlen+k\delta\strlen)}{2^{\strlen}}\sum_{x,y,z,\D',w}\norm{\alpha_{x,y,z,\D',w}}_2^2\\
&\leq \frac{9(q\delta\strlen+k\delta\strlen)}{2^{\strlen}}.
\end{align*}
\end{proof}

\claimpathbad*
\begin{proof}
Consider any basis state  $|(x_1,y_1),\ldots,(x_k,y_k),z\rangle\otimes|\D \rangle \in {\spath_{s+1}}$. Then $G_\D$ contains a path $P$ of length $s+1$. If $P$ does not end at any of the nodes $v_{x_i}$ for all $1\leq i\leq k$, then $\D \in \path_{s,k}(x_1,\ldots, x_k)$ and it immediately follows that $|(x_1,y_1),\ldots,(x_k,y_k),z\rangle\otimes|\D \rangle \in \sbad_{s,k}$. On the other hand, if $P$ ends at a node $v_{x_i}$ for some $i$, then consider a subpath $P'$ of $P$ excluding the node $v_{x_i}$ and an incident edge from $P$. Then it is clear that the length of $P'$ is $s$. If $P'$ does not end at any of the nodes $v_{x_j}$ for all $1\leq j\leq k$, then repeating the same argument, $\D \in \path_{s,k}(x_1,\ldots, x_k)$ and it immediately follows that $|(x_1,y_1),\ldots,(x_k,y_k),z\rangle\otimes|\D \rangle \in \sbad_{s,k}$. Otherwise, $P'$ should end at some node $v_{x_j}$ for some $1\leq j\leq k$ with $j\ne i$. Then we have that $\substring(\D(x_j),x_i)=1$ and $G_\D$ contains a path $P'$ of length $s$ ending at $v_{x_j}$, which implies that $\D\in\contain_{s,j}(x_1,\ldots,x_k)$. Thus, it follows that $|(x_1,y_1),\ldots,(x_k,y_k),z\rangle\otimes|\D \rangle \in \sbad_{s,k}$. Taken together, in any case we have $|(x_1,y_1),\ldots,(x_k,y_k),z\rangle\otimes|\D \rangle \in \sbad_{s,k}$, which implies that $\spath_{s+1}\subseteq \sbad_{s,k}$.
\end{proof}

\lemlucky*
\begin{proof}
\deleted{The following has been moved from the main body. From here:}
Similar to \lemref{lem:step:negl}, we consider the projection $\qvec{\phi'}=\cphso\qvec{\phi}$ onto orthogonal spaces as follows:
\begin{itemize}
\item We first define $P$ to be the projection onto the span of basis states $\qvec{x,y,z}\otimes\qvec{\D}\in\slucky_s$.
\item Next we define $Q$ to be the projection onto states $\qvec{x,y,z}\otimes\qvec{\D}$ such that $\qvec{x,y,z}\otimes\qvec{\D}\not\in\slucky_s$, $y\ne 0$, and $\D(x)=\bot$. Intuitively, $Q$ represents the projection on states with potentially lucky databases where $\cphso$ will affect $\D$ and the value of $x$ has not been specified in $\D$.
\item We then define $R$ to be the projection onto states $\qvec{x,y,z}\otimes\qvec{\D}$ such that $\qvec{x,y,z}\otimes\qvec{\D}\not\in\slucky_s$, $y\ne 0$, and $\D(x)\ne\bot$, so that the value of $x$ has been specified in databases corresponding to these states.
\item Finally, we define $S$ to be the projection onto states $\qvec{x,y,z}\otimes\qvec{\D}$ such that $\qvec{x,y,z}\otimes\qvec{\D}\not\in\slucky_s$ and $y=0$.
\end{itemize}
Since $P,Q,R,S$ project onto disjoint states that span the entirety of $\qvec{\phi'}$ then we have $P+Q+R+S=\mathbb{I}$, where $\mathbb{I}$ denotes the identity operator. We analyze how $P$ acts on these components separately. For $\cphso (P\qvec{\phi})$  it is easy to verify that $\norm{P\circ\cphso(P\qvec{\phi})}_2 \leq \norm{\cphso(P\qvec{\phi})}_2\leq \norm{P\qvec{\phi}}_2 $. \deleted{See Appendix E for the formal proof of Lemma 32}.
\begin{restatable}{lemma}{lemPP}
\lemlab{lem:PP}
$\norm{P\circ\cphso(P\qvec{\phi})}_2\leq \norm{P\qvec{\phi}}_2$.
\end{restatable}
To analyze how the projection $P$ acts on $\cphso  (Q \qvec{\phi})$, we note that $\cphso( \qvec{x,y,z}\otimes\qvec{\D}) = \qvec{x,y,z}\otimes 2^{-\lambda/2} \sum_{w} (-1)^{y \cdot w}  \qvec{\D \cup (x,w)}$ for any basis state in the support of $Q\qvec{\psi}$. 
Then we observe that the only way that the database $\D\cup\{(x,w)\}$ is in $\lucky_s$ is either (1) $x$ parses into $x=\chi\| (N+1)\| \ell_\varepsilon$ for an arbitrary string $\chi$ and $\ell_{\varepsilon} \in \{0,1\}^\lambda$ and the response $w$ is a lucky for the Merkle tree $\colmt_\D(\chi,\ell_{\varepsilon})$ rooted at $\ell_{\varepsilon}$, in which case $w\in\lucky(\D,\chi,\ell_{\varepsilon})$, or (2) $x$ parses into $x = \chi \| v \| r$ for some $v \leq N$ and $r \in \{0,1\}^{\leq n \lambda}$ and the new pair $(x,w)$ extends some prior incomplete Merkle tree of the form $\colmt_\D(\chi,\ell_{\varepsilon})$ in which case $w\in\pre(\D)$. We then bound the size of the set $\lucky(\D,\chi,\ell_\varepsilon)$ and $\pre(\D)$ for each case by applying \corref{cor:greenchal} and take the maximum to bound the value of $\norm{P \circ\cphso (Q \qvec{\phi})}_2^2$. 
We thus show the following\deleted{ in Appendix E}:
\begin{restatable}{lemma}{lemPQ}
\lemlab{lem:PQ}
$\norm{P \circ\cphso (Q \qvec{\phi})}_2^2  \leq (1-\alpha)^{\lfloor \lambda/n \rfloor }$.
\end{restatable}
\noindent We next consider how $P$ acts upon the basis states of $\cphso  (R \qvec{\phi})$.
The algebraic manipulation is similar to \cite{C:Zhandry19} and \lemref{lem:PRi} though we also depend on \corref{cor:greenchal} when upper bounding the norm\replaced{.}{;} \deleted{see Appendix E for the formal proof.}
\begin{restatable}{lemma}{lemPR}
\lemlab{lem:PR}
$\norm{P\circ\cphso(R\qvec{\phi})}_2^2\leq 9(1-\alpha)^{\lfloor \strlen/n\rfloor}.$
\end{restatable}
\noindent \added{The formal proof of} \lemref{lem:PP}, \lemref{lem:PQ}, \added{and} \lemref{lem:PR} \added{can be found below this proof.} Finally, we bound the projection of $P$ onto the states of $\cphso(S\qvec{\psi})$:

\begin{lemma}\lemlab{lem:PS}
$\norm{P\circ\cphso(S\qvec{\phi})}_2=0$.
\end{lemma}
\begin{proof}
We observe that for any basis state $\qvec{x,y,z}\otimes\qvec{\D}$ in the support of $S \qvec{\phi}$ we have $\cphso (\qvec{x,y,z}\otimes\qvec{\D}) =  \qvec{x,y,z}\otimes\qvec{\D} $ where $ \qvec{x,y,z}\otimes\qvec{\D} \not \in \slucky_s$. Thus, we have that $\norm{P\circ\cphso(S\qvec{\phi})}_2=0$.
\end{proof}
Thus from \lemref{lem:PP}, \lemref{lem:PQ}, \lemref{lem:PR}, \lemref{lem:PS}, and by triangle inequality, we have
\begin{align*}
\norm{P\circ\cphso\qvec{\phi}}_2 &\leq \norm{P\circ\cphso(P\qvec{\phi})}_2 + \norm{P\circ\cphso(Q\qvec{\phi})}_2\\
&\qquad+ \norm{P\circ\cphso(R\qvec{\phi})}_2 + \norm{P\circ\cphso(S\qvec{\phi})}_2 \\
&\leq \norm{P\qvec{\phi}}_2 + (1-\alpha)^{\frac{\lfloor \lambda/n \rfloor}{2} } + 3(1-\alpha)^{\frac{\lfloor \strlen/n\rfloor}{2}} \\
&\leq \moment(\qvec{\phi},\slucky_s) +  4(1-\alpha)^{\frac{\lfloor \lambda/n \rfloor}{2} }.
\end{align*}
Since we have that $\norm{P\circ\cphso\qvec{\phi}}_2 = \moment(\cphso\qvec{\phi},\slucky_s)$, we complete the proof.\deleted{To here.}
\end{proof}

\lemPP*
\begin{proof}
We first note that 
\[\cphso=\stddecomp\circ\cphso'\circ\stddecomp\circ\increase,\]
where every operation is unitary with the exception of the operator $\increase$. Thus,
\[\norm{\cphso(P\qvec{\phi})}_2=\norm{\stddecomp\circ\cphso'\circ\stddecomp(\qvec{X})}_2,\]
where $\qvec{X}:=\increase(P\qvec{\phi})$. Since $\stddecomp\circ\cphso'\circ\stddecomp$ is unitary, we have 
\[\norm{\stddecomp\circ\cphso'\circ\stddecomp(\qvec{X})}_2\leq\norm{\qvec{X}}_2.\]
Finally, we can observe that the operator $\increase$ (though not unitary) preserves norms, i.e., for any $\qvec{Y}=\sum_{\vec{x},\vec{y},z,\D}\alpha_{\vec{x},\vec{y},z,\D}\qvec{\vec{x},\vec{y},z}\otimes\qvec{\D}$, we have
\[ \increase(\qvec{Y}) = \sum_{\vec{x},\vec{y},z,\D}\alpha_{\vec{x},\vec{y},z,\D}\qvec{\vec{x},\vec{y},z}\otimes\qvec{\D}\qvec{\bot,0^\strlen} \mbox{  , and  }\]\[ \norm{\increase(\qvec{Y})}_2^2 = \sum_{\vec{x},\vec{y},z,\D}\alpha_{\vec{x},\vec{y},z,\D}^2 = \norm{\qvec{Y}}_2^2 \ .  \]
Plugging in $\qvec{Y}= P \qvec{\phi}$ we have $\norm{\qvec{X}}_2=\norm{\increase(P\qvec{\phi})}_2 = \norm{P\qvec{\phi}}_2$. Taken together, we have \[\norm{P\circ\cphso(P\qvec{\phi})}_2 \le \norm{\cphso(P\qvec{\phi})}_2  \le\norm{\qvec{X}}_2=\norm{\increase(P\qvec{\phi})}_2 = \norm{P\qvec{\phi}}_2 \ .\]
\end{proof}

\lemPQ*
\begin{proof}
If $y\ne 0$, $\D(x)=\bot$, and  $\qvec{x,y,z}\otimes\qvec{\D}\not\in\slucky_s$ then we have
\[ \cphso( \qvec{x,y,z}\otimes\qvec{\D}) = \qvec{x,y,z}\otimes 2^{-\lambda/2} \sum_{w} (-1)^{y \cdot w}  \qvec{\D \cup (x,w)} \ . \]

Fixing $\D$ such that $\D(x)=\bot$ and $\D \not \in \lucky_s$ we now focus on upper bounding the number of strings $w$ such that $\D\cup\{(x,w)\} \in \lucky_s$.  Since $\D \not \in \lucky_s$
we observe that the only way that the database $\D\cup\{(x,w)\}$ is in $\lucky_s$ is if one of the following cases hold:
\begin{itemize}
\item $x$ parses into $x = \chi \| (N+1) \| \ell_{\varepsilon}$ for an arbitrary string $\chi$ and $\ell_{\varepsilon} \in \{0,1\}^\lambda$ and the response $w$ is a lucky for the Merkle tree $\colmt_\D(\chi,\ell_{\varepsilon})$ rooted at $\ell_{\varepsilon}$, in which case $w\in\lucky(\D,\chi,\ell_{\varepsilon})$, or
\item  $x$ parses into $x = \chi \| v \| r$ for some $v \leq N$ and $r \in \{0,1\}^{\leq n \lambda}$ and  the new pair $(x,w)$ extends some prior incomplete Merkle tree of the form  $\colmt_\D(\chi,\ell_{\varepsilon})$. In this case we must have $w\in\pre(\D)$ since the  $\D(\chi \| (N+1) \| \ell_{\varepsilon})$ must already exist for the Merkle tree $\colmt_\D(\chi,\ell_{\varepsilon})$  to be lucky.
\end{itemize}
Note that if $x$ does not parse properly then for all responses $\D\cup\{(x,w)\} \not \in \lucky_s$ so for each $\D$ have \[ \left| \left\{w: \D\cup\{(x,w)\} \in \lucky_s  \right\}  \right|  = \max\left\{ \left| \lucky(\D,\chi,\ell_{\varepsilon})\right| ,   \left|\pre(\D) \right| \right\} \ .\] Applying \corref{cor:greenchal}  with $k=\lfloor \strlen/n\rfloor$ and $k'=\strlen - nk$ we have
\[ |\lucky(\D,\chi,\ell_\varepsilon)| \leq 2^\strlen(1-\alpha)^{\lfloor \strlen/n\rfloor}. \]  
Similarly, we observe  that $\left|\pre(\D) \right| \leq q \strlen (n+1)$ since $\D$ has at most $q$ entries. To see this note that $\D$ has at most $q$ entries of the form $(x,w)$ and each string $x$ has length at most $(n+1) \strlen$. Thus, for each $x$ there are at most $\tau-\strlen$ consecutive substrings of length $\strlen$. Therefore, for any $\D \not \in \lucky_s$ with at most $q$ entries we have \[  \left| \left\{w: \D\cup\{(x,w)\} \in \lucky_s  \right\}  \right|  \leq  \max\left\{ 2^\strlen(1-\alpha)^{\lfloor \strlen/n\rfloor},   q \strlen (n+1) \right\} \leq  2^\strlen(1-\alpha)^{\lfloor \strlen/n\rfloor} \ . \]
For convenience let $W_{\D,x} := \{ w: \D\cup\{(x,w)\} \in \lucky_s\}$.
 Let \[Q \qvec{\phi} = \sum_{x,y,z,\D} \alpha_{x,y,z,\D} \qvec{x,y,z} \otimes \qvec{\D},\] then  \[ P\circ\cphso( Q \qvec{\phi}) =  \sum_{x,y,z,\D} \alpha_{x,y,z,\D} \qvec{x,y,z} \otimes 2^{-\lambda/2} \sum_{w \in W_{\D,x}}  (-1)^{y \cdot w} \qvec{\D} \ . \] Thus,   
\begin{align*}
\norm{P \circ\cphso (Q \qvec{\phi})}_2^2  &\leq \sum_{x,y,z,\D} \alpha_{x,y,z,\D}^2 2^{-\strlen} |W_{\D,x}|  \\
& \leq (1-\alpha)^{\lfloor \lambda/n \rfloor } \sum_{x,y,z,\D} \alpha_{x,y,z,\D}^2 \\
& \leq (1-\alpha)^{\lfloor \lambda/n \rfloor }  \ .
\end{align*}
\end{proof}

\lemPR*
\begin{proof}
Similar to the algebraic manipulation from \cite{C:Zhandry19}, we have that if $\D'$ is the database $\D$ with $x$ removed, then $\cphso(\qvec{x,y,z}\otimes\qvec{\D'\cup (x,w)})$ can be written as
\begin{align*}
\qvec{x,y,z}\otimes\Big((-1)^{y\cdot w}&\left( \qvec{\D'\cup(x,w)}+2^{-\strlen/2}\qvec{\D'} \right)\\
&+ 2^{-\strlen} \sum_{w'} (1-(-1)^{y\cdot w}-(-1)^{y\cdot w'})\qvec{\D'\cup(x,w')}\Big).
\end{align*}
Observe that $\D=\D'\cup\{(x,w)\}\not\in\lucky_s$ implies $\D'\not\in\lucky_s$.  
Thus, we can simplify our above equation after applying the projection $P$:
\begin{align*}
P\circ&\cphso(\qvec{x,y,z}\otimes\qvec{\D'\cup(x,w)})\\
&= \qvec{x,y,z}\otimes \left( 2^{-\strlen} \sum_{\substack{w': \D'\cup\{(x,w')\}\in \\ \lucky_s}} (1-(-1)^{y\cdot w}-(-1)^{y\cdot w'})\qvec{\D'\cup(x,w')} \right).
\end{align*}
If we write 
\[R\qvec{\phi}=\sum_{x,y,z,\D',w} \alpha_{x,y,z,\D',w} \qvec{x,y,z}\otimes \qvec{\D'\cup(x,w)},\]
then $\norm{P\circ\cphso(R\qvec{\phi})}_2^2$ can be upper bounded by:
\[ \frac{1}{4^\strlen}\sum_{x,y,z,\D'}\sum_{\substack{w': \D'\cup\{(x,w')\}\in \\ \lucky_s}} \norm{\sum_w \alpha_{x,y,z,\D',w} (1-(-1)^{y\cdot w}-(-1)^{y\cdot w'})} _2^2, \]
since changing either $x,y,z,\D'$ or $w'$ results in a distinct basis state. 
Now we once again use a classical counting argument to upper bound the number of strings $w'$ such that $\D'\cup\{(x,w')\}\in\lucky_s$ which will be the same as (2). We can once again argue that for any $\D'$ this quantity is upper bounded by \[ \max\left\{ \left| \pre(\D')\right| , \max_{\chi, \ell_{\varepsilon}} \left|\lucky(\D',\chi,\ell_{\varepsilon}) \right| \right\} \leq 2^{\strlen}(1-\alpha)^{\lfloor \strlen/n\rfloor} \ . \]
Hence, the number of such strings $w'$ is upper bounded by $2^{\strlen}(1-\alpha)^{\lfloor \strlen/n\rfloor}$ and by Cauchy-Schwarz inequality on the sum of $w$ we have that
\begin{align*}
\norm{P\circ\cphso(R\qvec{\phi})}_2^2 &\leq \frac{1}{4^\strlen} \sum_{x,y,z,\D'} 2^{\strlen}(1-\alpha)^{\lfloor \strlen/n\rfloor}\norm{\sum_w \alpha_{x,y,z,D',w} \cdot 3  }_2^2 \\
&\leq \frac{2^{\strlen}(1-\alpha)^{\lfloor \strlen/n\rfloor}}{4^\strlen} \sum_{x,y,z,\D',w} 2^\strlen \norm{\alpha_{x,y,z,D',w} \cdot 3}_2^2 \\
&= 9(1-\alpha)^{\lfloor \strlen/n\rfloor} \sum_{x,y,z,\D',w} \norm{\alpha_{x,y,z,D',w} }_2^2 \\
&\leq 9(1-\alpha)^{\lfloor \strlen/n\rfloor}.
\end{align*}
\end{proof}

\lemreveal*
\begin{proof}
\added{Let $\qvec{\psi_0}$ be the initial state and let $U_r$ represent a unitary transform applied by $\A$ in between each query to the quantum oracle. Then we define $\qvec{\psi_r}= U_r \circ \cphso\qvec{\psi_{r-1}}$ for each round $r \in[N-1]$. 
Thus, the attacker $\A$ yields a sequence of states $\qvec{\psi_0},\ldots, \qvec{\psi_{N-1}}$.  
We remark that $U_r$ may only operate on the  registers $|x,y,z\rangle$ and cannot impact the compressed oracle $\D$, e.g., $U_r \left( |x',y',z'\rangle \otimes | \D \rangle \right) = \sum_{x,y,z} \alpha_{x,y,z} |x,y,z\rangle \otimes |\D\rangle$. 
Thus, 
\[\moment(U_r \circ \cphso\qvec{\psi_{r-1}}, \slucky_{s}) = \moment(\cphso\qvec{\psi_{r-1}}, \slucky_{s}),\] 
so we can effectively ignore the intermediate unitary transform $U_r$ in our analysis below. Now we can apply the previous lemma to conclude that 
  \[ \moment(\qvec{\psi_{i}}, \slucky_{s}) \leq  4(1-\alpha)^{\frac{\lfloor \lambda/n \rfloor}{2} }+  \moment(\qvec{\psi_{i-1}}, \slucky_s)  \ . \] 
By the triangle inequality we have  \[ \moment(\qvec{\psi_{N-1}}, \slucky_{s}) \leq 4q(1-\alpha)^{\frac{\lfloor \lambda/n \rfloor}{2} }  \ .\]
Hence, the probability $p'$ of measuring a database $\D \in \lucky_s$ for $s=N(1-\alpha)$ is at most $\left(4q(1-\alpha)^{\frac{\lfloor \lambda/n \rfloor}{2} }\right) ^2 = 16q^2(1-\alpha)^{\lfloor \lambda/n \rfloor}$. }
\end{proof}

%% file: usefulbounds.tex
\section{Useful Bounds}
\applab{sec:bounds}
In the classical parallel random oracle model (\pROM), running time is measured in terms of the number of rounds of random oracle queries~\cite{STOC:AlwSer15}, e.g., a round $i$ of computation ends when the attacker $\mathcal{A}^{\H(\cdot)}$ outputs a list $Q_i = \left(q_1^i,\ldots, q_{k_i}^i\right)$ of random oracle queries and a new initial state $\sigma_{i+1}$ for the next round of computation. 
During the next round of computation, the attacker will be $\mathcal{A}^{\H(\cdot)}$ with the initial state $\sigma_{i+1}$ as well as the answers to the random oracle queries made in the last round $A_i =  \left(\H\left(q_1^i\right),\ldots, \H\left(q_{k_i}^i\right)\right)$. 
In this model, an attacker is allowed to perform arbitrary computation (apart from querying the random oracle) in-between rounds for ``free'', i.e., even if it takes the attacker time $2^{\strlen}$ to compute the next batch of random oracle queries this still only counts as a single round. 
Security proofs in the \pROM tend to be information theoretic in nature, e.g., one can show that any attacker making at most $q$ random oracle queries, each query $q_i$ having length $|q_i| \leq \delta\strlen$, can successfully produce a valid $\H$-sequence of length $s$ in at most $s-1$ rounds with probability at most $\delta\strlen(q^2+ qs)/2^{\strlen}$~\cite{EC:CohPie18}. 
The permissive view that an attacker can perform arbitrary computation for free in between computation rounds is justified because it only makes the lower-bounds stronger.

We now recall two results that bound the Euclidean distance between a sequence of quantum queries with access to different oracles. 
The \emph{total variation distance} between two random variables $p$ and $q$ drawn from a discrete space $\Omega$ with corresponding probability mass functions $0\le p(x),q(x)\le 1$ is defined to be the quantity $\frac{1}{2}\sum_{x\in\Omega}|p(x)-q(x)|$. 

\begin{lemma}[\cite{BennettBBV97} Theorem 3.1]
\lemlab{lem:tvd}
Two unit-length superpositions that are within Euclidean distance $\eps$ will give samples from distributions that are within total variation distance at most $4\eps$ upon observation.
\end{lemma}

Let $|\phi_t\rangle=\sum_{x,y,z} \alpha_{x,y,z} |x,y,z\rangle$ be some state at time $t$. Fixing a query $x'$ we can define $q_{x'}(|\phi_t\rangle) = \sum_{y,z} \alpha_{x',y,z}^2$ as the query magnitude of states on which the query $x'$ is being submitted to the random oracle. 

\begin{lemma}[\cite{BennettBBV97} Theorem 3.3]
\lemlab{lem:dist}
Let $\A_Q$ be a quantum algorithm with runtime $T$ and access to oracle $\o$. 
Let $\eps>0$ and $S\subseteq[1,T]\times\{0,1\}^N$ be a set of time-string pairs with $\sum_{(t,x')\in S} q_{x'}(|\phi_t\rangle)\le\eps$. 
If $\o'$ is an oracle that answers each query $x'$ at time $t$ with $(t,x') \in S$ with an arbitrary string $h_{t,x'}$ (which does not have to be consistent with $\o$), then the Euclidean distance between the final states of $\A_Q$ with access to $\o$ and $\A_Q$ with access to $\o'$ is at most $\sqrt{T\eps}$.
\end{lemma}

Observe the dependency on running time $T$ in \lemref{lem:dist}. 
Whereas classic adversaries may not be able to gain information over time without additional queries to the random oracle, quantum adversaries can repeatedly increase the magnitude of the desired state, such as using Grover's algorithm. 

We require a modification of \lemref{lem:dist} to argue the indistinguishability of the time evolution of two superpositions with small Euclidean distance. 
The proof is similar to that of \lemref{lem:dist} in~\cite{BennettBBV97}.
\begin{lemma}
\lemlab{lem:dist:sets}
Let $|\psi_0\rangle,\ldots,|\psi_{T-1}\rangle$ be a sequence of superpositions such that each $|\psi_i\rangle$ is obtained from applying a unitary time evolution operator $U_i$ on $|\psi_{i-1}\rangle$, i.e., $|\psi_i\rangle=U_i|\psi_{i-1}\rangle$. 
Let $|\phi_0\rangle,\ldots,|\phi_{T-1}\rangle$ be a sequence of superpositions such that $|\phi_0\rangle = |\psi_0\rangle$ and for each $i \geq 1$ we have $|\phi_i\rangle = U_i |\phi_{i-1}\rangle + |E_i\rangle$ where $\norm{|E_i\rangle}_2^2 \le \epsilon$ for all $i < T$, i.e.,  $|E_i\rangle=\sum_x\alpha_{x,i}|x\rangle$ with $\sum_x\alpha_{x,i}^2\le\eps$. 
Then the Euclidean distance between $|\psi_T\rangle$ and $|\phi_T\rangle$ is at most $T\sqrt{\eps}$. 
\end{lemma}
\begin{proof}
We have $|\phi_i\rangle=U_i|\phi_{i-1}\rangle+|E_i\rangle$. 
We can write $|\psi_1\rangle = |\phi_1\rangle + |E_1\rangle$ and $\qvec{\psi_2} = U_2 (\qvec{\phi_1} + \qvec{E_1}) = \qvec{\phi_2} + \qvec{E_2} + U_2  \qvec{E_1}$ and similarly $\qvec{\psi_3} = \qvec{\phi_3} + \qvec{E_3} + U_3 \qvec{E_2} + U_3 \circ U_2 \qvec{E_1}$. 
In general, $\qvec{\psi_{i}} = \qvec{\phi_i} + \sum_{j=1}^{i} (U_{i}\circ \ldots \circ U_{j+1}) \qvec{E_j}$.  
Define $\qvec{\overline{E}_j} = U_{T-1}\circ \ldots \circ U_{j+1}\qvec{E_j}$ so that $\qvec{\psi_{i }}- \qvec{\phi_i} = \sum_{j=1}^{T-1} \qvec{\overline{E}_j}$ and let $\alpha_{x,i}$ the the associated amplitudes for $\qvec{\overline{E}_j}$, i.e., such that $\qvec{\overline{E}_j} = \sum_x \alpha_{x,i} |x\rangle$. 
Applying Cauchy-Schwarz inequality, we now have \[ \norm{\qvec{\psi_{T-1}}-\qvec{\phi_{T-1}}}_2^2 = \sum_x \left( \sum_{i=1}^{T-1} \alpha_{x,i} \right)^2 \leq T \sum_x \sum_{i=1}^{T-1} \alpha_{x,i}^2 \leq T^2 \epsilon. \] 
Thus, the Euclidean distance between the final states is at most $T\sqrt{\eps}$. 
\end{proof}

\section{$\H$-Sequences and Proofs of Sequential Work in the Classical Random Oracle Model}
\seclab{app:hseq}
Cohen and Pietrzak~\cite{EC:CohPie18} provide a construction for proofs of sequential work and showed that any classical attacker that successfully fools a verification algorithm by claiming a false proof of sequential work with non-negligible probability must produce a long $\H$-sequence. 

For example, consider the example of using the following labeling rule to obtain labels for nodes of a directed acyclic graph. 
\begin{definition}[Labeling]
\deflab{def:lab}
Let $\Sigma = \{0,1\}^\strlen$.  
Given a directed acyclic graph $G=(V,E)$, we define the labeling $\ell_v:\{0,1\}^*\rightarrow\Sigma$ of a node $v$ with input $\chi$ by 
\[\ell_v(\chi)=
\begin{cases}
\H(\chi,1),&\indeg(v)=0,\text{ and} \\
\H(\chi,v, \ell_{v_1}(\chi),\cdots,\ell_{v_d}(\chi)),& 0< \indeg(v)=d,
\end{cases}\]
where $v_1,\ldots,v_d$ are the parents of vertex $v$ in $G$, according to some predetermined topological order. 
We omit the dependency on the input $\chi$ if the context is clear. 
\end{definition}
Cohen and Pietrzak~\cite{EC:CohPie18} provide a construction for proofs of sequential work by fixing labels $\ell_1,\ell_2,\ldots,\ell_v$, i.e., via Merkle tree, and then checking the labels for local consistency. 
\begin{definition}[Green/Red node]\deflab{def:color}
For a fixed labeling $\ell_1,\ell_2,\ldots,\ell_v$ and a statement $\chi$, a node $v\in V$ with parents $v_1,\ldots, v_d$ is {\em green} if $\ell_{v}=\H(\chi,v,\ell_{v_1}, \ell_{v_2}, \cdots , \ell_{v_d})$. 
A node that is not green is called a {\em red} node.
\end{definition}
Similar techniques for using green/red nodes to check for local consistency were also used in~\cite{BlockiGGZ19}. 
Note that correctness of the fixed labeling $\ell_i$ for an input $\chi$ is not required, i.e., we do not require $\ell_i=\ell_i(\chi)$. 

The local testing procedure of Cohen and Pietrzak~\cite{EC:CohPie18} rejects with high probability if there is no long path of $\Omega(n)$ green nodes, but accepts if all nodes are green. 
\begin{lemma}[\cite{EC:CohPie18}]
For a fixed labeling $\ell_1,\ell_2,\ldots,\ell_v$, any path of green nodes corresponds to an $\H$-sequence.
\end{lemma}
Hence, a classical adversary that can construct a fictitious proof of sequential work could also construct an $\H$-sequence. 
\cite{EC:CohPie18} also proved that an attacker running in sequential time $T$ cannot produce an $\H$-sequence, except with negligible probability.